\documentclass[12pt]{article} 
\usepackage{cite}
\usepackage{wrapfig}
\usepackage{amssymb}
\usepackage{amsfonts}
\usepackage{authblk}
\usepackage{amsmath}
\usepackage{lineno}
\usepackage{color}
\usepackage{xcolor}
\usepackage{multirow}
\usepackage{longtable}
\usepackage{lscape}
\usepackage[T2A]{fontenc}
\usepackage[utf8x]{inputenc}
\usepackage{hyperref}
\usepackage{bigstrut}

\usepackage{ifpdf}
\ifpdf
  \usepackage[pdftex]{graphicx}
  \usepackage{epstopdf}
\else
  \usepackage{graphicx}
\fi
\DeclareGraphicsExtensions{.png,.pdf,.jpg,.mps,.eps}
\graphicspath{{./figs/}}

\begin{document}

\title{Low-energy interactions of mesons with participation of the first radially excited states in $U(3) \times U(3)$ NJL model}

\author[1]{M. K. Volkov \footnote{volkov@theor.jinr.ru}}
\author[1]{A.A. Pivovarov \footnote{tex$\_$k@mail.ru}}
\author[1,2,3]{K. Nurlan \footnote{nurlan@theor.jinr.ru}}

\affil[1]{\small Bogoliubov Laboratory of Theoretical Physics, JINR, 
                 141980 Dubna, Moscow region, Russia}
\affil[2]{\small The Institute of Nuclear Physics, Almaty, 050032, Kazakhstan}
\affil[3]{\small Al-Farabi Kazakh National University, 
                  Almaty, 050040 Kazakhstan}
                  
\date{}

\maketitle

\abstract{
The $U(3) \times U(3)$ chiral symmetric NJL model describing pseudoscalar, vector, and axial vector mesons in both the ground state and first radially excited states is shortly presented in this review. In this model, it is possible to describe a large number of low-energy interactions of mesons, $\tau$ lepton decays into mesons and processes of meson production in electron-positron annihilations in satisfactory agreement with the experiments. In describing a number of processes, it turned out to be necessary to take into account the interactions of mesons in the final state. 
}

\section{Introduction}
Studies of meson interactions at low energies are of great interest for investigations of both the nature of the intrinsic properties and their interaction with each other. Such studies are carried out from both the theoretical and experimental points of view. Among the experimental centers, we can note world scientific centers such as VEPP-2000 (Budker INP, Novosibirsk), UNK (IHEP, Protvino), BaBar (SLAC, USA ), Belle (KEK, Japan), BES III (BEPC II, China), LEP (CERN), etc. 

For a theoretical description of processes in the interested energy region below $\sim$ 2 GeV, it is impossible unfortunately to use the well-developed QCD perturbation theory due to a large value of the coupling constant. Therefore, as a rule, various phenomenological theories are applied here. These theories are based on the use of characteristic symmetries to which the corresponding interactions of elementary particles conform. One of the main symmetries of this kind is chiral symmetry. These symmetries were widely used even before the construction of the fundamental QCD theory \cite{Sakurai:1960ju, Gell-Mann:1962yej, Weinberg:1966fm, Wess:1967jq, Gell-Mann:1968hlm, Gasiorowicz:1969kn, Coleman:1969sm, Callan:1969sn}. 

   Among various models closely related to internal symmetries of strong interactions, the Nambu -- Jona-Lasinio model, proposed in 1961 \cite{Nambu:1961tp}, can be considered very successful. In the quark language, this model was first formulated in the papers \cite{Eguchi:1976iz, Kikkawa:1976fe}. Later, it was actively developed by many authors since the 80s in such works as  \cite{Ebert:1982pk, Volkov:1984kq, Hatsuda:1984jm, Hatsuda:1985eb, Volkov:1986zb, Ebert:1985kz, Vogl:1991qt, Klevansky:1992qe, Volkov:1993jw, Hatsuda:1994pi, Ebert:1994mf, Buballa:2003qv, Volkov:2005kw}. All these works are quite close to each other and differ mainly in different definitions of the internal parameters. In this review, we will use the version of the NJL model, described in the works \cite{Volkov:1984kq, Volkov:1986zb, Volkov:1993jw, Ebert:1994mf, Volkov:2005kw}.
    
    In the low-energy region an important role in addition to the ground state of mesons, is also played by the first radial excitations. This especially concerns the processes of meson production in colliding $e^+e^-$ beams as well as $\tau$  lepton decays. An important role in the description of these processes is played by channels containing mesons in both the ground and first radially excited states. Therefore, for a satisfactory description of these processes, the $U(3) \times U(3)$ symmetric extended NJL model was formulated in 1997 \cite{Volkov:1996br, Volkov:1996fk}. This model was used to describe many processes involving radially excited mesons \cite{Volkov:1997dd}. But this model turned out to be especially useful for describing $\tau$ lepton decays. Note that taking into account the higher degrees of excitation of intermediate mesons can only lead to insignificant corrections within the accuracy of the model. This is a consequence of the fact that the masses of mesons with higher-order radial excitation, as a rule, turn out to be heavy and, in particular, heavier than the $\tau$ lepton mass.
    
    The structure of the review is organized as follows. In the next Section \ref{NJL_st}, we will present the $U(3) \times U(3)$ chiral NJL model that describes only the ground meson states. The Lagrangians of the quark-meson interactions will be obtained and the values of the constituent and current masses of the $u$, $d$ and $s$ quarks will be found. Taking into account the t 'Hooft interaction, the mixing angle of the $\eta$ and $\eta'$ mesons will be defined. When describing kaons, the possibility of mixing $K_{1A}$ and $K_{1B}$ mesons will be taken into account.  

    In Section \ref{NJL_ext}, the extended NJL model will be described, the mixing angles of the ground and first radially excited states of mesons will be found, and the interaction Lagrangians of radially excited mesons with quarks will be constructed.
                                                                 
    In the fourth Section, a table containing the decay widths of radially excited mesons, calculated by proposed the extended NJL model will given, and a number of meson production processes in $e^+e^-$ beams will also be considered. Section \ref{NJL_tau} will demonstrate the calculations of the $\tau$ lepton decays widths with the production of one meson and two mesons with two pseudoscalar or pseudoscalar and vector particles. In the last Section \ref{NJL_dis}, a brief discussion of the proposed models and the results will be given. 
    
\section{The standard Nambu -- Jona-Lasinio model }
\label{NJL_st}
    \subsection{The $SU(2)\times SU(2)$ NJL model}
        \subsubsection{ Lagrangian for scalar, pseudoscalar, vector and axial-vector mesons } 
            In this section, we will show the construction of the standard NJL model following the works \cite{Volkov:1984kq, Volkov:1986zb, Volkov:1993jw, Volkov:2005kw}. The quark Lagrangian containing the four-quark interaction motivated by the fundamental QCD theory in the local approximation has the form
            \begin{eqnarray}
            \label{L_4-quark}
            \mathcal{L}(\bar{q},q)=\bar{q}(x)(i\hat{\partial}_x-m_0)q(x)+\frac{G_1}{2}\left((\bar{q}(x)
             q(x))^2+(\bar{q}(x) i \tau_a \gamma_5 q(x))^2 \right) \\ \nonumber
             -\frac{G_2}{2}\left((\bar{q}(x) \gamma_\mu \tau_a
            q(x))^2+(\bar{q}(x) \gamma_\mu \gamma_5 \tau_a  q(x))^2 \right), \label{lagr_4q}
            \end{eqnarray}
            where $\bar{q}(x)=\{u(x),d(x)\}$ are the $u$ and $d$ quark fields, $m_0$ is the current quark mass matrix, $G_{1}$, $G_{2}$ are the four-quark coupling constants and $\tau_a$ are the Pauli matrices.

            After the bozonization, this Lagrangian takes the form 
            \begin{eqnarray}
            \mathcal{L'}(\bar{q},q,\sigma,\pi,\rho,a_1)=\bar{q}(x)(i\hat{\partial}_x-m_0+\sigma(x) + i  \gamma_5
            \tau_a\pi_a(x) + \gamma_\mu \tau_a \rho^\mu_a(x) 
            \\ \nonumber
            + \gamma_\mu \gamma_5 \tau_a {a_1}^\mu_a )q(x) -\frac{(\sigma(x))^2 + (\pi_a(x))^2 }{2 G_1} +\frac{({\rho}^\mu_a(x)) ^2 + ({a_1}^\mu_a(x)) ^2}{2 G_2}.
            \label{fourQLagr}
            \end{eqnarray}
              
            It is easy to verify that in this expression the vacuum expectation of the scalar field is not equal to zero, which requires a redefinition of the vacuum. This redefinition of the scalar field is achieved by subtracting from it the nonzero vacuum expectation of the scalar field ${<\sigma>}_0 \neq 0$ and adding this value to the current quark mass: $\sigma'= \sigma - {<\sigma>}_0$. The described actions lead to spontaneous breaking of chiral symmetry. The values of the current and constituent quark masses are determined by the gap equation 
            \begin{eqnarray}
            \left.\frac{\delta \mathcal{L'}}{\delta
            \sigma^\prime}\right|_{\sigma^\prime=0}=0,\quad\Rightarrow\quad m_0=m(1- 8 G_{1}
            I_1(m) ) \label{gapNJL}
            \end{eqnarray}

            In the one-loop quark approximation (Figure \ref{fig:oneloop}) we obtain the following free Lagrangian for scalar and pseudoscalar meson fields:    
                   
            \begin{figure}[tb]
            \center{\resizebox{0.6\textwidth}{!}{\includegraphics{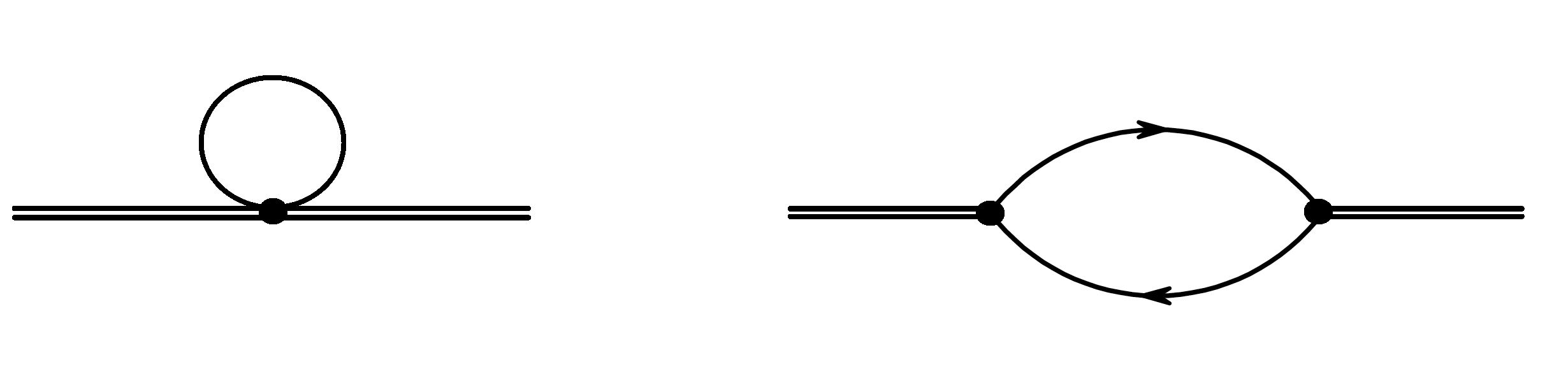}}}
            \caption{ Diagrams describing the mass and renormalization of $\pi$ and $\sigma$ mesons }
            \label{fig:oneloop}
            \end{figure}
            
            \begin{eqnarray}
            &&\mathcal{L^{\text{f}}}_{\sigma, \pi} = \left( -\frac{1}{2G_{1}} + 4I_1(m) 
            +2p^2I_2(m) \right)(\pi_a(p)\pi_a(-p) + \sigma'(p)\sigma'(-p))
            \\ \nonumber
            && - 8m^2 I_2(m) \sigma'(p)\sigma'(-p) = \frac{1}{2}(p^2-M^2_\pi)\pi^R_a(p)\pi^R_a(-p) + \frac{1}{2}(p^2-M^2_\sigma)\sigma^R(p)\sigma^R(-p),
            \nonumber \\
            &&\qquad \pi^R_a(p)=g_{\pi}\pi_a(p), \quad \sigma^R(p)=g_{\sigma}\sigma(p). \nonumber
            \end{eqnarray}
           
            Similarly, for the free Lagrangian of vector and axial-vector mesons, we obtain 
            \begin{eqnarray}
            &&\mathcal{L^{\text{f}}}_{\rho, a_1} = \left( -\frac{1}{2G_{2}}+ \sqrt{\frac{2}{3}}p^2I_2(m) \right) \times \left(g_{\mu\nu} - \frac{p_{\mu}p_{\nu}}{p^2}\right) 
            \\ \nonumber
            && \times (\rho^{\mu}_a(p)\rho^{\nu}_a(-p) + {a_1}^{\mu}_a(p){a_1}^{\nu}_a(-p))+\sqrt{6}I_2(m){a_1}^{\mu}_a(p){a_1}^{\nu}_a(-p).
            \end{eqnarray}
               
            Here, the field renormalization constants are expressed in terms of the logarithmic divergent integral
            \begin{eqnarray}
            \label{grho}
            g_{\pi}=g_{\sigma}=\sqrt{\frac{1}{4I_2(m)}}, \qquad g_{\rho}=g_{a_1}=\sqrt{\frac{3}{2I_2(m)}}.
            \end{eqnarray}    
            
            From the obtained formulas, the connection between the constants $g_{\sigma}$ and $g_{\rho}$ follows
            \begin{eqnarray}
            \label{pirho}
            g_{\rho} = \sqrt{6}g_{\sigma}.
            \end{eqnarray}  
            
            The mass formulas for the mesons $\pi$ and $\rho$ have the form 
            \begin{eqnarray}
            \label{pi_rho_mass}
            M^2_\pi=g_{\pi}^2 \left(\frac{1}{G_{1}}-8I_1(m) \right),\qquad
            M^2_{\rho} = \frac{g^2_{\rho}}{4G_{2}}.
            \end{eqnarray}

            The constants of the four-quark interactions $G_1$ and $G_2$ will be determined from these formulas using the experimental values of the $\pi$ and $\rho$ mesons masses in Section \ref{num_NJL}.

            Expressions for quadratically and logarithmically diverging integrals arising when considering quark loops have the form 
            \begin{eqnarray}
            \label{int_12}
            && I_1(m)=-i\frac{N_c}{(2 \pi)^4} \int \frac{ \Theta(\Lambda^2_4+k^2)}{m^2-k^2} \cdot d^4 k
            =\frac{N_c}{(4 \pi)^2}\left[\Lambda^2_4-m^2\ln\left(\frac{\Lambda^2_4}{m^2}+1\right)\right],\\
            && I_2(m)=-i\frac{N_c}{(2 \pi)^4} \int \frac{ \Theta(\Lambda^2_4+k^2)}
            {(m^2-k^2)^2} \cdot d^4 k =\frac{N_c}{(4
            \pi)^2}\left[\ln\left(\frac{\Lambda^2_4}{m^2}+1\right)-\left(1+\frac{m^2}{\Lambda^2_4}\right)^{-1}\right],
            \nonumber
            \end{eqnarray}
            where the integrals are given in Euclidean space; $N_c$ is the number of colors in QCD and $\Lambda_4$ is the cutoff parameter. The numerical values of the model parameters included in these integrals will be indicated below.  
                 
            Due to the existence of transitions between pseudoscalar and axial-vector mesons, non-diagonal terms appear in the Lagrangian. This leads to additional renormalization of the pion field, which is absent in scalar fields. This renormalization has the form 
            \begin{eqnarray}
            \label{g_pi}
            g_{\pi}=\sqrt{Z_\pi} g_{\sigma},\qquad Z_\pi=\left( 1 - \frac{6m^2}{M_{a_1}^2}\right)^{-1}.
            \end{eqnarray}
      
            As a result, for the interaction Lagrangian of quarks with $\pi$, $\rho$ and $a_1$ mesons, we obtain  
            	\begin{eqnarray}
            	\label{L3}
            && \Delta{\mathcal L} = \bar{q}\biggl[ig_{\pi}\gamma^5 \left( \tau_3\pi^{0} + \tau_{+}\pi^{+} + \tau_{-}\pi^{-} \right) 
            + \frac{g_\rho}{2}\gamma^{\mu} \left(\tau_3{\rho}^0_{\mu} + \tau_{-}{\rho}^{-}_{\mu}+\tau_{+}{\rho}^{+}_{\mu}\right)  \nonumber
              \\ && \qquad  
             + \frac{g_\rho}{2}\gamma^{\mu}\gamma^5 \left(\tau_3{a^0_1}_{\mu} + \tau_{-}{a^-_{1\mu}}+\tau_{+}{a^+_{1\mu}}\right) \biggr]q,
            	\end{eqnarray} 
            where $\tau_{\pm} = (\tau_{1} \pm i \tau_{2})/\sqrt{2}$.
            
        \subsubsection{Numerical estimates of the $SU(2) \times SU(2)$ model parameters}
         \label{num_NJL}
            Let us now determine the main parameters of the model: the masses of the constituent light quarks $m_u \approx m_d$, the ultraviolet cut-off parameter $\Lambda_4$ and the constants $G_1$, $G_2$. To determine the masses of the constituent quarks and the ultraviolet cut-off parameter $\Lambda_4$ of the quark loops, two equations will be used from the experimental values of the decay widths $\pi \to \mu \nu$ ($F_\pi=92.4$ MeV) and the strong decay $\rho \to \pi\pi$ ($g_{\rho}=6.0$) \cite{Volkov:2021fmi}.

            When calculating quark loops in our model, we will use the lowest order in the expansion of $1/{N_c}$, and also take into account only terms with minimum powers in external momenta. Under this condition is it possible to maintain the chiral-symmetric structure of the meson interaction Lagrangian at low energies \cite{Volkov:1986zb}. 

            The decay $\pi \to \mu \nu$ in the NJL model with $\pi - a_1$ transitions is described by the following amplitudes   
            \begin{eqnarray}
            T^1_{\pi \to \mu \nu} = i G_F V_{ud} Z_{\pi} \frac{m_u}{g_{\pi}} L_{\mu} p^{\mu}, \quad T^2_{\pi \to \mu \nu} = - \frac{6m_u^2}{M^2_{a_1}}T^1_{\pi \to \mu \nu}, \nonumber \\
            T_{\pi \to \mu \nu} = T^1_{\pi \to \mu \nu} + T^2_{\pi \to \mu \nu}= i G_F V_{ud}  F_{\pi} L_{\mu} p^{\mu}.
            \end{eqnarray}
            where $G_F$ is the Fermi constant, $V_{ud}$ is the element of the Cabibbo - Kobayashi - Maskawa matrix and $L_\mu$ is the lepton current.
            
            In this case, we obtain the Goldberger-Treiman relation for the weak decay constant 
            \begin{eqnarray}
            g_\pi=\frac{m_u}{F_{\pi}}.
            \end{eqnarray}

            The vector coupling constant $g_{\rho}$ will be determined from the strong decay width $\Gamma_{exp}(\rho \to \pi\pi) = (149.1 \pm 0.8)$ MeV \cite{Volkov:2021fmi, ParticleDataGroup:2020ssz}
            \begin{equation}
            \label{rhowidth}
            \Gamma_{\rho\to\pi\pi} =\frac{g^2_{\rho}}{48\pi }\, M_\rho \left(1 - \frac{4M^2_\pi}{M^2_\rho} \right)^\frac{3}{2}  \Rightarrow  g_{\rho}=6.0.
            \end{equation}     

            Using the relationship between the constants (\ref{pirho}) and (\ref{g_pi}), we obtain the following equation for the constituent quark mass:
            \begin{eqnarray}
            m^2_u = \frac{{M}^2_{a_1}}{12}\left[ 1- \sqrt{1- {\left({\frac{2 g_{\rho} F_{\pi}}{M_{a_1}}}\right)}^2} \right] \Rightarrow  m_u= 270 \textrm{ MeV}.
            \end{eqnarray}

            The cutoff parameter $\Lambda_4 = 1265$ MeV is found using the constant $g_{\rho}$, which is expressed through the integral $I_2(m_{u})$ (\ref{grho}). It is interesting to note that while using the value $g_{\rho} = 6.0$ this cut-off parameter have turned out to be noticeable more than in other versions of NJL model \cite{Vogl:1991qt, Klevansky:1992qe}. As it will be shown below this allow to include the first radially excited meson states within the $U(3) \times U(3)$ chiral symmetric extended NJL model (see Section \ref{NJL_ext}).

            The values of the constants $G_1$ and $G_2$ are determined from the equations for the masses of the pion and vector $\rho$ mesons (\ref{pi_rho_mass})
            \begin{eqnarray}
            G_1 = 4.743 \cdot 10^{-6} MeV^{-2}, \qquad G_2 = 14.9 \cdot 10^{-6} MeV^{-2}.
            \end{eqnarray}

            The value of the current quark mass is found from the gap Equation (\ref{gapNJL}), which gives $m_u^0 = 2.9$ MeV. 

            We now turn to the description of the $\omega \to \pi\pi$ decay, which will allow us to estimate the difference between the $u$ and $d$ quark masses. The amplitude of this decay contains contributions from the strong and electromagnetic transitions $\omega \to \rho$ \cite{Volkov:1986zb}        
            \begin{eqnarray}	
            &&\mathcal{M}_{\omega \to \pi \pi} = [B_1 + B_2 ] e_{\mu}(p_{\omega}) \left(p_{+} - p_{-}\right)^{\mu},  
            \end{eqnarray}	
            where
            \begin{eqnarray}
            B_1 = \frac{8(\pi \alpha_{\rho})^{3/2}M^2_{\omega}}{3(M^2_{\rho}-M^2_{\omega}+iM_{\omega}\Gamma_{\rho})}
            \left[ I_{2}(m_u) - I_{2}(m_d) \right],
            \end{eqnarray}
            \begin{eqnarray}
            B_2 = -\sqrt{\frac{\pi}{\alpha_{\rho}}} \frac{2 \alpha_{em} M^2_{\rho}}{3(M^2_{\rho}-M^2_{\omega}+iM_{\omega}\Gamma_{\rho})}.
            \end{eqnarray}	 
            
            Here $\alpha_{\rho}= g^2_\rho/{4\pi}$, $\alpha_{em}= 1/137$, $M_\rho$ and $M_\omega$ are vector meson masses and $p_{+}$, $p_{-}$ are the momenta of $\pi^+$ and $\pi^-$ mesons. 
            
            The decay width $\omega \to \pi\pi$ is described by formula (\ref{rhowidth}) by replacing $g_\rho \to (B_1+B_2)$. Using the experimental value for the branching fraction $Br(\omega \to \pi\pi) = (1.53 +0.11,-0.13)\times10^{-2}$ \cite{ParticleDataGroup:2020ssz}, we can estimate the mass difference between $u$ and $d$ quarks 
            \begin{eqnarray}
                m_{d} - m_{u} = 4 MeV.
            \end{eqnarray}

    \subsection{The $U(3)\times U(3)$ NJL model}  
    \label{NJL_U3U3}
        When extending the model for the group $U(3) \times U(3)$, the Pauli matrices $\tau_i (i=1,2,3)$ are replaced by the Gell-Mann matrices $\lambda_i (i=0, ... , 8)$, where $\lambda_{0} = \sqrt{\frac{2}{3}}{\bf 1}$ is the matrix proportional to the identity one. Instead of quark doublets $\bar{q} = \left( \bar{u}, \bar{d}\right)$ we have quark triplets $\bar{q} = \left( \bar{u}, \bar{d}, \bar{s}\right)$, and the current mass matrix $m_0$ is replaced by the matrix        
        \begin{eqnarray}
        m^0 = \left(\begin{array}{ccc}
        m_u^0 & 0  & 0 \\
        0 & m_d^0 & 0 \\
        0 & 0  & m_s^0
        \end{array}\right).
        \end{eqnarray}

        Pseudoscalar, vector and axial-vector mesons are introduced into the $U(3) \times U(3)$ model in the same way as it was done in the $SU(2) \times SU(2)$ one.
        
        As a result, after bosonization of the quark Lagrangian, we obtain the following mass formulas for mesons containing $s$ quarks: 
        \begin{eqnarray}
        \label{Mk}
       && M^2_K=g^2_K\left[{1\over G_{1}} - 4[I_1(m_u)+I_1(m_s)]\right]+Z_{K}(m_s-m_u)^2,\\
       && M_{K^\star}^2=\frac{g_{K^\star}^2}{4G_{2}}+\frac{3}{2}(m_s-m_u)^2
        ,\,\,\,M_\phi^2=\frac{g_\phi^2}{4G_{2}}.
        \end{eqnarray}
   
        The mass formulas for non-strange mesons remain unchanged.

        The renormalization constants of the vector fields $g_{K^\star}$ and $g_\phi$ are defined in terms of the integrals 
        \begin{eqnarray}
        g_{K^\star} = \sqrt{\frac{3}{2I_{2}(m_u,m_s)}}, \quad  g_\phi = \sqrt{\frac{3}{2I_2(m_s)}},
        \end{eqnarray}
        where
        \begin{eqnarray}
            I_{2}(m_u,m_s) &=&-i\frac{N_c}{(2 \pi)^4} \int \frac{ \Theta(\Lambda^2_4+k^2)}
            {(m_u^2-k^2)(m_s^2-k^2)} \cdot d^4 k \nonumber\\
            &=&\frac{N_c}{(4\pi)^2} \frac{1}{m_s - m_u}\left[m_s^2\ln\left(\frac{\Lambda^2_4}{m_s^2}+1\right)-m_u^2\ln\left(\frac{\Lambda^2_4}{m_u^2}+1\right)\right]
        \end{eqnarray}

        When deriving the free Lagrangian for the kaon, we take into account additional renormalization of the kaon field ($Z_K$) due to possible transitions between the pseudoscalar and axial-vector mesons. Here it is necessary to take into account two physical axial-vector states ${K}_{1}(1270)$ and ${K}_{1}(1400)$. They are the result of mixing of two axial-vector strange mesons ${K}_{1A}$, ${K}_{1B}$ and related to each other by the following relations: 
        \begin{displaymath} \label{eq_1}
        {K}_{1}(1270)=\sin(\beta){K}_{1A} + \cos(\beta){K}_{1B},
        \end{displaymath}
        \begin{equation}
        {K}_{1}(1400)=\cos(\beta){K}_{1A} - \sin(\beta){K}_{1B}.
        \end{equation}

        In the case of the chiral group $SU(2) \times SU(2)$, the axial-vector mesons $a_1(1260)$ from the nonet $^{3}P_{1}$ and $b_1(1235)$ from the nonet $^{1}P_{1}$ do not mix with each other. This is a consequence of the fulfillment of chiral symmetry and the proximity of the constituent  $u$ and $d$ quark masses. Therefore, when describing the $\pi - {a}_{1}$ transitions, the pion has only one partner among the axial-vector mesons, namely ${a}_{1}(1260)$ \cite{Volkov:2019awd}.
        
        At the same time, for the $U(3)\times U(3)$ group, due to a large mass difference between $m_s$ and $m_d \approx m_u$, chiral symmetry is noticeably broken and the states $K_{1A}$ and $K_{1B}$ begin to mix with each other with a coupling constant proportional to the mass difference $m_s - m_u$ \cite{Volkov:2019awd}. As a consequence, physically observed axial-vector mesons begin to have masses $M_{{K}_{1}(1270)}=1253 \pm 7$ MeV and $M_{{K}_{1}(1400)}=1403 \pm 7$ MeV \cite{ParticleDataGroup:2020ssz}. The NJL model describes only the meson $K_{1A}$. Further we will denote it just $K_{1}$. As a result, for the renormalization constant, we obtain   
        \begin{equation}
        g_{K} = \sqrt{\frac{Z_{K}}{4I_{2}(m_{u},m_{s})}},
        \end{equation}
        where
        \begin{equation}
        Z_{K} = \left(1 - \frac{3}{2}(m_{s}+m_{u})^{2}\left(  
        \frac{{\sin}^{2}(\beta)}{M^{2}_{K_{1}(1270)}}+
        \frac{{\cos}^{2}(\beta)}{M^{2}_{K_{1}(1400)}}
        \right)\right)^{-1},
        \end{equation}
        where $\beta=57^{\circ}$ \cite{Volkov:2019awd}. 

        Equation (\ref{Mk}) can be used to determine the mass of the $s$ quark $m_s = 420$ MeV, which leads to agreement of the charged kaon mass with the experimental value  \cite{Volkov:2019yhy}. For the $s$ quark current mass we obtain the value $m^0_s=76$ MeV. 
        
    \subsection{The 't Hooft interaction}  
    \label{tHooft}
        After the introduction of a heavier $s$ quark into our model, in the framework of the $U(3)\times U(3)$ group, ideal mixing occurs between the eighth representative of the pseudoscalar meson octet and the pseudoscalar singlet meson. Moreover, one of these states contains only $u$ and $d$ quarks, while the other contains only $s$ quarks, which contradicts the experimental data. To solve this problem, it is also necessary to take into account the 't Hooft interaction, which leads to mixing of the states containing light $u, d$ quarks with the heavier $s$ quark \cite{Volkov:1999yi, Volkov:1998ax}. Taking into account the 't Hooft interaction allows us to correctly describe the masses of the pseudoscalar mesons $\eta$ and $\eta'$. This interaction has the form \cite{tHooft:1976rip} 
        \begin{eqnarray}
        \mathcal{L}^{tH} =- K \left( {\det}[{\bar q}(1+\gamma_5)q]+{\det}[{\bar q}(1-\gamma_5)q]
        \right), \label{Ldet}
        \end{eqnarray}
        where $\bar{q}=\{\bar{u},\bar{d},\bar{s}\}$ are the antiquark fields, $ m ^ 0 $ is the diagonal matrix of current quark masses $m^0_u$, $m^0_d$, $m^0_s$ $(m^0_u \approx m^0_d)$ and $K$ is the 't Hooft constant. 
   
        To describe the $ U (3) \times U (3) $ NJL model, we can combine the six-quark interaction with the initial four-quark interaction. In this case, we use the method of separating the main four-quark interaction from the 't Hooft six-quark interaction. The details of these procedures are fairly well described in many papers, in particular in the works \cite{Vogl:1991qt, Klevansky:1992qe, Volkov:2005kw, Volkov:1998ax}. Therefore, omitting the details, we write the new Lagrangian in the following form (see papers \cite{Volkov:2005kw, Volkov:1998ax}):
        \begin{eqnarray}
        &&\mathcal{L} = {\bar q}(i{\hat \partial} - m^0)q + {\frac{1}{2}} \sum_{i=1}^9[G_i^{(-)}
        ({\bar q}{\lambda^\prime}_i q)^2 +G_i^{(+)}({\bar q}i{\gamma}_5{\lambda^\prime}_i q)^2] 
        \nonumber \\
        &&\qquad+ G^{(-)}_{us}({\bar q}{\lambda}_u q)({\bar q}{\lambda}_s q)  + G_{us}^{(+)}({\bar
        q}i{\gamma}_5{\lambda}_u q)({\bar q}i {\gamma}_5{\lambda}_s q)~, \label{LGus}
        \end{eqnarray}
        where
        \begin{eqnarray} &&{\lambda^\prime}_i={\lambda}_i ~~~ (i=1,...,7),~~~\lambda^\prime_8 = \lambda_u = ({\sqrt 2}
        \lambda_0 + \lambda_8)/{\sqrt 3},\nonumber\\
        &&\lambda^\prime_9 = \lambda_s = (-\lambda_0 + {\sqrt 2}\lambda_8)/{\sqrt 3}, \nonumber \\
        &&G_1^{(\pm)}=G_2^{(\pm)}=G_3^{(\pm)}= G \pm 4Km_sI_1(m_s), \nonumber \\
        &&G_4^{(\pm)}=G_5^{(\pm)}=G_6^{(\pm)}=G_7^{(\pm)}= G \pm 4Km_uI_1(m_u),
        \nonumber \\
        &&G_u^{(\pm)}= G \mp 4Km_sI_1(m_s), ~~~ G_s^{(\pm)}= G, ~~~ G_{us}^{(\pm)}= \pm 4{\sqrt
        2}Km_uI_1(m_u). \label{DefG}
        \end{eqnarray}

        For the $\eta$ and $\eta'$ mesons masses, taking into account the mixing of $u, d$ and $s$ quarks caused by the 't Hooft interaction, we obtain the following formulas:    
        \begin{eqnarray}
        M^2_{(\eta,\eta')}={1\over 2}\left[ M^P_{ss} + M^P_{uu} \mp
        {\sqrt {(M^P_{ss}-M^P_{uu})^2 + 4(M^P_{us})^2}} \right],
        \end{eqnarray}
        where
        \begin{eqnarray}
        && M^P_{uu}= g_{\eta_u}^2\left({1\over 2}(T^P)^{-1}_{uu} - 8I_1(m_u) \right),
        \nonumber \\
        && M^P_{ss}= g_{\eta_s}^2\left({1\over 2}(T^P)^{-1}_{ss}-8I_1(m_s) \right), \nonumber
        \end{eqnarray}
        \begin{eqnarray}
        && M^P_{us}= {1\over 2}g_{\eta_u}g_{\eta_s}(T^P)^{-1}_{us}, \nonumber \label{Mpuu} \\
        && M^S_{uu}= g_{\sigma_u}^2\left({1\over 2}(T^S)^{-1}_{uu} - 8I_1(m_u) \right)
        +4m^2_u , \nonumber \\
        && M^S_{ss}= g_{\sigma_s}^2\left({1\over 2}(T^S)^{-1}_{ss}-8I_1(m_s)
        \right) + 4m^2_s , \label{Msuu}\nonumber \\
        && M^S_{us}= {1\over 2}g_{\sigma_u}g_{\sigma_s}(T^S)^{-1}_{us}. \nonumber\\
        &&g_{\sigma_u}=g_{\sigma},~~~g_{\sigma_s}=[4I_2(m_s)]^{-1/2},
        ~~~g_{\eta_u}=g_{\pi},~~~g_{\eta_s}=Z^{1/2}g_{\sigma_s},\nonumber 
        \end{eqnarray}
        \begin{eqnarray}
        &&T^{P(S)}=\frac{1}{2}  \left(
        \begin{array}{cc}
        G_u^{(\pm)}    & G_{us}^{(\pm)} \\
        G_{us}^{(\pm)}  & G_s^{(\pm)}
        \end{array}
        \right) \label{Tps}.
        \end{eqnarray}
        \begin{eqnarray}
        &&\eta_s=\eta \cos(\bar{\alpha})+\eta' \sin (\bar{\alpha}), \\ \nonumber
        &&\eta_u=-\eta \sin (\bar{\alpha})+\eta' \cos (\bar{\alpha}), \quad  \bar{\alpha}=\alpha - \alpha_0,
        \end{eqnarray}
        where $\alpha_0 \approx 35.5 ^{\circ}$ is the ideal mixing angle and $\alpha$ is the singlet-octet mixing angle 
        \begin{eqnarray}
        \tan(2\bar{\alpha}) = \frac{2M^P_{us}}{-M^P_{ss}+M^P_{uu}}.
        \end{eqnarray}
          
        The best agreement with the experimental values for the $\eta$ and $\eta'$ mesons masses can be obtained for the value of the angle $\alpha= -19.0^\circ$
         \begin{eqnarray}
        M_{\eta}=527 \textrm{ MeV}, \quad M_{\eta'}=1004 \textrm{ MeV}.
        \end{eqnarray}

        For the 't Hooft constant we get the value $K=13.0 \textrm{ GeV}^{-5}$. 
        Here we will not discuss scalar mesons, since the contribution of scalar mesons to the decays we are considering turns out to be negligible. In addition, a detailed description of the scalar sector in the NJL model can be found in our previous works \cite{Volkov:2000vy, Volkov:2001ns}. And, finally, at the moment there are a number of works describing scalar mesons taking into account their tetraquark state \cite{Achasov:1999wv, Agaev:2018fvz, Lee:2019bwi}, while in the NJL model considered here, tetraquarks are not taken into account. 

        As a result, in the framework of the $U(3)\times U(3)$ NJL model, we obtain the following interaction Lagrangian of quarks with strange mesons 
        \begin{eqnarray}
        && \Delta L_{int} = 
        \bar{q} \biggl[ \frac{g_{K_{1}}}{2} \gamma^{\mu}\gamma^{5} \left(\sum_{j=\pm,0} \lambda_{j}^{K} K^{j}_{1\mu} + \lambda_{0}^{\bar{K}} \bar{K}^{0}_{1\mu}\right) + \frac{g_{K^{*}}}{2} \gamma^{\mu} \left(\sum_{j=\pm,0} \lambda_{j}^{K} K^{*j}_{\mu} + \lambda_{0}^{\bar{K}} \bar{K}^{*0}_{\mu}\right) \nonumber
         \\ && \qquad  \nonumber
        + i g_{K} \gamma^{5} \left(\sum_{j=\pm,0} \lambda_{j}^{K} K^{j} + \lambda_{0}^{\bar{K}} \bar{K}^{0}\right)  
        + i \sin(\bar{\alpha}) g_{\eta_{u}} \gamma^{5}  \lambda_{u} \eta + i \cos(\bar{\alpha}) g_{\eta_{s}} \gamma^{5}  \lambda_{s} \eta \nonumber
         \\ && \qquad 
         \quad + i \cos(\bar{\alpha}) g_{\eta_{u}} \gamma^{5}  \lambda_{u} \eta' - i \sin(\bar{\alpha}) g_{\eta_{s}} \gamma^{5}  \lambda_{s} \eta' + \frac{g_\phi}{2} \gamma^\mu \lambda_s \phi_\mu
        \biggl]q,
        \end{eqnarray}
        where 
        \begin{eqnarray}
        \label{lambda}
            \lambda^{K}_{\pm} = \frac{\lambda_{4} \pm i \lambda_{5}}{\sqrt{2}}, \quad \lambda^{K}_{0} = \frac{\lambda_{6} + i \lambda_{7}}{\sqrt{2}}, \quad \lambda^{\bar{K}}_{0} = \frac{\lambda_{6} - i \lambda_{7}}{\sqrt{2}},
        \end{eqnarray}
        where matrices $\lambda_{u}$ and $\lambda_{s}$ are defined in (\ref{DefG}).

        The parameters used in this model, as noted in the introduction, differ markedly from the parameters used in other versions of the NJL model \cite{Klevansky:1992qe}. This difference causes our cut-off parameter to significantly exceed the cut-off parameter used in \cite{Klevansky:1992qe}. This circumstance allows us, within the framework of the $U(3) \times U(3)$ chiral symmetric model, to describe not only the four main mesonic nonets but also their first radial excitations. Taking into account intermediate mesons in the ground and first radially excited states in $\tau$ lepton decays turns out to be essential, while higher excitations play a less important role and can be neglected within the framework of the model's accuracy. In the next Section, we will show how, using the simplest form factor of the lowest order in momenta, one can describe the first radial excitations of mesons without going beyond the limits of the admissible breaking of chiral symmetry allowed by the requirement of the PCAC theorem. 

The precision of the NJL model is determined on the basis of partial conservation of the axial current (PCAC). In the case of $U(3)$ symmetry, it can be determined with the ratio ${M^2_{K}}/{M^2_{\Sigma}} \approx 17\%$ \cite{Vainshtein:1970zm}. There are a large number of other sources of uncertainties. Therefore, we use our previous results to estimate the error of the model. Without any exotic states the average uncertainty can be estimated at the level of 10\%. This error obtained from the real calculation covers all possible sources including PCAC and can be absorbed by it. Therefore, we estimate the uncertainty of this model at the level of 17\%. 
    
\section{The extended NJL model}
\label{NJL_ext}
The extended NJL model was formulated in the works \cite{Volkov:1996br, Volkov:1996fk}. As it has been mentioned above, due to the fact that the cut-off parameter is close to the masses of the first radially excited meson states, it turns out possible to include these states to the $U(3) \times U(3)$ chiral symmetric NJL model. To take into account the excited states of mesons, it is more convenient to rewrite the initial four-quark Lagrangian (\ref{L_4-quark}) in terms of the current interactions and redefine them: 
    \begin{eqnarray}
        \mathcal{L}_{int}(\bar{q},q) = \sum_{i = 1}^{9} \sum_{m = 1}^{2} \left\{ \frac{G_{1}}{2} \left[j^{s}_{im}(x)j^{s}_{im}(x) + j^{p}_{im}(x)j^{p}_{im}(x)\right] \right.\nonumber\\
         \left.- \frac{G_{2}}{2} \left[j^{v}_{im\mu}(x)j^{v\mu}_{im}(x) + j^{a}_{im\mu}(x)j^{a\mu}_{im}(x)\right] \right\},
    \end{eqnarray}
    where $G_{1}, G_{2}$ are the four-quark coupling constants and $j(x)$ are the scalar, pseudoscalar, vector and axial-vector currents: 
    \begin{eqnarray}
        j^{n}_{im}(x) = \int d^{4}x_{1} \int \bar{q}(x) F^{n}_{im}(x, x_{1}, x_{2})q(x) d^{4}x_{2}.
    \end{eqnarray}
    
   Here $F^{n}_{im}(x, x_{1}, x_{2})$, $n = s, p, v, a$ are the scalar, pseudoscalar, vector and axial-vector form factors. In the case of the standard NJL model, they are equal to the $\delta$ functions in the coordinates that remove the integrals. For further reasoning, it is convenient to reduce them to momentum representation: 
    \begin{eqnarray}
        F^{n}_{im}(x, x_{1}, x_{2}) = \int \frac{d^{4}p}{\left(2\pi\right)^{4}} \int F^{n}_{im}(k, p) e^{\frac{i}{2}\left[(p + k)(x - x_{1}) + (p - k)(x - x_{2})\right]} \frac{d^{4}k}{\left(2\pi\right)^{4}},
    \end{eqnarray}
    where $k$ and $p$ are the relative and external momenta of the quark-antiquark pair. The momentum $k$ can be written in transverse form: 
    \begin{eqnarray}
        k_{\perp} = k - \frac{(k, p)}{p^{2}}p.
    \end{eqnarray}

    In this case, in the rest frame of the meson formed by these quarks, the momentum $k$ can be represented in three-dimensional form $k_{\perp} = \textbf{k}$. 
      
    The form factors $F$ can describe mesons in the ground ($m=1$) and first radially excited states ($m=2$) and, in the momentum representation take the form 
    \begin{eqnarray}
        F^{s}_{i1}(\textbf{k}, p) & = & 1, \nonumber\\
        F^{p}_{i1}(\textbf{k}, p) & = & i \gamma^{5} \lambda_{i}, \nonumber\\
        F^{v}_{i1}(\textbf{k}, p) & = & \gamma^{\mu} \lambda_{i}, \nonumber\\
        F^{a}_{i1}(\textbf{k}, p) & = & \gamma^{\mu} \gamma^{5} \lambda_{i}, \nonumber\\
        F^{s}_{i2}(\textbf{k}, p) & = & c_{s} f_{i}(\textbf{k}), \nonumber\\
        F^{p}_{i2}(\textbf{k}, p) & = & i \gamma^{5} \lambda_{i} c_{p} f_{i}(\textbf{k}), \nonumber\\
        F^{v}_{i2}(\textbf{k}, p) & = & \gamma^{\mu} \lambda_{i} c_{v} f_{i}(\textbf{k}), \nonumber\\
        F^{a}_{i2}(\textbf{k}, p) & = & \gamma^{\mu} \gamma^{5} \lambda_{i} c_{a} f_{i}(\textbf{k}).
    \end{eqnarray}

    Here $c_{s}$, $c_{p}$, $c_{v}$ and $c_{a}$ are the coefficients of the form factors of the excited meson states. The functions $f_{i}(\textbf{k})$ have the form of a quadratic polynomial in the relative momentum of quarks in the meson: 
    \begin{eqnarray}
        f_{i}(\textbf{k}) = 1 + d_{i}\textbf{k}^{2}.
    \end{eqnarray}

	The slope parameter $d_{i}$ is unambiguously fixed from the requirement that the introduction of excited states does not change the value of the quark condensate, i.e. so that the gap Equation (\ref{gapNJL}) remains unchanged. This condition is provided by the requirement 
    \begin{eqnarray}
        i N_{c} m \int \frac{f_{i}(\textbf{k})}{m^{2} - k^{2}} \Theta(\Lambda_{3}  - |\textbf{k}|) \frac{d^{4}k}{(2\pi)^{4}} = 0,
    \end{eqnarray}
    where $\Lambda_{3} = 1030$ MeV is the three-dimensional cutoff parameter. It is fixed by the four-dimensional cutoff parameter defined for the standard NJL model, based on the requirement that the values of the integrals $I_{2}$ do not change. As a result, we obtain three values for the slope parameters depending on the quark composition of the corresponding meson: 
    \begin{eqnarray}
        d_{uu} = -1.784 \textrm{ GeV}^{-2}, \quad d_{us} = -1.761 \textrm{ GeV}^{-2}, \quad d_{ss} = -1.737 \textrm{ GeV}^{-2}.
    \end{eqnarray}
    
 The proximity of these parameters to each other contributes to the conservation of chiral symmetry after the introduction of an excited states. 
    \subsection{ Pseudoscalar mesons}
        As a result of the bosonization, the free Lagrangian of pion fields in the one-loop approximation after renormalization takes the form: 
        \begin{eqnarray}
        \label{L_pi_ext}
            \mathcal{L}(\pi_{1}, \pi_{2}) = \frac{p^{2}}{2} \left(\pi_{1}^{2} + 2R_{\pi}\pi_{1}\pi_{2} + \pi_{2}^{2}\right) - \frac{M_{\pi_{1}}^{2}}{2} \pi_{1}^{2} - \frac{M_{\pi_{2}}^{2}}{2} \pi_{2}^{2},
        \end{eqnarray}
        where $p$ is the meson momentum, 
        \begin{eqnarray}
            R_{\pi} = \frac{I_{2}^{f}(m_{u})}{\sqrt{Z_{\pi}I_{2}(m_{u})I_{2}^{ff}(m_{u})}}.
        \end{eqnarray}    
        
        The masses of nonphysical mesons are determined as follows: 
        \begin{eqnarray}
           && M_{\pi_{1}}^{2} = g_{\pi}^{2} \left(\frac{1}{G_{1}} - 8 I_{1}(m_{u})\right), \nonumber\\
           && M_{\pi_{2}}^{2} = g_{\hat{\pi}}^{2} \left(\frac{1}{G_{1}} - 8 I_{1}^{ff}(m_{u})\right),
        \end{eqnarray}
        where $g_{\pi}, g_{\hat{\pi}}$ are the pion renormalization constants: 
        \begin{eqnarray}
            g_{\pi} = \sqrt{\frac{Z_{\pi}}{4I_{2}(m_{u})}}, \quad g_{\hat{\pi}} = \sqrt{\frac{1}{4I_{2}^{ff}(m_{u})}},
        \end{eqnarray}
        where $I_{2}^{ff}(m_u)$ is the integral of the form (\ref{int_12}) with two form factors in the numerator.
        
        Here for the constant $g_{\hat{\pi}}$, the $\pi-a_{1}$ transitions are not taken into account due to their small contribution. 
        
    	The resulting Lagrangian (\ref{L_pi_ext}) is non-diagonal. Its diagonalization leads to the following Lagrangian: 
        \begin{eqnarray}
            \mathcal{L}(\pi, \hat{\pi}) = \frac{p^{2}}{2} \left(\pi^{2} + \hat{\pi}^{2}\right) - \frac{M_{\pi}^{2}}{2} \pi^{2} - \frac{M_{\hat{\pi}}^{2}}{2} \hat{\pi}^{2}.
        \end{eqnarray}

        The masses of physical mesons are expressed in terms of non-physical masses as follows: 
        \begin{eqnarray}
            M_{\pi}^{2} = \frac{1}{2\left(1 - R_{\pi}^{2}\right)} \left[M_{\pi_{1}}^{2} + M_{\pi_{2}}^{2} - \sqrt{\left(M_{\pi_{1}}^{2} - M_{\pi_{2}}^{2}\right)^{2} + \left(2M_{\pi_{1}}M_{\pi_{2}}R_{\pi}\right)^{2}}\right],\nonumber\\
            M_{\hat{\pi}}^{2} = \frac{1}{2\left(1 - R_{\pi}^{2}\right)} \left[M_{\pi_{1}}^{2} + M_{\pi_{2}}^{2} + \sqrt{\left(M_{\pi_{1}}^{2} - M_{\pi_{2}}^{2}\right)^{2} + \left(2M_{\pi_{1}}M_{\pi_{2}}R_{\pi}\right)^{2}}\right].
        \end{eqnarray}

        In this case, new meson fields were obtained as a result of the transformation 
        \begin{eqnarray}
            \pi_{1} = \frac{\sqrt{Z_{\pi}}}{\sin{\left(2\theta_{\pi}^{0}\right)}} \left[\pi\sin{\left(\theta_{\pi} + \theta_{\pi}^{0}\right)} - \hat{\pi}\cos{\left(\theta_{\pi} + \theta_{\pi}^{0}\right)}\right],\nonumber\\
            \pi_{2} = \frac{1}{\sin{\left(2\theta_{\pi}^{0}\right)}} \left[\pi\sin{\left(\theta_{\pi} - \theta_{\pi}^{0}\right)} - \hat{\pi}\cos{\left(\theta_{\pi} - \theta_{\pi}^{0}\right)}\right].
        \end{eqnarray}

        The mixing angles are defined as follows: 
        \begin{eqnarray}
          &&  \sin{\theta_{\pi}^{0}} = \sqrt{\frac{1 + R_{\pi}}{2}}, \nonumber\\
          &&  \tan\left(2\theta_{\pi} - \pi\right) = \sqrt{\frac{1}{R_{\pi}^{2}} - 1} \frac{M_{\pi_{1}}^{2} - M_{\pi_{2}}^{2}}{M_{\pi_{1}}^{2} + M_{\pi_{2}}^{2}}.
        \end{eqnarray}

        These formulas lead to the following mixing angles for pions: 
        \begin{eqnarray}
            \theta_{\pi} = 59.48^{\circ}, \quad \theta_{\pi}^{0} = 59.12^{\circ}.
        \end{eqnarray}

        As a result, the quark-meson Lagrangian takes the form: 
        \begin{eqnarray}
            \mathcal{L}(q, \pi) = \bar{q} i\gamma^{5}\sum_{i = \pm, 0} \lambda_{i}^{\pi}\left(A_{\pi}\pi^{i} + A_{\hat{\pi}}\hat{\pi}^{i}\right)q,
        \end{eqnarray}
        where
        \begin{eqnarray}
            A_{\pi} = \frac{1}{\sin{\left(2\theta_{\pi}^{0}\right)}} \left[g_{\pi} \sin\left(\theta_{\pi} + \theta_{\pi}^{0}\right) + g_{\hat{\pi}} f_{uu}(\textbf{k}^{2}) \sin\left(\theta_{\pi} - \theta_{\pi}^{0}\right)\right], \nonumber\\
            A_{\hat{\pi}} = \frac{-1}{\sin{\left(2\theta_{\pi}^{0}\right)}} \left[g_{\pi} \cos\left(\theta_{\pi} + \theta_{\pi}^{0}\right) + g_{\hat{\pi}} f_{uu}(\textbf{k}^{2}) \cos\left(\theta_{\pi} - \theta_{\pi}^{0}\right)\right],
        \end{eqnarray}
       where $\lambda^{\pi}$ are linear combinations of Gell-Mann matrices: 
        \begin{eqnarray}
        \label{lambda_pi}
            \lambda^{\pi}_{\pm} = \frac{\lambda_{1} \pm i \lambda_{2}}{\sqrt{2}}, \quad \lambda^{\pi}_{0} = \lambda_{3}.
        \end{eqnarray}

        Reasoning in a similar way and replacing one light quark with an $s$ quark, one can obtain the quark-meson Lagrangian for kaons \cite{Volkov:1996fk}:
        \begin{eqnarray}
            \mathcal{L}(q, K) = \bar{q} i\gamma^{5}\left[\sum_{i = \pm,0} \lambda_{i}^{K}\left(A_{K}K^{j} + A_{\hat{K}}\hat{K}^{j}\right) 
            + \lambda^{\bar{K}^{0}}\left(A_{K}\bar{K}^{0} + A_{\hat{K}}\hat{\bar{K}}^{0}\right)\right]q,
        \end{eqnarray}
        where
        \begin{eqnarray}
            A_{K} = \frac{1}{\sin{\left(2\theta_{K}^{0}\right)}} \left[g_{K} \sin\left(\theta_{K} + \theta_{K}^{0}\right) + g_{\hat{K}} f_{us}(\textbf{k}^{2}) \sin\left(\theta_{K} - \theta_{K}^{0}\right)\right], \nonumber\\
            A_{\hat{K}} = \frac{-1}{\sin{\left(2\theta_{K}^{0}\right)}} \left[g_{K} \cos\left(\theta_{K} + \theta_{K}^{0}\right) + g_{\hat{K}} f_{us}(\textbf{k}^{2}) \cos\left(\theta_{K} - \theta_{K}^{0}\right)\right].
        \end{eqnarray}

        The coupling constants have the form
        \begin{eqnarray}
            g_{K} = \sqrt{\frac{Z_{K}}{4I_{2}(m_{u}, m_{s})}}, \quad
            g_{\hat{K}} = \sqrt{\frac{1}{4I_{2}^{ff}(m_{u}, m_{s})}}.
        \end{eqnarray}

        The matrices $\lambda^{K}$ are defined in (\ref{lambda}). 
        For the mixing angles of kaons, we get the values 
        \begin{eqnarray}
            \theta_{K} = 58.11^{\circ}, \quad \theta_{K}^{0} = 55.52^{\circ}.
        \end{eqnarray}

	After bosonization and renormalization in the one-loop approximation for the last two particles of the pseudoscalar nonet and their first radial excitations, taking into account the 't Hooft interaction, we can obtain the Lagrangian of the following form: 
        \begin{eqnarray}
            \mathcal{L}(\varphi_{1}^{8}, \varphi_{2}^{8}, \varphi_{1}^{9}, \varphi_{2}^{9}) = \frac{p^{2}}{2} \left(\left(\varphi_{1}^{8}\right)^{2} + \left(\varphi_{2}^{8}\right)^{2} + \left(\varphi_{1}^{9}\right)^{2} + \left(\varphi_{2}^{9}\right)^{2} + 2R_{u}\varphi_{1}^{8}\varphi_{2}^{8} \right.\nonumber\\
          \left.  + 2R_{s}\varphi_{1}^{9}\varphi_{2}^{9} + 2\frac{g_{s}g_{\hat{s}}}{G_{us}^{+}}\varphi_{1}^{8}\varphi_{1}^{9}\right) - \frac{M_{\varphi_{1}^{8}}^{2}}{2} - \frac{M_{\varphi_{2}^{8}}^{2}}{2} - \frac{M_{\varphi_{1}^{9}}^{2}}{2} - \frac{M_{\varphi_{2}^{9}}^{2}}{2},
        \end{eqnarray}
        where $R_{u} = R_{\pi}$, $G_{us}^{+}$ is defined in (\ref{DefG}),
        \begin{eqnarray}
            g_{s} = \sqrt{\frac{1}{4I_{2}(m_{s})}}, \quad g_{\hat{s}} = \sqrt{\frac{1}{4I_{2}^{ff}(m_{s})}}, \quad  R_{s} = \frac{I_{2}^{f}(m_{s})}{\sqrt{I_{2}(m_{s})I_{2}^{ff}(m_{s})}}.
        \end{eqnarray}

        In this case, the diagonalization of the free Lagrangian is performed not analytically but numerically due to the fact that four states take part in this. 
        
        As a result, the quark-meson Lagrangian for four physical mesons takes the form: 
        \begin{eqnarray}
            \mathcal{L}(q, \eta) = \bar{q} i\gamma^{5} \sum_{i = u, s} \lambda_{i} \left[A^{i}_{\eta}\eta + A^{i}_{\eta'}\eta' + A^{i}_{\hat{\eta}}\hat{\eta} + A^{i}_{\hat{\eta}'}\hat{\eta}'\right]q,
        \end{eqnarray}
        where
        \begin{eqnarray}
            A^{u}_{M} & = & g_{\pi} a^{u}_{1M} + g_{\hat{\pi}} a^{u}_{2M} f_{uu}(\textbf{k}^{2}), \nonumber\\
            A^{s}_{M} & = & g_{s} a^{s}_{1M} + g_{\hat{s}} a^{s}_{2M} f_{ss}(\textbf{k}^{2}).
        \end{eqnarray}
        
        Here $M$ stands for $\eta$, $\eta'$, $\hat{\eta}$ or $\hat{\eta}'$ meson. The values of the mixing ($a$) parameters are shown in the Table \ref{tab:eta}. The $\eta'$ meson corresponds to the physical state $\eta'(958)$ and the $\hat{\eta}$, $\hat{\eta}'$ mesons correspond to the first radial excitation mesons $\eta$ and $\eta'$.
      
\begin{table}[h!]
\begin{center}
\begin{tabular}{ccccc}
\hline
\textbf{}	& \textbf{$\eta$}	& \textbf{$\hat{\eta}$}     & \textbf{$\eta'$}       & \textbf{$\hat{\eta}'$}       \\
\hline
$a^{u}_{1}$		& 0.71			& 0.62            &-0.32             & 0.56    \\
$a^{u}_{2}$		& 0.11			& -0.87           & -0.48            & -0.54   \\
$a^{s}_{1}$               & 0.62                        & 0.19            & 0.56             & -0.67 \\
$a^{s}_{2}$               & 0.06                       & -0.66           & 0.3               & 0.82 \\
\hline
\end{tabular}
\end{center}
\caption{Mixing parameters of $\eta$ mesons.}
\label{tab:eta}
\end{table}

        The matrices $\lambda_{u}$ and $\lambda_{s}$ are defined in (\ref{DefG}).
        
        \subsection{Vector and axial-vector mesons }
        Let us consider the vector sector of the extended NJL model in the example of $\rho$ mesons.
        After renormalization, the free Lagrangian of $\rho$ mesons has nondiagonal form: 
        \begin{eqnarray}
          \mathcal{L}(\rho_{1}, \rho_{2}) = -\frac{1}{2} \left(g^{\mu\nu}p^{2} - p^{\mu}p^{\nu}\right) \left(\rho_{1\mu}\rho_{1\nu} + 2R_{\rho}\rho_{1\mu}\rho_{2\nu} + \rho_{2\mu}\rho_{2\nu}\right) \nonumber\\
             + \frac{M_{\rho_{1}}^{2}}{2} \rho_{1\mu}\rho_{1}^{\mu} + \frac{M_{\rho_{1}}^{2}}{2}\rho_{2\mu}\rho_{2}^{\mu},
        \end{eqnarray}
        where
        \begin{eqnarray}
            R_{\rho} = \frac{I_{2}^{f}(m_{u})}{\sqrt{I_{2}(m_{u})I_{2}^{ff}(m_{u})}}.
        \end{eqnarray}
        
        Non-physical masses are expressed by the formulas 
        \begin{eqnarray}
            M_{\rho_{1}}^{2} = \frac{g_{\rho}^{2}}{4G_{2}}, \quad
            M_{\rho_{2}}^{2} = \frac{g_{\hat{\rho}}^{2}}{4G_{2}}.
        \end{eqnarray}
        
        The interaction constant $g_{\rho}$ is defined in (\ref{pirho}),
        \begin{eqnarray}
             g_{\hat{\rho}} = \sqrt{\frac{3}{2I_2^{ff}(m_{u})}}
        \end{eqnarray}

        The free Lagrangian is diagonalized in the same way as in the pseudoscalar sector using the mixing angles 
        \begin{eqnarray}
           && \sin{\theta_{\rho}^{0}}= \sqrt{\frac{1 + R_{\rho}}{2}}, \nonumber\\
           && \tan\left(2\theta_{\rho} - \pi\right)= \sqrt{\frac{1}{R_{\rho}^{2}} - 1} \frac{ M_{\rho_{1}}^{2} -M_{\rho_{2}}^{2}}{ M_{\rho_{1}}^{2} + M_{\rho_{2}}^{2}}.
        \end{eqnarray}

        For the rest of the vector mesons, the reasoning is similar. In the case of the $K^{*}$ meson, one light quark is replaced by the $s$ quark, and in the case of the $\phi$ meson, both light quarks are replaced by the $s$ quarks. 

        Then the quark-meson Lagrangian for the vector fields takes the form: 
        \begin{eqnarray}
            \mathcal{L}(q, \rho, \omega, \phi, K^{*}) = \bar{q} \frac{1}{2} \gamma^{\mu}\biggl[\sum_{i = \pm, 0} \lambda_{i}^{\pi}\left(A_{\rho}\rho^{i}_{\mu} + A_{\hat{\rho}}\hat{\rho}^{i}_{\mu}\right) + \lambda_{u}\left(A_{\omega}\omega_{\mu} + A_{\hat{\omega}}\hat{\omega}_{\mu}\right) + \lambda_{s}A_{\phi}\phi_{\mu} \nonumber\\
             + \lambda_{s}A_{\hat{\phi}}\hat{\phi}_{\mu} + \sum_{i = \pm,0} \lambda_{i}^{K}\left(A_{K^{*}}K^{i*}_{\mu} + A_{\hat{K}^{*}}\hat{K}^{i*}_{\mu}\right)
             + \lambda^{\bar{K}^{0}}\left(A_{K^{*}}\bar{K}^{*0}_{\mu} + A_{\hat{K}^{*}}\hat{\bar{K}}^{*0}_{\mu}\right)\biggl]q,
        \end{eqnarray}
        where
        \begin{eqnarray}
            A_{M} =  \frac{1}{\sin{\left(2\theta_{M}^{0}\right)}} \left[g_{M} \sin\left(\theta_{M} + \theta_{M}^{0}\right) + g_{\hat{M}} f_{M}(\textbf{k}^{2}) \sin\left(\theta_{M} - \theta_{M}^{0}\right)\right], \nonumber\\
            A_{\hat{M}} = \frac{-1}{\sin{\left(2\theta_{M}^{0}\right)}} \left[g_{M} \cos\left(\theta_{M} + \theta_{M}^{0}\right) + g_{\hat{M}} f_{M}(\textbf{k}^{2}) \cos\left(\theta_{M} - \theta_{M}^{0}\right)\right].
        \end{eqnarray}

        Here $M = \rho, \omega, \phi, K^{*}$. The mixing angles take the following values: 
        \begin{eqnarray}
            \theta_{\rho} = \theta_{\omega} = 81.8^{\circ}, &\quad& \theta_{\rho}^{0} = \theta_{\omega}^{0} = 61.5^{\circ}, \nonumber\\
            \theta_{\phi} = 68.4^{\circ}, &\quad& \theta_{\phi}^{0} = 57.13^{\circ}, \nonumber\\
            \theta_{K^{*}} = 84.74^{\circ}, &\quad& \theta_{K^{*}}^{0} = 59.56^{\circ}.
        \end{eqnarray}

        The matrices $\lambda_{u}$ and $\lambda_{s}$ are defined in (\ref{DefG}), the matrices $\lambda^{K}$ are defined in (\ref{lambda}) and the matrices $\lambda^{\pi}$ are defined in (\ref{lambda_pi}).

        For axial vector mesons, renormalization yields the same $I_{2}$ integrals as for vector mesons. These integrals are included in the definition of nonphysical meson masses, through which, in turn, the mixing angles are determined. This gives grounds to use the same parameters for axial vector mesons as in the vector case
        \begin{eqnarray}
            \mathcal{L}(q, a_{1}, K_{1}) = \bar{q} \frac{1}{2}\gamma^{\mu}\gamma^{5}\biggl[\sum_{i = \pm, 0} \lambda_{i}^{\pi}\left(A_{\rho}a_{1\mu}^{i} + A_{\hat{\rho}}\hat{a}_{1\mu}^{i}\right) + \sum_{i = \pm,0} \lambda_{i}^{K}\left(A_{K^{*}}K_{1\mu}^{i} + A_{\hat{K}^{*}}\hat{K}_{1\mu}^{i}\right) \nonumber\\
              + \lambda^{\bar{K}^{0}}\left(A_{K^{*}}\bar{K}_{1\mu}^{0} + A_{\hat{K}^{*}}\hat{\bar{K}}_{1\mu}^{0}\right)\biggl]q.
        \end{eqnarray}

        In the axial vector case, we do not consider isoscalar states for two reasons. First, there are difficulties in describing the mixing of these states. Second, they are not needed to describe the processes considered in this review. 

\section{Strong decays of radially excited mesons and meson production in $e^+e^-$ collisions at low energies}
\label{NJL_ee}
The formulated extended version of the NJL model made it possible to describe various low-energy meson interaction processes with the participation of radially excited states. A number of processes calculated in the extended model were described in detail in the review \cite{Volkov:1999yi} and original papers \cite{Volkov:1998fa, Volkov:1999qb, Volkov:2000ry}. Brief results of these calculations are presented in Table \ref{table_1}. 

\begin{table}[h!]
\begin{center}
\begin{tabular}{ccc}
\hline
\textbf{Decays}	& \textbf{Decay width in the extended NJL model, MeV}	& \textbf{Experiment, MeV}  \\
\hline
$\hat{\pi} \to \rho \pi$ & $220$ & 200-600 \cite{ParticleDataGroup:2020ssz}\\
$\hat{\rho} \to 2\pi$ & $22$ & --  \\
$\hat{\rho} \to \omega \pi$ & $75$ &52-78 \cite{Clegg:1993mt}\\
$\hat{\omega} \to \rho \pi$ & $225$ & $174 \pm 60$ \cite{Clegg:1993mt}\\
$\hat{K}^{*} \to K^* \pi$ & $90$ & $<95.52 \pm 8.64$ \cite{ParticleDataGroup:2020ssz} \\
$\hat{K}^{*} \to K\pi$ & $20$ & $15.3 \pm 3.0$  \cite{ParticleDataGroup:2020ssz} \\
$\hat{K} \to K^{*} \pi$ & $90$ & $\sim 109$  \cite{ParticleDataGroup:2020ssz} \\
$\hat{K} \to K \rho$ & $50$ & $\sim 34$ \cite{ParticleDataGroup:2020ssz} \\
$\hat{\phi} \to K^{*} K$ & $90$ & -- \\
$\hat{\phi} \to \bar{K}K$ & $10$ & --- \\
\hline
\end{tabular}
\end{center}
\caption{Strong decay widths calculated in the extended NJL model}
\label{table_1}
\end{table}

These results in Table \ref{table_1} are in satisfactory agreement with the experimental data within the precision of the model (see Section \ref{NJL_st}). The dominant decays of the excited mesons $\hat{\pi}$, $\hat{\rho}$, $\hat{\omega}$, $\hat{K^*}$ and $\hat{\phi}$ are the decays $\hat{\pi} \to \rho \pi$, $\hat{\rho} \to \omega \pi$, $\hat{\omega} \to \rho \pi$, $\hat{K}^{*} \to K^* \pi$ and $\hat{\phi} \to K^{*} K$, which go through the triangle quark loops of the anomaly type. The decays of the type $\hat{\rho} \to 2\pi$, $\hat{K}^{*} \to K\pi$ and $\hat{\phi} \to \bar{K}K$, going through the other (not anomaly type) quark diagrams, have smaller strong decay widths. So one can see that our model satisfactorily describes not only the weak-decay coupling constants of the radially excited mesons but also their decay widths. We would like to emphasize that there were not used any additional parameter for description of the decays. 

The extended NJL model also makes it possible to describe quite satisfactorily a whole series of meson production processes in colliding electron-positron beams at low energies (<2 GeV). 

Among other forms of meson interaction with the participation of radially excited states, the study of $\tau$ lepton decays is of particular interest. A more detailed discussion of these processes in the NJL model will be given in the next Section \ref{NJL_tau}. Here we will focus on the description of some meson production processes in colliding electron-positron beams described in the framework of the proposed extended NJL model. 

\subsection{Processes $e^+e^- \to [\pi, \pi(1300)] \gamma$}
In the extended NJL model, the processes $e^+e^- \to [\pi, \pi(1300)] \gamma$ ($e^+e^- \to \pi \gamma$ and $e^+e^- \to \pi(1300) \gamma$) were described in \cite{Arbuzov:2011fv}. The main role in this process is played by the channels in both the ground and first radially excited states with vector mesons $\rho$, $\omega$, $\phi$, $\hat{\rho}$ and $\hat{\omega}$. The corresponding amplitude takes the form 
        \begin{eqnarray} 
            \mathcal{M} = l^\mu e^{*\lambda}(p_{\gamma})
            \varepsilon_{\mu\lambda\alpha\beta}\frac{p_\pi^\alpha p_\gamma^\beta}{m s}
            \bigl\{B_{\gamma}+B_{\rho+\omega+\phi}+B_{\hat{\rho}+\hat{\omega}}\bigr\},
        \end{eqnarray}
where $s=(p_1(e^+) + p_2(e^-))^2$ and $l_\mu = \bar{e}\gamma_\mu e$ is the lepton current, $e^{*\lambda}(p_{\gamma})$ is the photon polarization vector, $\varepsilon_{\mu\lambda\alpha\beta}$ is an antisymmetric tensor arising from the fact that this process, like most of the other in this review of $e^+ e^-$ annihilation processes, contains an anomalous quark triangle in which divergent integrals do not arise. 

The contact diagram contribution reads
\begin{eqnarray}
B_{\gamma} = 2 V_{\gamma^*\pi^0\gamma}(s).
\end{eqnarray}
The sum of the contributions of the $\rho$ and $\omega$ mesons has the form 
\begin{eqnarray}
B_{\rho+\omega+\phi} = \biggl\{\frac{C_\rho}{g_\rho} \frac{s}{s-M_\rho^2+iM_\rho\Gamma_\rho}
+\frac{C_\rho}{g_\rho} \frac{s}{s-M_\omega^2+iM_\omega\Gamma_\omega}
\nonumber \\
+\frac{C_\phi}{g_\phi} \frac{s\sqrt{2}\sin\theta_{\omega\phi}}{s-M_{\phi}^2+iM_{\phi}\Gamma_{\phi}}
\biggr\}
V_{\rho\pi^0\gamma}(s),
\end{eqnarray}  
where the constants $C_\rho$, $C_\phi$ describe the transition $\gamma \to \rho(\omega, \phi)$ through the quark loop         \begin{eqnarray}
        \label{C_const}
            C_{M}= \frac{1}{\sin{\left(2\theta_{M}^{0}\right)}} \left[\sin{\left(\theta_{M} + \theta_{M}^{0}\right)} + R_{M} \sin{\left(\theta_{M} - \theta_{M}^{0}\right)}\right], \nonumber\\
            C_{\hat{M}} = \frac{-1}{\sin{\left(2\theta_{M}^{0}\right)}} \left[\cos{\left(\theta_{M} + \theta_{M}^{0}\right)} + R_{M} \cos{\left(\theta_{M} - \theta_{M}^{0}\right)}\right],
\end{eqnarray}
where $M$ denotes the corresponding meson. The mixing angles $\theta$ and ratios of integrals $R$ for the different types of mesons are defined in Section \ref{NJL_ext}. 

At the vertex with the $\phi$ meson, we take into account mixing $\omega - \phi$ \cite{Arbuzov:2011fv}. 

The contributions from the channels with the excited $\hat{\rho}$ and $\hat{\omega}$ mesons take the form 
\begin{eqnarray}
            && B_{\hat{\rho}+\hat{\omega}} =
            \frac{C_{\hat{\rho}}}{g_{\rho}}
            \biggl\{\frac{s}{s-M_{\hat{\rho}}^2+iM_{\hat{\rho}}\Gamma_{\hat{\rho}}}
            +\frac{s}{s-M_{\hat{\omega}}^2+iM_{\hat{\omega}}\Gamma_{\hat{\omega}}}
            \biggr\}
            \times
             V_{\hat{\rho}\pi^0\gamma}(s).
\end{eqnarray}

The explicit expressions for the vertices $V_{\gamma^*\pi^0\gamma}(s)$, $V_{\rho\pi^0\gamma}(s)$ and $V_{\hat{\rho}\pi^0\gamma}(s)$ can be found in \cite{Arbuzov:2011fv}. 
\begin{figure}[tb]
\center{\resizebox{0.7\textwidth}{!}{\includegraphics{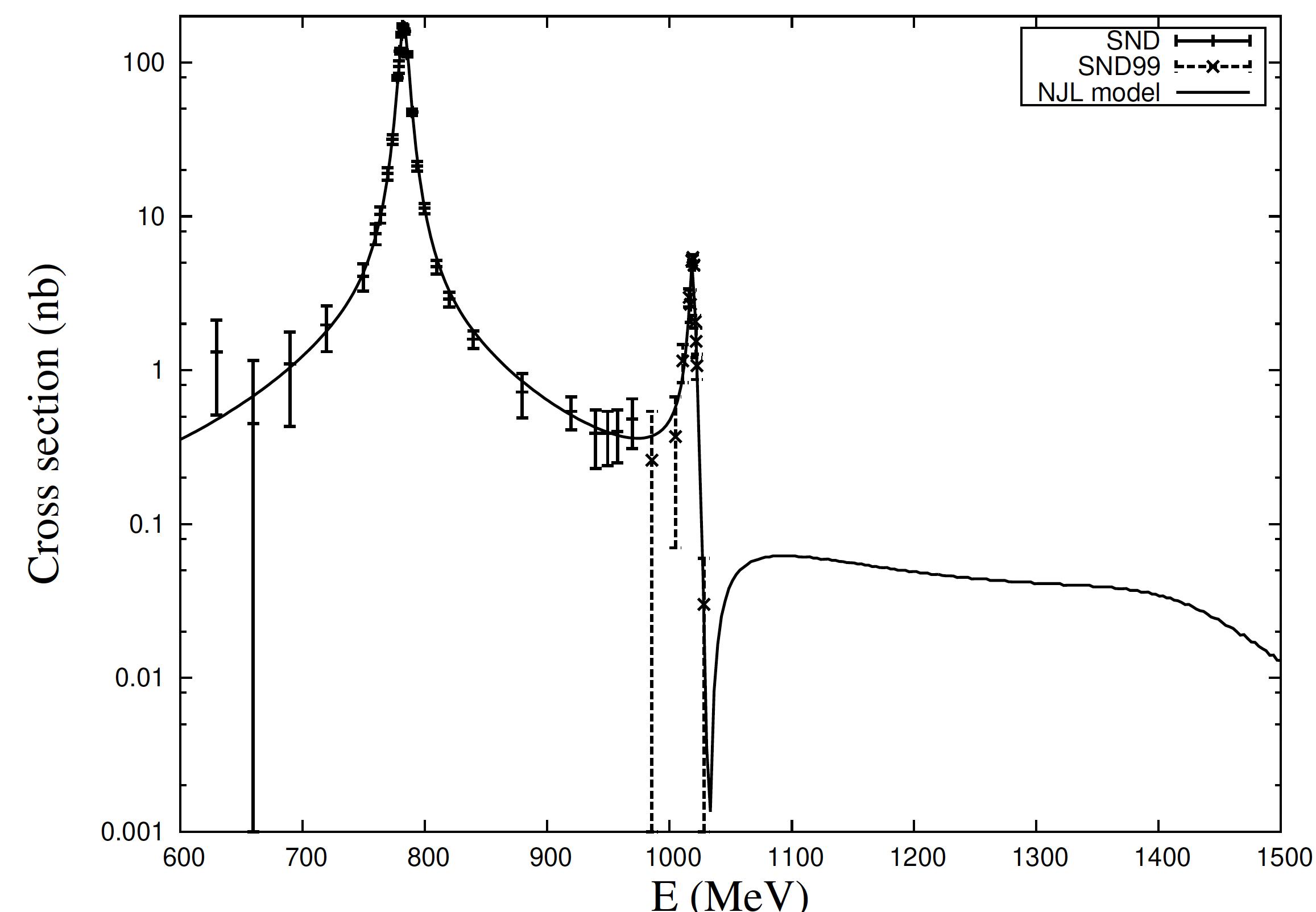}}}
\caption{Comparison of NJL predictions with experimental data for the process $e^+e^- \to \pi \gamma $}
\label{pigamma}
\end{figure}    
\begin{figure}[tb]
\center{\resizebox{0.7\textwidth}{!}{\includegraphics{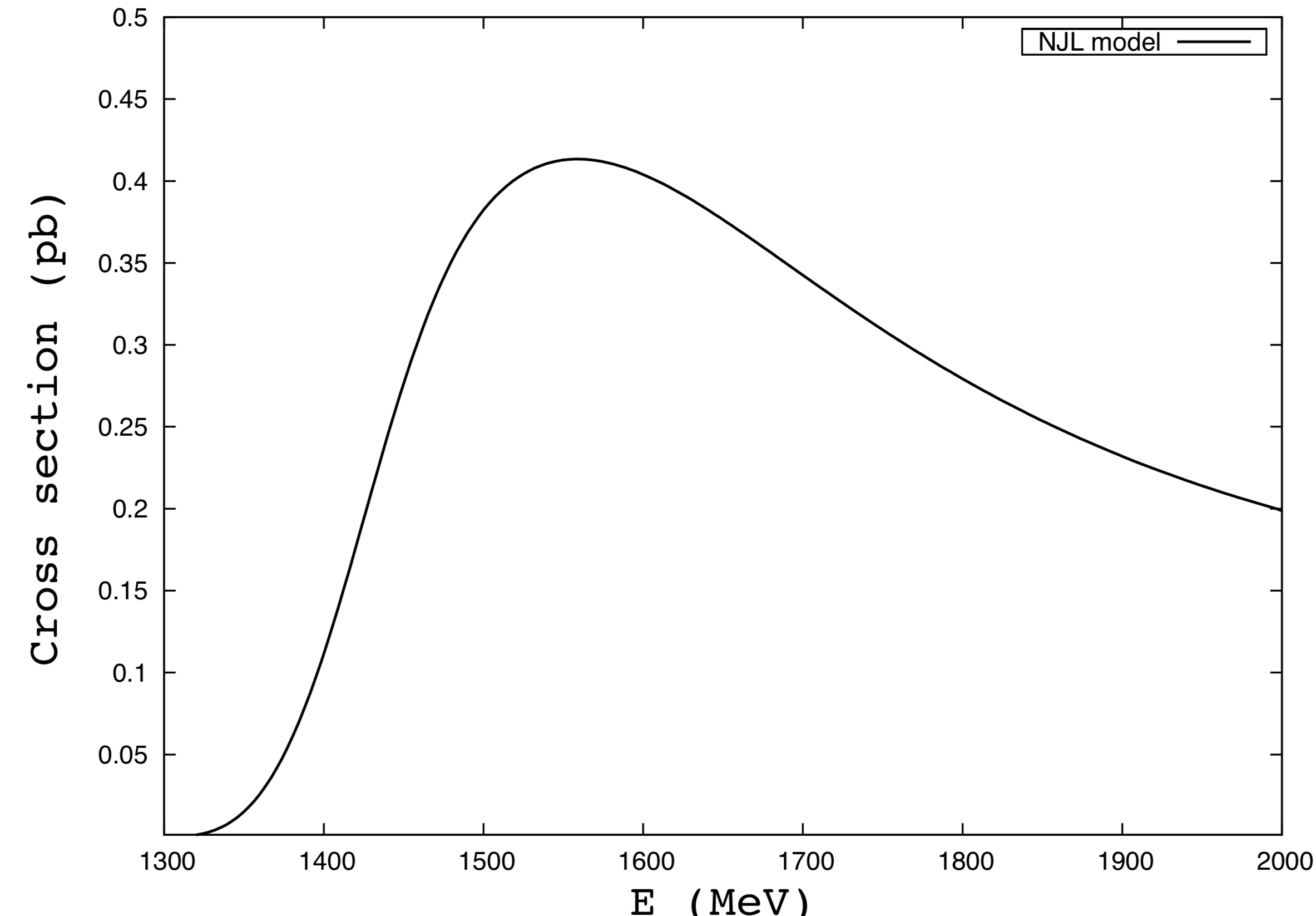}}}
\caption{Predictions of the NJL model for the cross sections of the process $e^+e^- \to \hat{\pi} \gamma $}
\label{pipgamma}
\end{figure} 

The cross section of the process under consideration can be calculated by the following formula:      
\begin{eqnarray}
 \label{sigma}
&& \sigma^{e^+e^-\to\pi\gamma}(s) =
\frac{\alpha^3_{em}}{24\pi^2s^3 F_\pi^2}\lambda^{3/2}(s,0,M_\pi^2)
\frac{1}{g_{\pi}^2}
 \times
|B_{\gamma}+B_{\rho+\omega+\phi}+B_{\hat{\rho}+\hat{\omega}}|^2,
\\ \nonumber
&& \lambda(s,0,M_\pi^2) = (s-M_\pi^2)^2-4M_{\pi}^2.
\end{eqnarray}

The obtained results of model calculations and the comparison with the experimental data of the SND collaboration \cite{Achasov:2000zd, CMD-2:2003bgh}, are presented in the Figure \ref{pigamma}. This Figure shows that theoretical predictions are in good agreement with experiments. The amplitude of the $\hat{\pi} \gamma$ production process has a similar structure; it is obtained by replacing the vertex $\pi \to \hat{\pi}$. The corresponding predictions of the NJL model are given in Figure \ref{pipgamma}. These predictions for future experiments can be tested at the $e^+e^-$ colliders. 

\subsection{Processes $e^+e^- \to \gamma [\eta, \eta', \eta(1295), \eta(1475)]$}
The process of electron-positron annihilation into $\eta_i \gamma$ meson pairs in the framework of the extended NJL model were described in the paper \cite{Ahmadov:2013ksa}. The structure of the amplitudes of these processes is close to the processes $e^+e^- \to [\pi, \hat{\pi}] \gamma$ considered above. In this amplitude, we take into account the contributions of the contact diagram and diagrams from mesons in both the ground and first radially excited states. Note that both $u$, $d$ and $s$ quark parts of $\eta_i$ mesons work here. The calculated amplitude takes the form 
        \begin{equation}
            \mathcal{M} = l^{\mu} e^{*\lambda}(p_{\gamma}) \varepsilon_{\mu\lambda\alpha\beta} \cdot  \frac{p_{\eta}^{\alpha}p_{\gamma}^{\beta}}
            {m s}\cdot \{B_{\gamma}+B_{\rho + \omega} + B_{\phi}+B_{\hat{\rho}+\hat{\omega}}+B_{\hat{\phi}}\}.
        \end{equation}
where $s=(p_+(e^+)+p_-(e^-))^2$. The expressions for contributions with different intermediate states read
\begin{eqnarray}
&&B_\gamma=\frac{2}{3}\left (5 \frac{16}{3}\pi^2 m_u V_{\gamma u} + \sqrt{2}\frac{16}{3}\pi^2 m_s V_{\gamma s} \right),\\ 
&&B_{\rho + \omega}= \left(\frac{3s}{m_{\rho}^2 - s - i\sqrt{s}\Gamma_\rho} + \frac{1}{3}\frac{s}{m_{\omega}^2 - s - i\sqrt{s}\Gamma_\omega}\right) \cdot \frac{C_{\rho}}{g_{\rho_1}} \left( \frac{16}{3}\pi^2 m_u V_{\rho} \right), \\
&&B_{\phi} = -\frac{2\sqrt 2}{3}\frac{s}{m_{\phi}^2  - s - i\sqrt{s}\Gamma_\phi} \frac{C_{\phi}}{g_{\phi_1}} \left( \frac{16}{3}\pi^2   m_s V_{\phi}\right), \\
&&B_{\hat{\rho} + \hat{\omega}}= \left(\frac{3s}{m_{\rho'}^2 - s - i\sqrt{s}\Gamma_{\hat{\rho}}(s)} + \frac{1}{3}\frac{s}{m_{\hat{\omega}}^2 - s  -i\sqrt{s}\Gamma_{\hat{\omega}}}\right) \cdot \frac{C_{\hat{\rho}}}{g_{\rho_1}} \left( \frac{16}{3}\pi^2 m_u V_{\hat{\rho}} \right) e^{i\pi}, \\
&&B_{\hat{\phi}} = -\frac{2\sqrt 2}{3}\frac{s}{m_{\hat{\phi}}^2 - s - i\sqrt{s}\Gamma_{\hat{\phi}}}\frac{C_{\hat{\phi}}}{g_{\phi_1}} \left ( \frac{16}{3}\pi^2 m_s V_{\hat{\phi}}\right),
\end{eqnarray}
 where the coefficients $C_{V}$ describe the photon transitions into vector mesons. 
 
The vertex values $V_{\gamma,\rho,\phi,\hat{\rho},\hat{\phi}} = V_{\gamma,\rho,\phi,\hat{\rho},\hat{\phi}}^{\eta,\eta',\hat\eta, \hat\eta'}$ can be found in \cite{Ahmadov:2013ksa}. The standard values for all masses and widths of mesons are taken from PDG \cite{ParticleDataGroup:2020ssz}. 

A number of experimental works have shown that when describing the production of mesons on colliding $e^+ e^-$ beams additional relative phase factors $e^{i \pi}$ can appear in intermediate states. Such factors are not described by the NJL model and introduced here in accordance with the experiment \cite{Achasov:2006dv}.

\begin{figure}[tb]
\center{\resizebox{0.8 \textwidth}{!}{\includegraphics{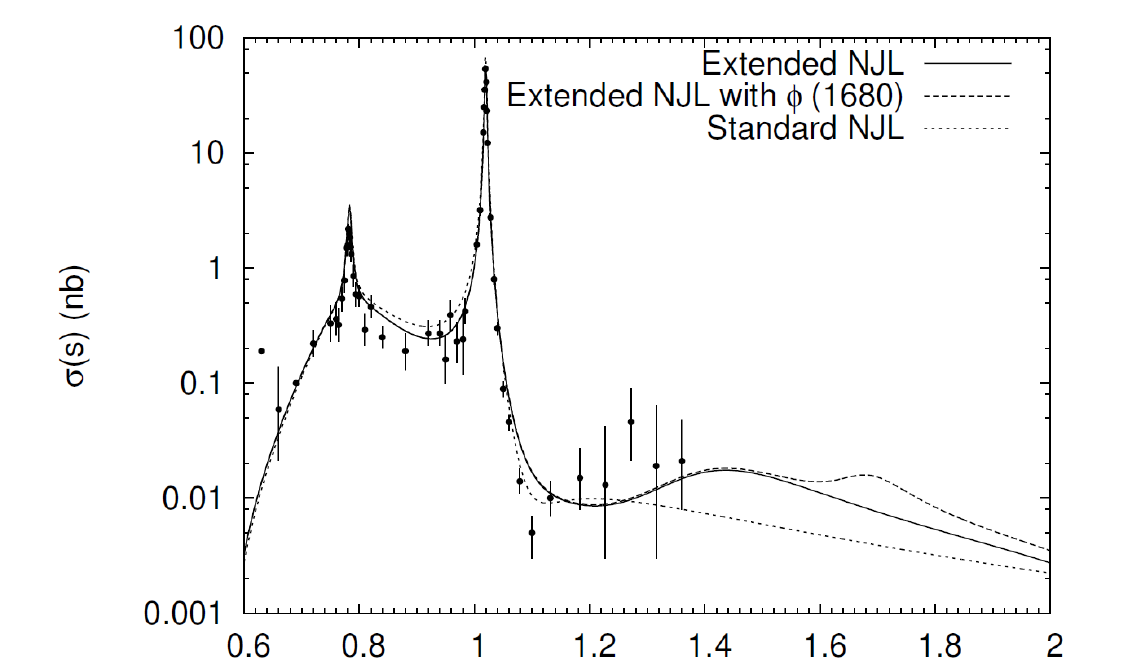}}}
\caption{Comparison of the NJL model predictions with the experiment~\cite{Achasov:2006dv} for the $e^+e^-\to\eta\gamma$ process}
\label{gamma_eta}
\end{figure}      
\begin{figure}[tb]
\center{\resizebox{0.8 \textwidth}{!}{\includegraphics{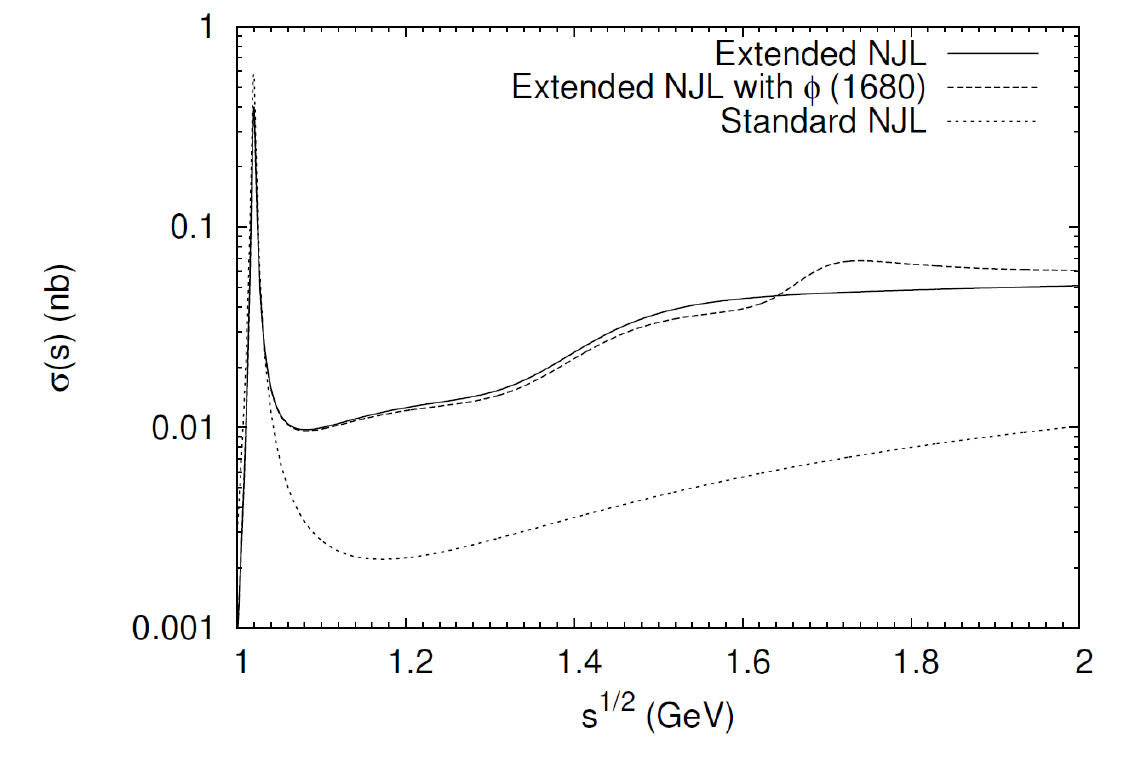}}}
\caption{Predictions for the $e^+e^-\to\eta'\gamma$ process given by the extended and standard NJL models}
\label{gamma_etap}
\end{figure} 

The cross section of the processes under consideration can be calculated by the following formula 
         \begin{equation}
            \sigma(s) = \frac{\alpha_{em}}{24\pi^2 s^3} \lambda^{3/2}(s,M,0)|\mathcal{M}|^2,
        \end{equation}
        where $\lambda(a,b,c) = (a-b-c)^2 - 4bc$, $M = M_\eta, M_{\eta'}, M_{\hat\eta}, M_{\hat\eta'}$.

\begin{figure}[tb]
\center{\resizebox{0.8 \textwidth}{!}{\includegraphics{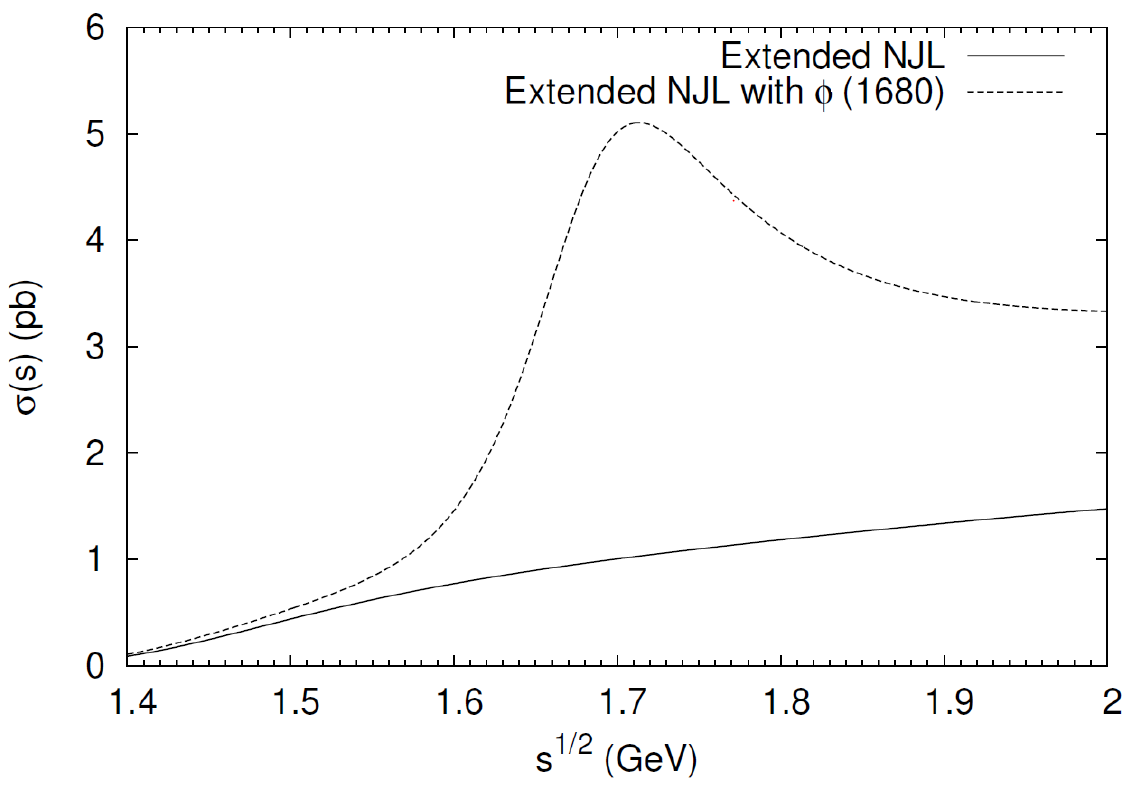}}}
\caption{Predictions for the $e^+e^-\to\eta(1295) \gamma$ process given by the extended NJL model}
\label{gamma_eta1295}
\end{figure} 
\begin{figure}[tb]
\center{\resizebox{0.8 \textwidth}{!}{\includegraphics{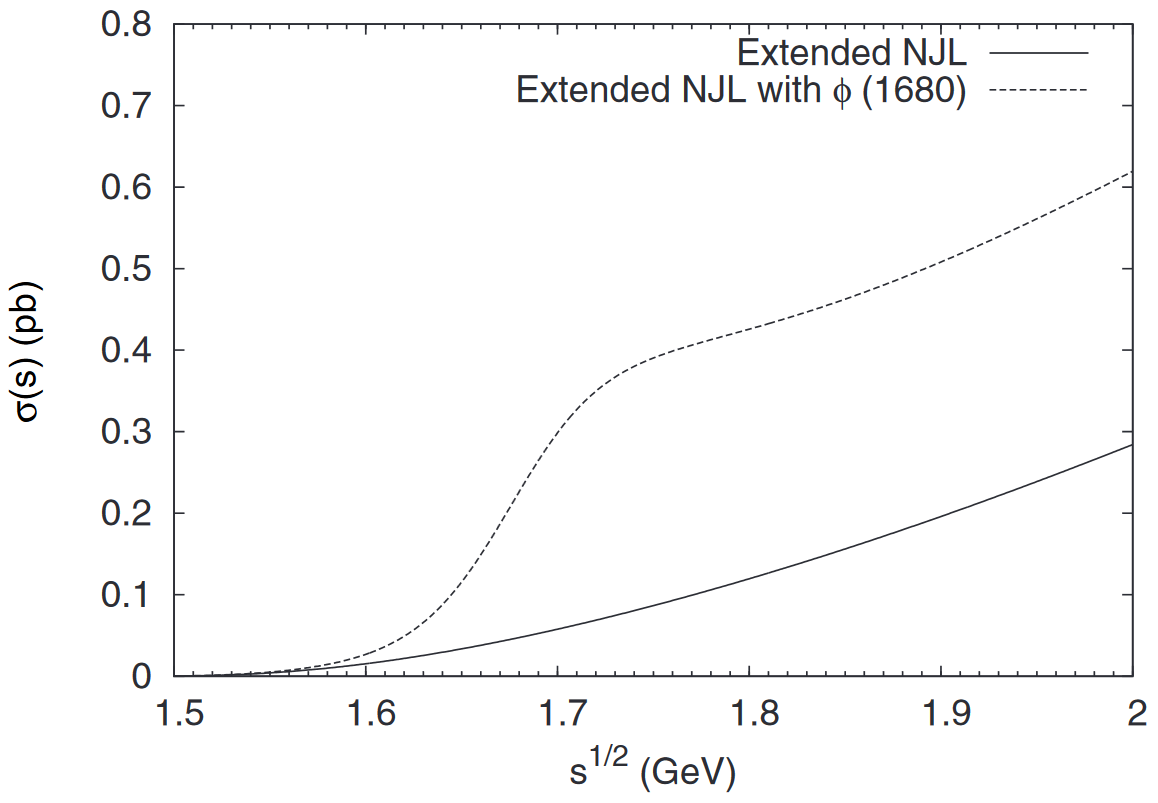}}}
\caption{Predictions for the $e^+e^-\to\eta(1475) \gamma$ process given by the extended NJL model}
\label{gamma_eta1475}
\end{figure} 

The results of numerical calculations for the cross-section are presented in Figures \ref{gamma_eta}, \ref{gamma_etap}, \ref{gamma_eta1295} and \ref{gamma_eta1475}. As we can see, the results of the extended NJL model for the process $e^+e^- \to \gamma \eta$ are in satisfactory agreement with the experimental data. Figure \ref{gamma_eta} shows two sharp peaks. The first peak corresponds to the contributions of the intermediate $\rho$, $\omega$ mesons and the second to the contribution of the $\phi$ meson. Contributions from radially excited states give an insignificant contribution after $1.5$ GeV for the process $e^+e^- \to \gamma \eta$. For the rest of the processes involving $\eta', \eta(1295), \eta(1475)$ mesons, we give predictions for future experiments. 

The process $e^+e^- \to \pi^+ \pi^-$ will be described in Section \ref{NJL_tau} devoted to the $\tau$ lepton decays, therefore, consider the process $e^+e^- \to K^+ K^-$ in the next.

\subsection{Process $e^+e^- \to K^+ K^-$}
Consider the process $e^+e^- \to K^+ K^-$, following the work \cite{Volkov:2018cqp}. This process was experimentally studied at the accelerator of the SLAC laboratory by the BaBar collaboration at Stanford \cite{BaBar:2013jqz} and at the VEPP-2000 collider at the Budker Institute of Nuclear Physics in Novosibirsk \cite{Achasov:2016lbc, Kozyrev:2017agm} at low energies (< 2 GeV). In the extended NJL model, we calculate this process by considering channels containing intermediate mesons $\phi$, $\hat{\phi}$, $\rho$, $\hat{\rho}$, $\omega$ and $\hat{\omega}$. For the corresponding amplitude in the extended NJL model, we obtain the following expression: 
\begin{eqnarray}
 \mathcal{M} = \frac{16 \pi \alpha_{em}}{s} l^{\mu} \biggl[B_{(\gamma)} + B_{(\rho + \hat{\rho})} + 
B_{(\omega + \hat{\omega})}+e^{i\pi}B_{(\phi + \hat{\phi})}\biggl]_{\mu\nu}(p_{K^{+}}-p_{K^{-}})^{\nu}\,
\end{eqnarray}
where $s = (p(e^{-}) + p(e^{+}))^2$, $l^{\mu} = \bar{e}\gamma^{\mu}e$ is the lepton current. 
The contribution from the diagram with an isolated photon reads 
\begin{equation}
B_{(\gamma)\mu\nu} =  g_{\mu\nu}I^{KK}_{11},
\end{equation}

The sum of the contributions of the vector mesons $V$ and $\hat{V}$ have the form 
\begin{eqnarray}
             B_{V + \hat{V}}^{\mu\nu} = r_{V}
            \biggl[\frac{C_{V}}{g_{V}}\frac{g^{\mu\nu}q^2 - q^{\mu}q^{\nu}}{M^{2}_{V} - q^2 - i\sqrt{q^2}\Gamma_{V}(q^2)} I^{VKK}_{11}  
            \nonumber \\
            + \frac{C_{\hat{V}}}{g_{V}}\frac{g^{\mu\nu}q^2 - q^{\mu}q^{\nu}}{M^{2}_{\hat{V}} - q^2 - i\sqrt{q^2}\Gamma_{\hat{V}}(q^2)} I^{\hat{V}KK}_{11}\biggl]\, ,
        \end{eqnarray}
where $V=\rho, \omega, \phi$ and $\hat{V}=\hat{\rho}, \hat{\omega}, \hat{\phi}$ are vector mesons, $q$ is momentum of colliding leptons and $q^2 = s$. The numerical coefficients are $r_{\rho}=1/2$, $r_{\omega}=1/6$, $r_{\phi}=1/3$. Here instead of the constant decay width $\Gamma_{V}$, we use $\Gamma(s)$, following the work \cite{BaBar:2013jqz}
\begin{equation}
\Gamma_{V}(s) = \Gamma_{V}\frac{s}{M^{2}_{V}}{\left(\frac{\beta(s, M_{K})}{\beta(M^{2}_{V}, M_{K})}\right)}^{3},
\end{equation}
where $\beta(s, M_{K})=\sqrt{1-{4M^2_{K}}/s}$.

  The integrals $I_{11}$ appear at the vertices of the intermediate meson decay into final states
       \begin{eqnarray}
        \label{integral_ext}
            I_{n_{1}n_{2}}^{M_{1}M_{2}\dots} = -i\frac{N_{c}}{\left(2\pi\right)^4} \int \frac{A_{M_{1}}A_{M_{2}}\dots}{\left(m_{u}^{2} - k^{2}\right)^{n_{1}}\left(m_{s}^{2} - k^{2}\right)^{n_{2}}} \Theta(\Lambda_{3}  - |\textbf{k}|) d^{4}k.
        \end{eqnarray}
        where $A_{M}$ are the vertices of the extended NJL model Lagrangian, defined for various mesons in Section \ref{NJL_ext}.

\begin{figure}[tb]
\center{\resizebox{0.7 \textwidth}{!}{\includegraphics{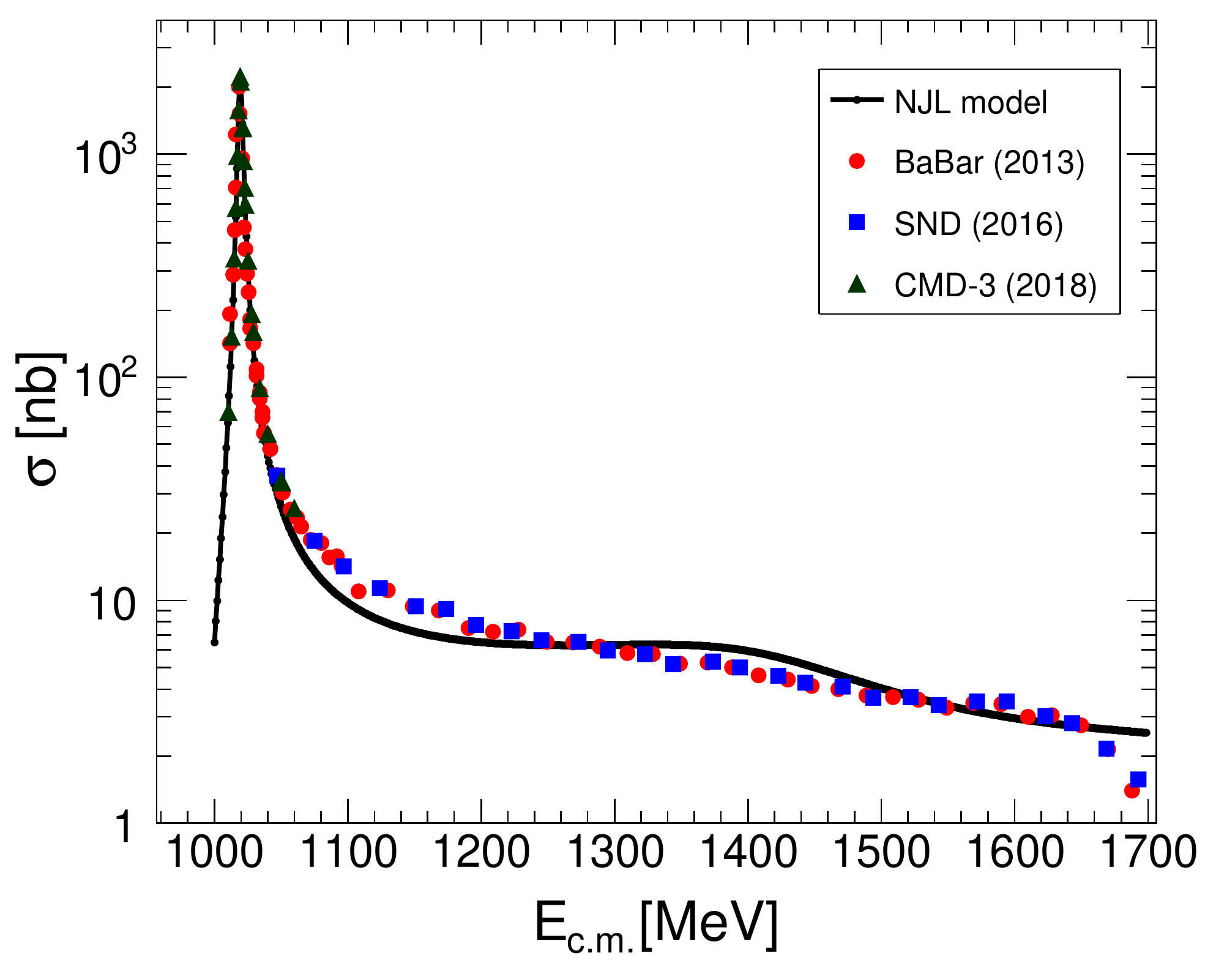}}}
\caption{Comparison of the NJL model predictions for the process $e^+e^- \to K^+K^-$ with experimental data
\cite{BaBar:2013jqz, Achasov:2016lbc, Kozyrev:2017agm}}
\label{KK_1}
\end{figure}      
\begin{figure}[tb]
\center{\resizebox{0.6 \textwidth}{!}{\includegraphics{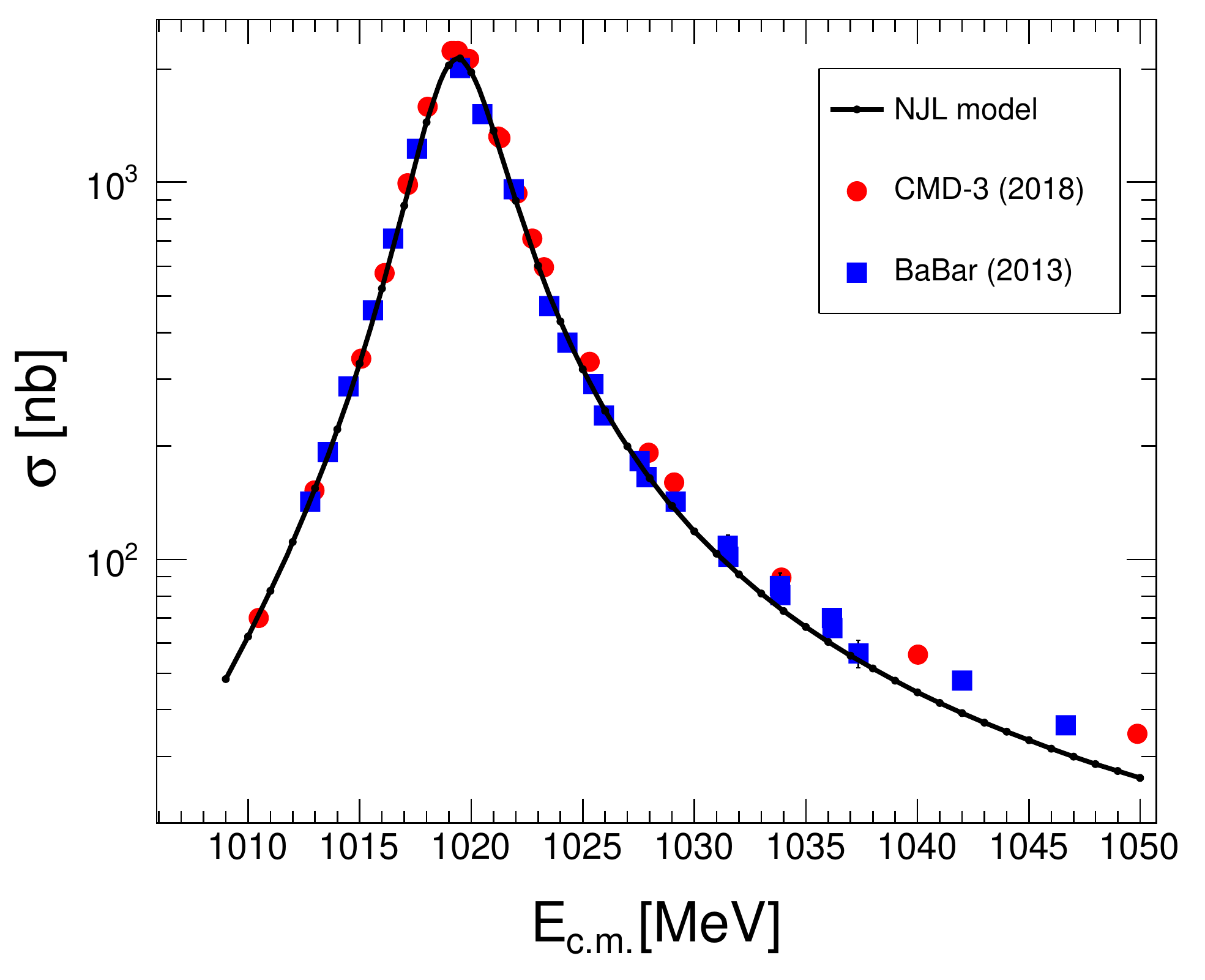}}}
\caption{Contribution of the $\phi$ meson resonance to the process $e^+e^- \to K^+ K^-$. The experimental points are taken from~\cite{BaBar:2013jqz, Kozyrev:2017agm}.}
\label{KK_2}
\end{figure} 

The prediction of the extended NJL model for the cross section of the process $e^+e^- \to K^+ K^-$ and the comparison with experimental data are shown in Figures \ref{KK_1} and \ref{KK_2}. As we can see, the model describes well the cross section for processes in the energy range 1-1.6 GeV in agreement with the SND, CMD-3, and BaBar experiments. At energies exceeding $1.6$ GeV, the second radially excited states of the vector mesons $\rho(1700)$ and $\omega(1650)$ play an important role. Since our model does not include these states, we cannot claim correct descriptions in the region above $1.6$ GeV. 

\subsection{Process $e^+e^- \to \pi \omega$}
Consider the process $e^+e^- \to \pi \omega$, following the work \cite{Arbuzov:2010xi}. This process was experimentally studied at energies up to 2 GeV in a number of experiments \cite{Achasov:2000wy, CMD-2:2003bgh, KLOE:2008woc, Li:2008xm}. 

In the NJL model, the process $e^+e^- \to \pi \omega$ is described by the channels with an isolated photon and intermediate vector mesons $\rho$ and $\hat{\rho}$. The process amplitude is calculated similarly to the process $e^+e^- \to \pi \gamma$ described above. The difference will be in the expression for the triangular vertex of the $\pi \omega$ pair production instead of $\pi \gamma$, namely $V_{\gamma^*\pi^0\omega}(s)$, $V_{\rho\pi^0\omega}(s)$ and $V_{\hat{\rho}\pi^0\omega}(s)$ \cite{Arbuzov:2010xi}. The total cross section of the process $e^+e^- \to \pi \omega$ in the NJL model is calculated by the following formula: 
\begin{eqnarray}
 \label{sigma}
&& \sigma(s) = \frac{3\alpha^2_{em}}{32\pi^3s^3}\lambda^{3/2}(s,M_\omega^2,M_\pi^2)
\frac{g_\rho^2}{F_\pi^2}|J^{(3)}|^2 
 \times
{\mathrm Br}(\omega\to\pi^0\gamma),
\\ \nonumber
&& \lambda(s,M_\omega^2,M_\pi^2) = (s-M_\omega^2-M_\pi^2)^2-4M_{\omega}^2M_{\pi}^2,
\end{eqnarray}
where
\begin{eqnarray}
\label{J3}
&& J^{(3)} = \biggl(1 - \frac{s}{s-M_\rho^2+iM_\rho\Gamma_\rho}\biggr)I^{(3)}_\gamma
+ R_{\rho} \frac{s}{s-M_{\hat{\rho}}^2+i\sqrt{s}\Gamma_{\hat{\rho}}(s)}I^{(3)}_{\hat{\rho}},
\end{eqnarray}
where the integrals $I^{(3)}_\gamma, I^{(3)}_{\hat{\rho}}$ with different degrees of the form factor are defined in \cite{Arbuzov:2010xi}. 

\begin{figure}
\includegraphics[scale = 1.2]{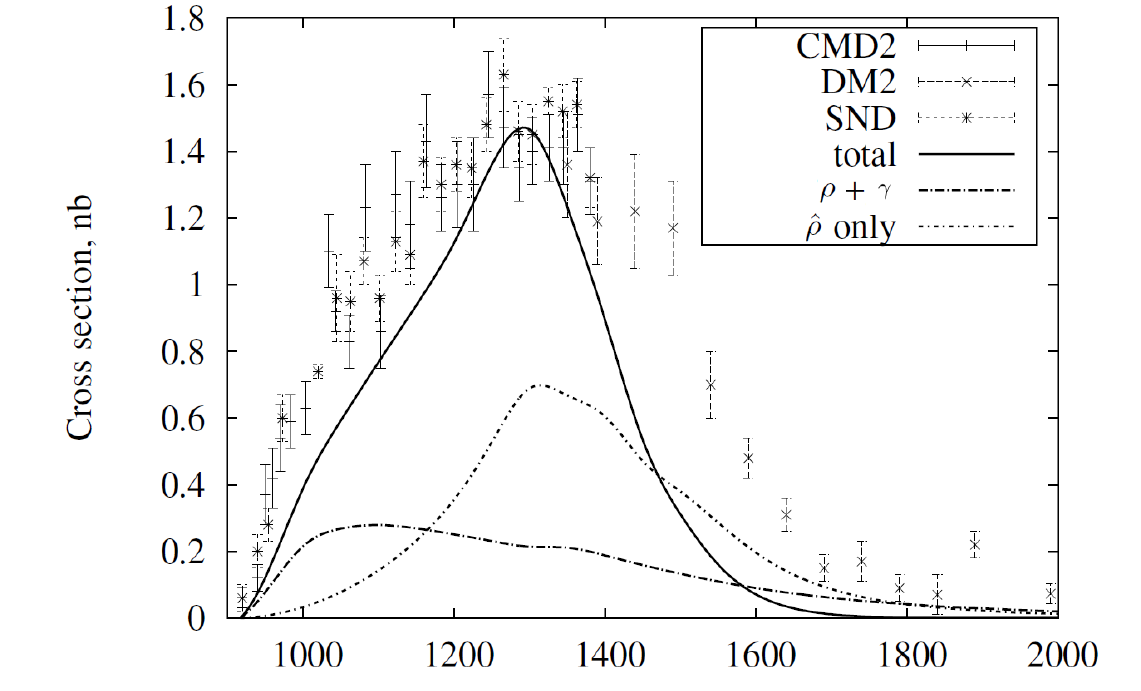}
\caption{Comparison of experimental results for $e^+e^-\to\pi^0\omega$
with the NJL model prediction (lines).
\label{opi}}
\end{figure}

In the Figure \ref{opi} we present a comparison of model predictions with experimental data \cite{Achasov:2000wy,CMD-2:2003bgh,DM2:1990npw}. We see that the model qualitatively describes the experiment at energies up to 2 GeV. In this case, the contribution of the channel with $\hat{\rho}$ mesons in the region $\sqrt{s} \sim M_{\hat{\rho}}$ dominates. At high energies, it is necessary to take into account the contributions from the channels with higher-order excited vector meson states. 

From a theoretical point of view, this process has been considered in a number of works by other authors. In the paper \cite{CMD-2:2003bgh}, the Vector Dominance Model (VDM) was used where the contributions of intermediate mesons $\rho$, $\rho(1450)$ and $\rho(1700)$ were taken into account. At the same time, the free parameters of the model were fitted according to the experiment of the same process.      

\subsection{Process $e^+e^- \to K^{*}K$}
\begin{figure}[h]
\center{\includegraphics[scale = 0.25]{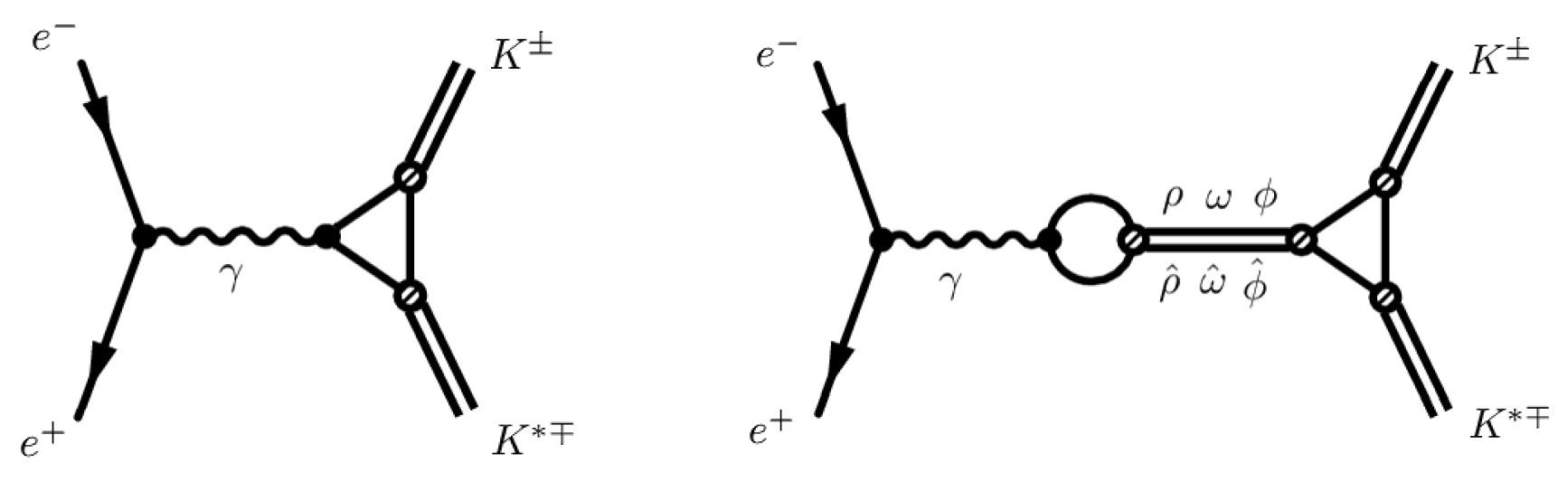}}
\caption{Contact diagram and diagram with intermediate vector mesons of the process $e^{+}e^{-} \to K^{*}K$}
\label{kvk_ee}
\end{figure}

In the extended NJL model, this process was described in the work \cite{Volkov:2016zdw}. The diagrams describing the process $e^+e^- \to K^{*}K$ are shown in Figure \ref{kvk_ee}. 

The process amplitude includes channels with intermediate mesons $\rho$, $\hat{\rho}$, $\omega$, $\hat{\omega}$, $\phi$ and $\hat{\phi}$. In the extended NJL model, for the amplitude of this process, we obtain 
\begin{equation}
 \mathcal{M} = \frac{8 \pi \alpha_{em}}{s} \left\{B_{\gamma} + B_{\rho + \hat{\rho}} + B_{\omega + \hat{\omega}} +
 B_{\phi + \hat{\phi}}\right\} l^{\mu} \varepsilon_{\mu\lambda\delta\sigma} e^{*\lambda}(p_{K^{*}}) p_{K}^{\delta}p_{K^{*}}^{\sigma},
\end{equation}

The contribution from the contact diagram takes the form 
\begin{equation}
B_{(\gamma)} = \frac{2}{3} \left[2m_{s}I^{K^{*}K}_{21} - m_{u}I^{K^{*}K}_{12}\right],
\end{equation}
where the integrals $I_{21}$ and $I_{12}$ are defined in (\ref{integral_ext}).

For contributions from the intermediate vector mesons $\rho$, $\hat{\rho}$, $\omega$ and $\hat{\omega}$ we obtain 
\begin{eqnarray}
             B_{(V + \hat{V})} = a_{V}
            \biggl[\frac{C_{V}}{g_{V}}\frac{s}{M^{2}_{V} - s - i\sqrt{s}\Gamma_{V}(s)} I^{VK^*K}_{21}  
            \nonumber \\
            + e^{i\pi} \frac{C_{\hat{V}}}{g_{V}}\frac{s}{M^{2}_{\hat{V}} - s - i\sqrt{s}\Gamma_{\hat{V}}(s)} I^{\hat{V}K^*K}_{21}\biggl]\, ,
\end{eqnarray}
where $a_\rho=m_s$ and $a_\omega=m_s/3$. The constants $C_{V}$ are defined in (\ref{C_const}).

The contribution from the mesons $\phi$ and $\hat{\phi}$ is obtained by replacing the integral $I^{V K^*K}_{21} \to I^{\phi K^* K}_{12}$, $I^{\hat{V} K^* K}_{21} \to I^{\hat{\phi}K^* K}_{12}$ and constants $a_\phi=-2m_u/3$. The values of the masses and widths of mesons are taken from PDG \cite{ParticleDataGroup:2020ssz}.
\begin{figure}[h]
\center{\includegraphics[scale = 0.9]{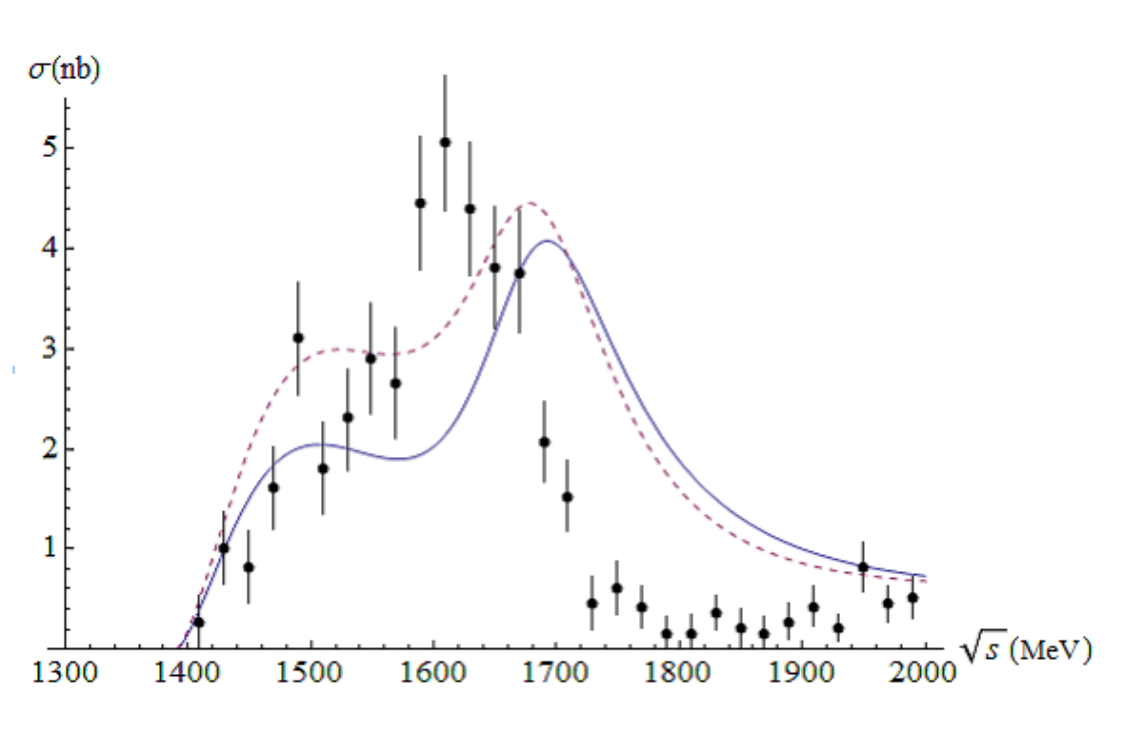}}
\caption{Cross section of the process $e^+e^- \to K^{*}K$ in the extended NJL model obtained with six intermediate meson states. Experimental points are taken from \cite{BaBar:2007ceh}.}
\label{Cross_kvk}
\end{figure} 
 
The calculated cross-section of the process $e^+e^- \to K^{*}K$ in the extended model is shown in Figure \ref{Cross_kvk}. The experimental points are taken from the paper of the BaBar collaboration \cite{BaBar:2007ceh}. Note that changing the value of the width $\Gamma_{\hat{\rho}}=340$ MeV provides a slight shift to the left and increases the theoretical peak. The corresponding section is shown with a dashed line. It is also important to note that the results of the NJL model were obtained without using additional arbitrary parameters.

\subsection{Process $e^+e^- \to \phi \eta$}
The amplitude of the process $e^+e^- \to \phi \eta$ in the extended NJL model is calculated by considering the channels only with the mesons $\phi$ and $\hat{\phi}$ \cite{Volkov:2016zdw}. This is due to the fact that the mesons $\phi$ and $\hat{\phi}$ consist of $s$ quarks, the $\eta$ meson contains both $u$, $d$, and $s$ quark structures. The amplitude of the process takes the following form:
        \begin{displaymath}
            \mathcal{M} = \frac{4 \pi \alpha_{em}}{s} \frac{8m_{s}}{3s} \left\{I^{\phi\eta}_{03}
            + \frac{C_{\phi}}{g_{\phi}} I^{\phi\phi\eta}_{03}
            \frac{s}{M^{2}_{\phi} - s - i\sqrt{s}\Gamma_{\phi}}\right.
            \end{displaymath}
            \begin{equation}
            \left. + e^{i\pi}\frac{C_{\hat{\phi}}}{g_{\phi}} I^{\hat{\phi}\phi\eta}_{03}
            \frac{s}{M^{2}_{\hat{\phi}} - s - i\sqrt{s}\Gamma_{\hat{\phi}}}\right\}
            l^{\mu} \varepsilon_{\mu\lambda\delta\sigma} e^{*\lambda}(p_\phi) p_{\eta}^{\delta}p_{\phi}^{\sigma}.
        \end{equation}
        
   Here the constants $C_{\phi}$ and $C_{\hat{\phi}}$ are defined in (\ref{C_const}); the integrals $I^{\phi\phi\eta}_{03}$ and $I^{\hat{\phi}\phi\eta}_{03}$ are defined in (\ref{integral_ext}).    
\begin{figure}[h]
\center{\includegraphics[scale = 0.9]{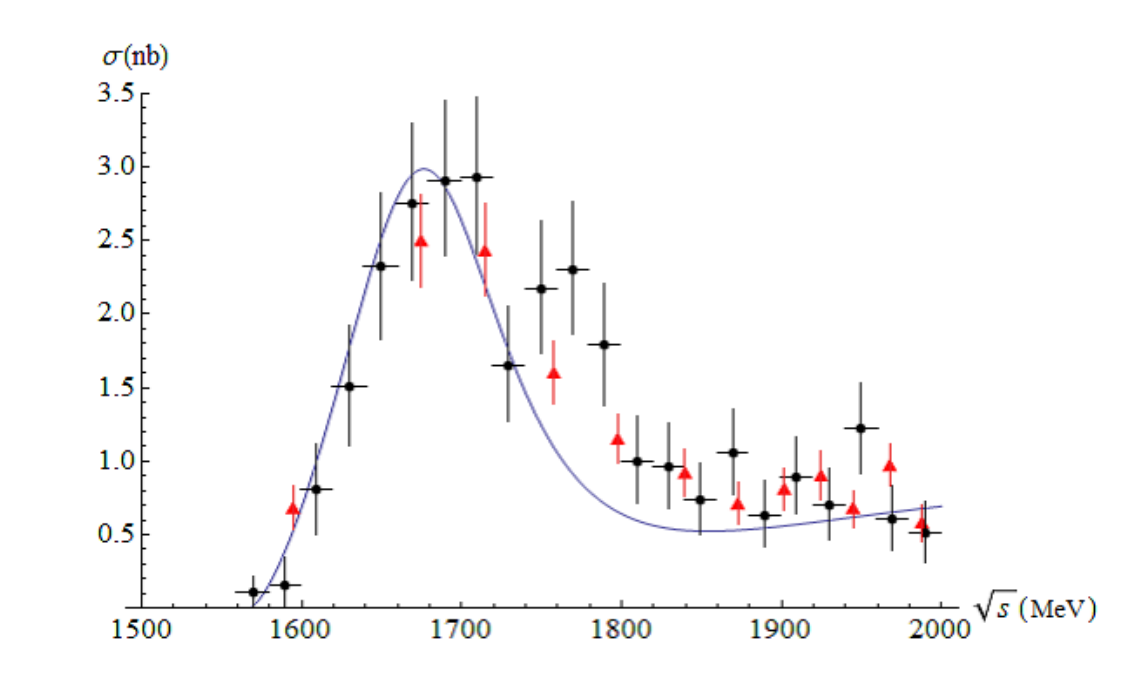}}
\caption{Comparison of the  $e^+e^- \to \phi \eta$ process total cross section with experiments. The solid line corresponds to the prediction of the extended NJL model. The BaBar \cite{BaBar:2007ceh} and CMD-3 \cite{Ivanov:2016blh} experimental data are given as separate points.}
\label{kvk_2}
\end{figure} 

The results of numerical calculations for the cross section of the process under consideration are shown in Figure \ref{kvk_2}. We compare the model predictions for the cross section with the data of the Babar \cite{BaBar:2007ceh} and CMD-3 \cite{Ivanov:2016blh} experiments. The plot shows that the main contribution to the cross section is given by the channel with the first radially excited meson $\hat{\phi}$. The results obtained show that the extended NJL model allows one to describe the total cross section of the $e^+e^- \to \phi \eta$ process in satisfactory agreement with experiments at energies up to 2 GeV. 

\subsection{Process $e^+e^- \to \phi \pi$} 
Consider the process $e^+e^- \to \phi \pi$, following the work \cite{Volkov:2020jor}. This process in the NJL model proceeds due to the $\omega$ and $\phi$ meson mixing. This mixing can be considered as interactions of the $\omega$ and $\phi$ mesons through kaon loops. Such a mechanism was described in the works \cite{Benayoun:2007cu, Benayoun:2009im}. Here we will not consider in detail the nature of mixing of these mesons. First of all, we calculate the value of the mixing angle $\alpha_{\omega\phi}$ using the decay $\phi \to \pi \gamma$, as it was done in the work \cite{Klingl:1996by}. 

The amplitude of the $\phi \to \pi \gamma$ decay in the NJL model takes the form 
\begin{eqnarray}
\mathcal{M} = \frac{3}{4} \frac{\sqrt{\alpha_{em}}}{\pi^{3/2} F_{\pi}} g_{\phi} \sin(\alpha_{\omega\phi}) \varepsilon^{\mu \nu \lambda \delta} e_{\mu}(p_{\phi}) e_{\nu}^{*}(p_{\gamma}) p_{\pi \lambda} p_{\gamma \delta},
\end{eqnarray}
where $\alpha_{em}$ is the electromagnetic interaction constant, $e_{\mu}^{*}(p_{\phi})$ and $e_{\nu}^{*}(p_{\gamma})$ are the polarization vectors of the $\phi(1020)$ meson and the photon. A similar amplitude has already been obtained earlier in the NJL model for the $\omega \to \pi \gamma$ process in \cite{Volkov:1986zb}. Using the experimental value of the width $\Gamma(\phi \to \pi^{0} \gamma)_{exp} = 5.5 \pm 0.2$ keV \cite{ParticleDataGroup:2020ssz}, we can fix the mixing angle which equals to $\alpha_{\omega\phi}=3.1^{\circ}$. 

The structure of the amplitude of the process $e^+e^- \to \phi \pi$ is close to the process $e^+e^- \to \omega \pi$ described in the NJL model \cite{Arbuzov:2010xi}. Only $\rho$ and $\hat{\rho}$ mesons will participate as intermediate mesons (along with the photon). The corresponding amplitude has the form 
         \begin{equation}
        \mathcal{M}(e^+e^- \to \phi \pi) = \frac{8\pi\alpha_{em}}{s}m_u \sin(\alpha_{\omega\phi}) (B_{c} + B_{\rho} + B_{\hat{\rho}}) l^{\mu} \varepsilon_{\mu \nu \lambda \delta} e^{*\nu}(p_{\phi}) p_{\phi}^{\lambda}  p_{\pi}^{\delta},
        	\end{equation}

The terms corresponding to the contributions from the contact diagram and the intermediate $\rho$ meson diagram are 
	\begin{equation}
		B_{c} = g_{\pi} I^{\omega}_{30},
	\end{equation}
	\begin{equation}
		B_{\rho} = \frac{C_{\rho} g_{\pi} I^{\rho\omega}_{30}}{g_{\rho}}\frac{s}{M_\rho^2 - s - i\sqrt{s}\Gamma_\rho(s)}.
	\end{equation}
	
The contribution to the amplitude from the intermediate $\hat{\rho}$ meson reads 
        	\begin{equation}
        	B_{\hat{\rho}} = \frac{C_{\hat{\rho}} g_{\pi} I^{\hat{\rho}\omega}_{30}}{g_{\rho}}\frac{s}{M_{\hat{\rho}}^2 - s - i\sqrt{s}\Gamma_{\hat{\rho}}(s)}.
        	\end{equation}
	
	Here the constants $C_{\rho}$ and $C_{\hat{\rho}}$ are defined in (\ref{C_const}). The integrals $I_{30}$ with different meson vertices are defined in (\ref{integral_ext}).
\begin{figure}[h!]
\center{\includegraphics[scale = 0.5]{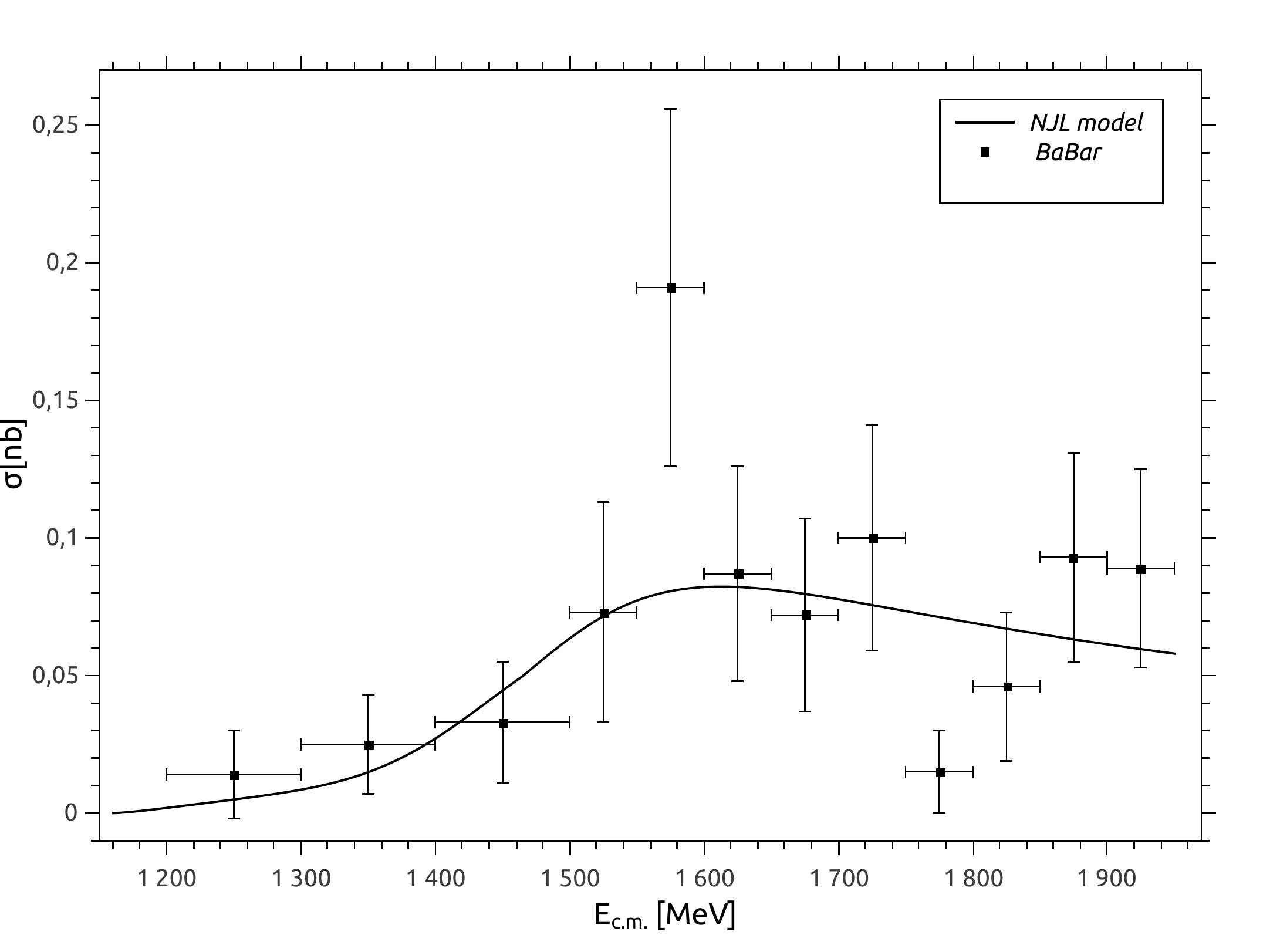}}
\caption{The total cross section of the process $e^+e^- \to \phi \pi$. The experimental points are taken from the work of the BaBar collaboration \cite{BaBar:2007ceh}}
\label{ee_phipi}
\end{figure}

A comparison of the total cross section of the process $e^+e^- \to \phi \pi$ with experimental data is shown in Figure \ref{ee_phipi}. As we can see, the results are in satisfactory agreement with the experimental data. 

\subsection{Processes $e^+e^- \to \gamma a_1, \gamma f_1$} 
In conclusion of the section devoted to the processes of $e^+e^-$ annihilation, we consider the processes $e^+ e^- \to \gamma[a_1(1260), f_1(1285)]$

Note that processes with the participation of axial-vector mesons have been insufficiently studied. The processes under consideration in the NJL model go through anomalous quark loops \cite{Osipov:2017ray}. First of all, let us describe radiative decays with the participation of axial-vector mesons $[\hat{\rho}, \hat{\omega}] \to \gamma [a_1, f_1]$. The decay width $\hat{\rho} \to \gamma f_1$ takes the form 
        \begin{eqnarray}
        \Gamma(\hat{\rho} \to \gamma f_1) = \frac{\alpha_{em}}{54} \frac{(M_{\hat{\rho}}^{2} + M_{f_{1}}^{2}) (M_{\hat{\rho}}^{2} - M_{f_{1}}^{2})^{3}}{M_{\hat{\rho}} M_{f_{1}}^{2}}   \left[-I_{3}^{(f_{1}\hat{\rho})} + 2m_{u}^{2}I_{4}^{(f_{1}\hat{\rho})} - m_{u}^{4}I_{5}^{(f_{1}\hat{\rho})}\right]^{2},
        \end{eqnarray}
where the explicit forms of the integrals $I_{3}^{(f_{1}\hat{\rho})}, I_{4}^{(f_{1}\hat{\rho})}$ and $I_{5}^{(f_{1}\hat{\rho})}$ can be found in \cite{Volkov:2017qag}. 
Similarly, one can obtain the widths of the related decays $\omega(1420)\to f_{1}(1285) \gamma$, $\rho(1450) \to a_{1}(1260) \gamma$ and $\omega(1420) \to a_{1}(1260) \gamma$. The results obtained for radiative decays are presented in Table \ref{Tab_3}.

 \begin{table}[h!]
\label{Tab_3}
\begin{center}
\begin{tabular}{ccc}
\hline
\textbf{}	& \textbf{$f_{1}(1285) \gamma$}	& \textbf{$a_{1}(1260) \gamma$}  \\
\hline
$\rho(1450) \to$   & 1.43 keV & 0.33 keV \\
$\omega(1420) \to$ & 0.07 keV &	1.63 keV \\
\hline
\end{tabular}
\end{center}
\caption{The predictions of the extended NJL model for radiative decay widths}
\end{table}

The amplitudes of the processes $e^+e^- \to \gamma a_1, \gamma f_1$ contain contributions from the contact diagram and the diagram with intermediate mesons $\rho$ and $\omega$ both in the ground and first radially excited states. The corresponding amplitude has the form 
\begin{eqnarray}
&& \mathcal{M}(e^{+}e^{-} \to f_1(1285)\gamma) = \frac{e^{3}}{s} l_{\nu} \left\{M_{f_{1}\rho}^{\mu\nu\lambda} + M_{f_{1}\omega}^{\mu\nu\lambda}\right\}  e(p_{f_{1}})_{\mu} e(p_{\gamma})_{\lambda}, \nonumber\\
&& \mathcal{M}(e^{+}e^{-} \to a_1(1260)\gamma) = \frac{e^{3}}{s} l_{\nu} \left\{M_{a_{1}\rho}^{\mu\nu\lambda} + M_{a_{1}\omega}^{\mu\nu\lambda}\right\}  e(p_{a_{1}})_{\mu} e(p_{\gamma})_{\lambda},
\end{eqnarray}
where
        \begin{eqnarray}
       && M_{f_{1}\rho}^{\mu\nu\lambda} = \frac{1}{2} \left\{I_{(f_{1})}^{\mu\nu\lambda}
        + \frac{C_{\rho}}{g_{\rho}} \frac{s}{M_{\rho}^{2} - s - i\sqrt{s}\Gamma_{\rho}}I_{(f_{1}\rho)}^{\mu\nu\lambda} + \frac{C_{\hat{\rho}}}{g_{\rho}} \frac{s}{M_{\hat{\rho}}^{2} - s - i\sqrt{s}\Gamma_{\hat{\rho}}}I_{(f_{1}\hat{\rho})}^{\mu\nu\lambda}\right\}, \nonumber\\
       && M_{f_{1}\omega}^{\mu\nu\lambda} = \frac{1}{18} \left\{I_{(f_{1})}^{\mu\nu\lambda}
        + \frac{C_{\omega}}{g_{\omega}} \frac{s}{M_{\omega}^{2} - s - i\sqrt{s}\Gamma_{\omega}}I_{(f_{1}\omega)}^{\mu\nu\lambda} + \frac{C_{\hat{\omega}}}{g_{\omega}} \frac{s}{M_{\hat{\omega}}^{2} - s - i\sqrt{s}\Gamma_{\hat{\omega}}}I_{(f_{1}\hat{\omega})}^{\mu\nu\lambda}\right\}, \nonumber
          \end{eqnarray}
          \begin{eqnarray}
       && M_{a_{1}\rho}^{\mu\nu\lambda} = \frac{1}{6} \left\{I_{(a_{1})}^{\mu\nu\lambda}
        + \frac{C_{\rho}}{g_{\rho}} \frac{s}{M_{\rho}^{2} - s - i\sqrt{s}\Gamma_{\rho}}I_{(a_{1}\rho)}^{\mu\nu\lambda} + \frac{C_{\hat{\rho}}}{g_{\rho}} \frac{s}{M_{\hat{\rho}}^{2} - s - i\sqrt{s}\Gamma_{\hat{\rho}}}I_{(a_{1}\hat{\rho})}^{\mu\nu\lambda}\right\}, \nonumber\\
       && M_{a_{1}\omega}^{\mu\nu\lambda} =\ \frac{1}{6} \left\{I_{(a_{1})}^{\mu\nu\lambda}
        + \frac{C_{\omega}}{g_{\omega}} \frac{s}{M_{\omega}^{2} - s - i\sqrt{s}\Gamma_{\omega}}I_{(a_{1}\omega)}^{\mu\nu\lambda} + \frac{C_{\hat{\omega}}}{g_{\omega}} \frac{s}{M_{\hat{\omega}}^{2} - s - i\sqrt{s}\Gamma_{\hat{\omega}}}I_{(a_{1}\hat{\omega})}^{\mu\nu\lambda}\right\},
        \end{eqnarray}
where the constants $C$ for different mesons are defined in (\ref{C_const}). The explicit form for integrals over the quark loops $I^{\mu \nu \lambda}$ for different mesons can be found in \cite{Volkov:2017qag}.

Here we have divided the amplitudes into the $\rho$ and $\omega$ meson channels combining the corresponding parts of the contact diagram with other components. 
\begin{figure}[h!]
\center{\includegraphics[scale = 0.65]{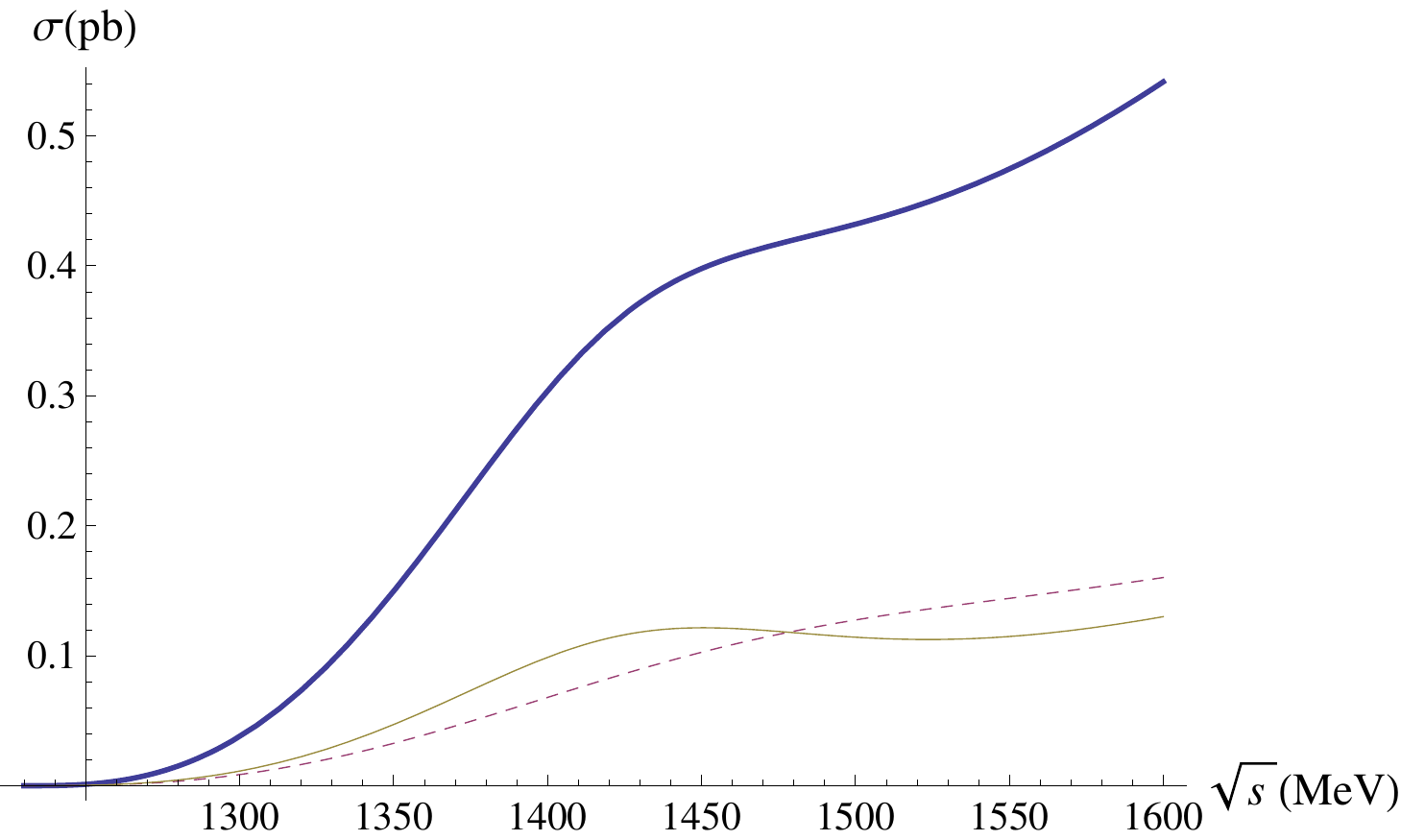}}
\caption{cross section of the process $e^+e^- \to \gamma a_1$. The thick line corresponds to the total cross section,
the dashed line corresponds to the $\rho$ meson channel (diagram with intermediate $\rho$ mesons + appropriate part of the contact diagram), the thin line corresponds to the $\omega$ meson channel (diagram with intermediate $\omega$ mesons + appropriate part of the contact diagram).}
\label{ee_a1_gamma}
\end{figure}
\begin{figure}[h!]
\center{\includegraphics[scale = 0.7]{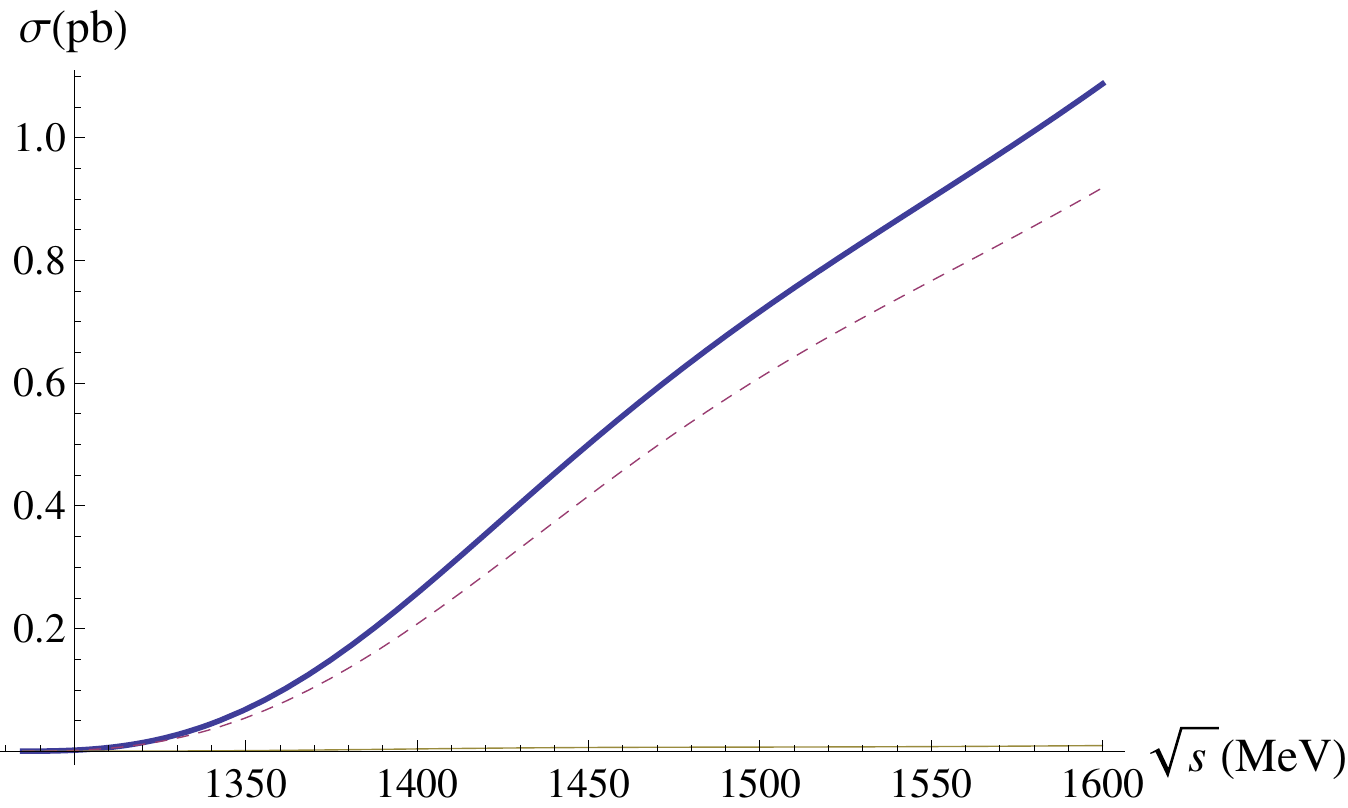}}
\caption{Cross section of the process $e^{+} e^{-} \to f_1 \gamma$. The bold line corresponds to the total cross section, the dotted and thin lines correspond to the contributions of the $\rho$ and $\omega$ mesons with a contact diagram. 
}
\label{ee_f1_gamma}
\end{figure}

As a result, the resulting cross section of the processes $e^+e^- \to \gamma [a_1, f_1]$, depending on the energy in the center-of-mass system of colliding leptons, are shown in Figures  \ref{ee_a1_gamma} and \ref{ee_f1_gamma}. The dashed lines correspond to the channel with the $\rho$ meson and the thin lines correspond to the $\omega$ meson channel. The bold lines show the total contribution. The channel from the $\omega$ mesons is two orders of magnitude lower and is almost invisible in the process  $e^{+} e^{-} \to f_1 \gamma$. 

Our calculations show that the channels with the $\rho$ and $\omega$ mesons make the same contributions to the process $e^+e^- \to \gamma a_1$, whereas in the process $e^+e^- \to \gamma f_1$ the channel with the $\rho$ mesons dominates. 

In the absence of the corresponding experimental data, our theoretical predictions can serve as a guide for future experiments and can be used to determine a physical program for further experimental studies at modern $e^+e^-$ colliders. Such experiments will allow a deeper understanding of the anomalous nature of hadronic interactions. 

\section{$\tau$ lepton decays}
\label{NJL_tau}
This Section describes some of the main $\tau$ lepton decays. The extended NJL model, which allows one to take into account the first radial excites meson states, turned out to be especially useful in the study of such processes. This is due to the value of the $\tau$ lepton mass ($m_{\tau}=1777$ MeV), which sets the energy limit for these decays. Higher excitations of mesons are, as a rule, above this energy limit and their contributions can be neglected.

An interesting feature of the processes with two pseudoscalar mesons in the final state is the need to take into account the corrections associated with the final state interactions. This requires going beyond the lower order of the $1/N_{c}$ expansion in which the NJL model is formulated. However, this interaction is not always significant. For example, in the process $e^+e^- \to K^+K^-$ considered above, there was no need to take into account the final state interactions. Such interactions will be considered in more detail in the section devoted to the $\tau$ lepton decay with two pseudoscalar particles in the final state.
\subsection{Two-particle $\tau$ lepton decays}
\subsubsection{The decays $\tau \to [\pi, \hat{\pi}] \nu_\tau$}
Lets consider the simplest $\tau \to [\pi, \hat{\pi}]  \nu_\tau$ decays. 

\begin{figure}[tb]
\center{\resizebox{0.9 \textwidth}{!}{\includegraphics{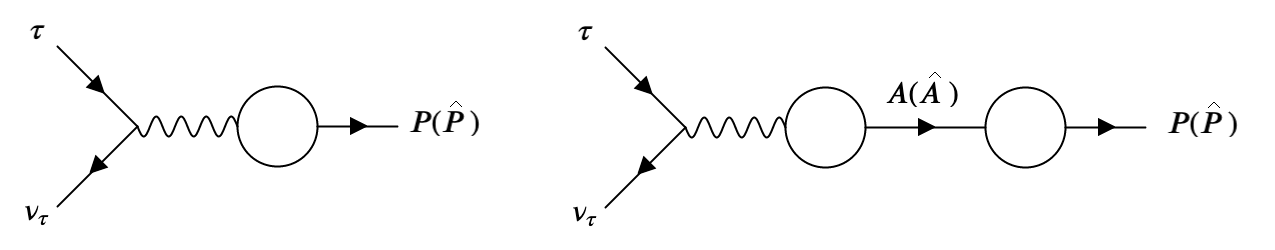}}}
\caption{Contact diagram and diagram with intermediate mesons describing decays $\tau \to P(\hat{P}) \nu_\tau$}
\label{p_cont}
\end{figure}

Diagrams describing the $\tau \to P(\hat{P}) \nu_\tau$ decays, where $P= \pi, K$ are shown in Figure \ref{p_cont}. As a result, using the quark-meson Lagrangians, we obtain the following expression for the decay amplitude $\tau \to \pi \nu_\tau$:
\begin{eqnarray}
&&\mathcal{M}_{\tau \to \pi \nu_\tau} = G_F L_\mu V_{ud} \frac{m_u}{g_\pi}\left[Z_\pi - Z_\pi \frac{6m^2_u}{M^2_{a_1}}\right] p^\mu = G_F V_{ud} F_\pi L_{\mu}p^\mu.
\end{eqnarray}

The first term in the amplitude describes the contribution from the contact diagram. The second term corresponds to the contribution from the intermediate channel with the $a_1$ meson. For the weak interaction constant, we obtain the value $F_{\pi}={m_u}/{g_{\pi}}=92.5$ MeV. This leads to full agreement of the $\tau \to \pi \nu_\tau$ decay width with the experiment $\Gamma_{exp}(\tau \to \pi \nu_\tau) = (2.46 \pm 0.01) \times 10^{-10}$ \cite{ParticleDataGroup:2020ssz}.

In the extended NJL model, we obtain the following three terms for the considered decay amplitude corresponding to the contact channel and channels with the intermediate $a_1$ and $\hat{a_1}$ mesons 
\begin{eqnarray}
&&\mathcal{M}_{\tau \to \pi \nu_\tau} = G_F L_\mu V_{ud}\biggl[ \frac{Z_\pi m_u}{g_\pi} \biggl[1- \frac{6m^2_u}{M^2_{a_1}} {\left[ \frac{\sin(\theta+\theta_0)}{\sin(2\theta_0)}+R_\rho \frac{\sin(\theta - \theta_0)}{\sin(2\theta_0)} \right]}^2 \biggl]
 \\ && \qquad  \nonumber
 -\frac{Z_\pi m_u}{g_\pi} \frac{6m^2_u}{M^2_{\hat{a}_1}} {\left[ \frac{\cos(\theta+\theta_0)}{\sin(2\theta_0)}+R_\rho \frac{\cos(\theta-\theta_0)}{\sin(2\theta_0)} \right]}^2 \biggl] p^\mu
\end{eqnarray}
where the mixing angles $\theta$ and $R_{\rho}$ are defined in Section \ref{NJL_ext}. 

The weak decay constant in the extended model takes the value $F_\pi=94.1$ MeV. Using this value for the decay width $\tau \to \pi \nu_\tau$, we obtain $\Gamma(\tau \to \pi \nu_\tau) = 2.52 \times 10^{-10}$ MeV. As we can see, the result is in satisfactory agreement with the experiment. 

Similar calculations can be performed using the vertices of the quark-meson Lagrangian for the decay $\tau \to \hat{\pi} \nu_\tau$. As a result, we obtain the following amplitude: 
            \begin{eqnarray}
            &&\mathcal{M}_{\tau \to \hat{\pi} \nu_\tau} = G_F V_{ud} F_{\hat{\pi}} L_{\mu}p^\mu,
            \end{eqnarray}
where the constant $F_{\hat{\pi}}$ in the extended NJL model takes the form 
            \begin{eqnarray}
            &&F_{\hat{\pi}} = C_{\hat{\pi}} \frac{Z_\pi m_u}{g_\pi} - \frac{C_\rho}{g_\rho} 4m_u I^{a_1 \hat{\pi}}_{20} \frac{6m^2_u}{M^2_{a_1} - M^2_{\hat{\pi}} - i \Gamma_{a_1}M_{a_1}} \left(1-\frac{M^2_{\hat{\pi}}}{M^2_{a_1}} \right)
             \\ && \qquad  \nonumber
             -\frac{C_{\hat{\rho}}}{g_\rho} 4m_u I^{\hat{a}_1 \hat{\pi}}_{20} \frac{6m^2_u}{M^2_{\hat{a}_1} - M^2_{\hat{\pi}} - i \Gamma_{\hat{a}_1}M_{\hat{a}_1}} \left(1-\frac{M^2_{\hat{\pi}}}{M^2_{\hat{a}_1}} \right) \approx 5 \ MeV,
            \end{eqnarray}
where $C_{\hat{\pi}}$, $C_{\rho}$, $C_{\hat{\rho}}$ are the constants describing the transitions of the $W$ boson to intermediate mesons (\ref{C_const}). In the propagators of axial-vector mesons, we take into account the gauge invariant form and widths of intermediate mesons $\Gamma_{a_1}= 250-600$ MeV, $\Gamma_{\hat{a}_1}= 254\pm40$ MeV \cite{ParticleDataGroup:2020ssz}. The integrals $I^{a_1 \hat{\pi}}_{20}$, $I^{\hat{a}_1 \hat{\pi}}_{20}$ correspond to the quark loops of the transitions $a_1 \to \hat{\pi}$, $\hat{a}_1 \to \hat{\pi}$ and defined in (\ref{integral_ext}). As a result, for the decay width and weak decay constant we obtain $\Gamma_{NJL} \left(\tau\rightarrow \hat{\pi} \nu_\tau\right)=1.04\times{10}^{-13}$ MeV, $F_{\hat{\pi}}=5$ MeV. This value does not exceed the experimentally established bound for the width $\Gamma_{exp}\left(\tau\rightarrow\hat{\pi}\nu_\tau\right)<4.31\times{10}^{-13}$ MeV obtained in the work \cite{CLEO:1999rzk}.

\subsubsection{The decays $\tau \to [K \hat{K}] \nu_\tau$}
The $\tau$ lepton decays into neutrino and strange mesons $K$ and $\hat{K}$ are similar in structure to the processes $\tau \to [\pi\ \hat{\pi}] \nu_\tau$. Only here the roles of intermediate mesons are played by strange mesons $K_1$, $\hat{K_1}$. The corresponding amplitude in the standard NJL model, taking into account the $K - K_{1A}$ transitions takes the form: 
\begin{eqnarray}
&&\mathcal{M}_{\tau \to K \nu_\tau} = \frac{G_F}{\sqrt{2}}L_\mu V_{us} \left( B_c + B_{K_1}\right) p^\mu,
\end{eqnarray}
where
\begin{eqnarray}
&&B_c = \sqrt{2}Z_K \frac{m_s+m_u}{2g _K},
\end{eqnarray}
\begin{eqnarray}
&&B_{K_1} = \sqrt{2}Z_K \frac{m_s+m_u}{2g _K} \left(1-Z_K \right).
\end{eqnarray}

Representing the hadronic current in the form $\sqrt{2}F_{K} p_\mu$, for the weak decay constant $F_K$ we obtain the following expression: 
\begin{eqnarray}
&&F_K = \frac{m_s+m_u}{2g _K}.
\end{eqnarray}

As a result, for the decay width $\tau\rightarrow K\nu_\tau$ and the weak interaction constant, we obtain $ \Gamma(\tau\to K\nu_\tau)=1.19\times{10}^{-11}\ MeV$ and $F_K= 95\ MeV$. The experimental data are $ \Gamma_{exp}(\tau\to K\nu_\tau)=(1.58 \pm 0.02)\times{10}^{-11}\ MeV$ and $F_K=110.2$ MeV \cite{ParticleDataGroup:2020ssz}. 

The calculated amplitude of the decay $\tau\rightarrow K\nu_\tau$ in the extended NJL model takes the form 
\begin{eqnarray}
&&\mathcal{M}_{\tau \to K \nu_\tau} = \frac{G_F}{\sqrt{2}}L_\mu V_{us}
\biggl[ Z_K \frac{m_s+m_u}{2g _K} C_K
  \\ && \qquad  \nonumber
 - \frac{6(m_s+m_u)^3}{2 M^2_{K_{1A}}} \frac{C_{K_1}}{g_{K_1}} I^{K_1K}_{11} 
 - e^{i\phi} \frac{6(m_s+m_u)^3}{2 M^2_{K_1(1650)}} \frac{{\hat{C}}_{K_1}}{g_{K_1}} I^{\hat{K}_1K}_{11}
\biggl]  p^\mu,
\end{eqnarray}
where the integrals $I^{K_1K}_{11}, I^{\hat{K}_1K}_{11}$ describe the transitions of intermediate mesons to the kaon \cite{Volkov:2019yhy}. The value $M^2_{K_{1A}}$ is defined as follows: 
            \begin{eqnarray}
            \label{Mk1A}
                M^2_{K_{1A}} = \left(\frac{\sin^2(\beta)}{M_{K_{1}(1270)}^2} + \frac{\cos^2(\beta)}{M_{K_{1}(1400)}^2}\right)^{-1/2},
            \end{eqnarray}
where the angle $\beta$ is defined in Section \ref{NJL_U3U3}. 

The extended NJL model does not claim to correctly describe the relative phases between the ground and excited states of intermediate mesons. Therefore, we are considering several versions for choosing the phase. 

Using the obtained hadronic current, similar calculations can be performed for the decays $\tau \to \hat{K} \nu_\tau$, $K \to \mu \nu_\mu$ and $\hat{K} \to \mu \nu_\mu$. The obtained theoretical estimates for the decay widths and weak decay constants are given in the Table \ref{Tab_d}.  

 \begin{table}[h!]
\begin{center}
\begin{tabular}{cccc}
\hline
 & \multicolumn{3}{c}{\textbf{Decay width in the extended NJL model, MeV}} \\
 \textbf{Decay}	& \textbf{$\phi = 0^\circ$}	& \textbf{$\phi = 180^\circ$}   & \textbf{$\phi = 102^\circ$} \\
\hline
$\tau \to K\nu_{\tau}$ & $1.30\times 10^{-11}$ & $1.77\times 10^{-11}$ & $1.59\times 10^{-11}$ \\
$\tau\to \hat{K} \nu_{\tau}$ & $2.09\times 10^{-14}$ & $2.16\times 10^{-13}$ & $1.42\times 10^{-13}$ \\
$K \to \mu \nu_\mu$ & $2.76\times 10^{-14}$ & $3.75\times 10^{-14}$& $3.37\times 10^{-14}$ \\
$\hat{K} \to \mu \nu_\mu$ & $1.14\times 10^{-15}$ & $1.18\times 10^{-14}$ & $7.81\times 10^{-15}$ \\
$F_{K}$ & $100$ & $116$ & $110.14$ \\
$F_{\hat{K}}$ & $11.3$ & $36.3$ & $29.54$ \\
\hline
\end{tabular}
\end{center}
\caption{Predictions of the extended NJL model for decay widths and weak constants}
\label{Tab_d}
\end{table}

Our calculations in the NJL model show that the dominant contribution in determining $F_{K}$ comes from the contact diagram and, accordingly, smaller contributions from intermediate axial-vector mesons.

In determining the decay width $\tau \to \hat{K} \nu_\tau$ and the constant $F_{\hat{K}}$, the contribution from the intermediate channel $K_1\left(1650\right)$ becomes commensurate with the contribution of ground states $K_{1}(1270)$ and $K_{1}(1400)$. Consequently, the phase of the excited $K_1\left(1650\right)$ meson plays an essential role. The best agreement with the experimental data was obtained in the extended NJL model with the phase $\phi=102^\circ$  of the intermediate axial-vector meson $K_1\left(1650\right)$.

\subsubsection{The decays $\tau \to [V, \hat{V}] \nu_\tau$, $\tau \to [A, \hat{A}] \nu_\tau$}
\begin{figure}[tb]
\center{\resizebox{0.45\textwidth}{!}{\includegraphics{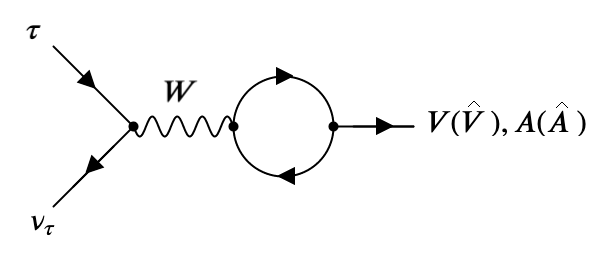}}}
\caption{Diagram, describing decays $\tau \to V (\hat{V}) \nu_\tau$ and $\tau \to A (\hat{A}) \nu_\tau$}
\label{v_cont}
\end{figure}

We now turn to the description of $\tau$ lepton decays into neutrino and one vector or axial-vector mesons both in the ground state and first radially excited states. These decays are described by the quark loop describing the transition of the $W$ boson into meson (Figure \ref{v_cont}).  

The obtained amplitude in the extended NJL model has the form \cite{Volkov:2016afr}
\begin{eqnarray}
&&\mathcal{M}_{\tau \to M(\hat{M}) \nu_\tau} = G_F L_\mu V_{ud(s)} \frac{C_{M(\hat{M})}}{g_M} \left( g^{\mu\nu}(p^2 -a_M) - p^\mu p^\nu \right) e^M_\nu, 
\end{eqnarray}
where M denotes the corresponding meson. The constants $a_M$ read 
\begin{eqnarray}
a_{\rho(\hat{\rho})} = 0, \quad a_{a_1({\hat{a}_1})}=6m^2_u, 
\end{eqnarray}
\begin{eqnarray}
 a_{K^*(\hat{K^*})}= \frac{3}{2}{\left( m_s-m_u\right)}^2, \quad a_{K_1(\hat{K}_1)} = \frac{3}{2}{\left( m_s + m_u \right)}^2.
\end{eqnarray}

Here the constants $C_{M(\hat{M})}$ are defined in (\ref{C_const}).

The obtained theoretical values for the decay widths and their experimental values are given in Table \ref{Tab2}.

\begin{table}[h!]
\begin{center}
\begin{tabular}{ccc}
\hline
\textbf{Decay}	&    \textbf{Decay width in the NJL model, MeV}	& \textbf{Experiment PDG \cite{ParticleDataGroup:2020ssz}, MeV}  \\
\hline
$\tau \to \rho \nu_\tau$ & $4.19 \times 10^{-10}$ & -- \\
$\tau \to \rho(1450) \nu_\tau$ & $4.79 \times 10^{-11}$ & --  \\
$\tau \to a_1(1260) \nu_\tau$ & $2.94 \times 10^{-10}$ & -- \\
$\tau \to a_1(1640) \nu_\tau$ & $6.96 \times 10^{-11}$ & --\\
$\tau \to K^*(892) \nu_\tau$ & $2.07 \times 10^{-11}$ & $(2.72 \pm 0.15) \times 10^{-11}$ \\
$\tau \to K^*(1410) \nu_\tau$ & $3.67 \times 10^{-12}$ & $3.4(+3.17, -2.27) \times 10^{-12}$ \\
$\tau \to K_1(1270) \nu_\tau$ & $0.57 \times 10^{-11}$ & $(1.06 \pm 0.24) \times 10^{-11}$ \\
$\tau \to K_1(1400) \nu_\tau$ & $2.4 \times 10^{-12}$ & $(3.86 \pm 5.90) \times 10^{-12}$ \\
$\tau \to K_1(1650) \nu_\tau$ & $3.34 \times 10^{-13}$ & --- \\
\hline
\end{tabular}
\end{center}
\caption{Predictions of the extended NJL model for $\tau$ lepton decays into vector and axial vector mesons}
\label{Tab2}
\end{table}

As a result, in the NJL model, $\tau$ lepton decays into neutrino and one meson (pseudoscalar, vector, axial vector)  were described both in the ground and radially excited states. The obtained results taking into account the model accuracy can be considered quite satisfactory. Note that the accuracy of the $U(3) \times U(3)$ model is $\pm 17\%$. In more complex $\tau$ decays, where the final products are neutrino and two mesons intermediate channels with the meson states described above, play a decisive role. 

    \subsection{The decays $\tau \to P P \nu_{\tau}$}
        \label{tau_PP}
        \subsubsection{The processes $\tau \to \pi \pi \nu_\tau$ and $e^+ e^- \to \pi^+ \pi^-$}
            The process $\tau \to \pi \pi \nu_\tau$ is the most probable decay mode of the $\tau$ lepton. In many theoretical works, the agreement with experiment is achieved by the phenomenological parameterization of the pion form factor and by fitting the results with experimental data \cite{Kuhn:1990ad, Bartos:2017oam, Miranda:2018cpf, Dai:2018thd}.
            
\begin{figure}[h]
        		\center{\includegraphics[scale = 0.25]{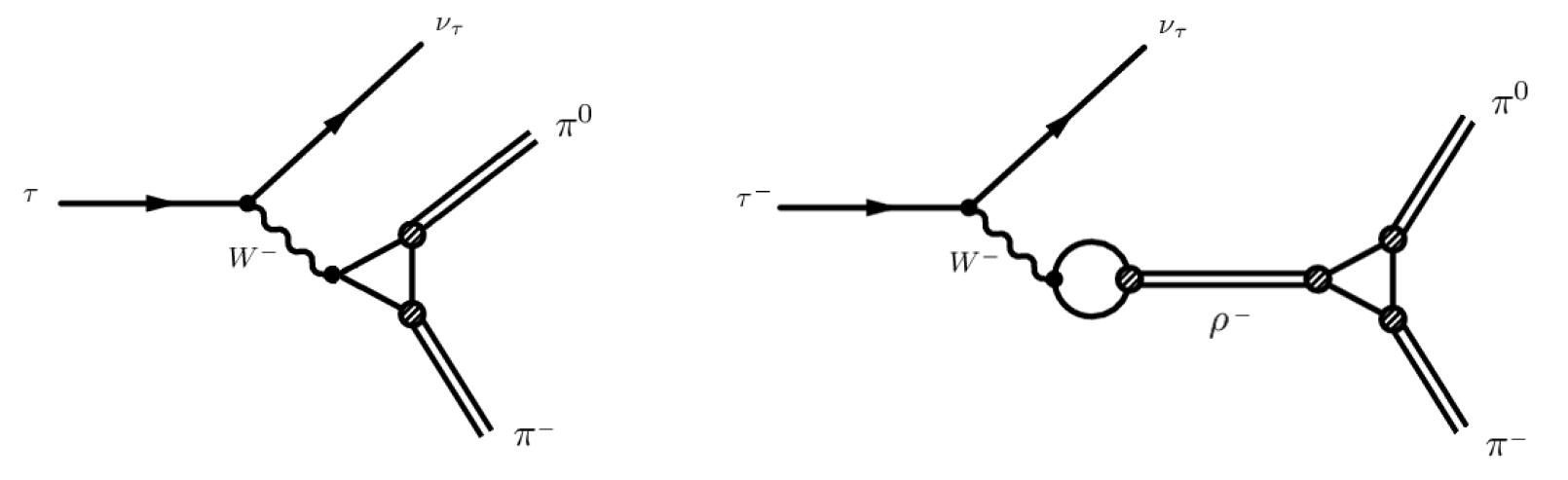}}
        		\caption{Contact diagram and diagram with intermediate $\rho$ meson of the process $\tau \to \pi \pi \nu_\tau$}
        		\label{tauContact}
        	\end{figure}

            The diagrams of the process $\tau \to \pi \pi \nu_\tau$ are shown in Figure ~\ref{tauContact}.
        	
        	The threshold of the final pions production is lower than the value of the intermediate $\rho$ meson mass. That is why the diagram with the pointed meson gives the main contribution. The contribution of the diagram with the excited meson state is negligible. This allows us to restrict ourselves to the standard NJL model when considering this process.
        	
        	The calculation result of the process $\tau \to \pi \pi \nu_\tau$ in the framework of the NJL model is 30\% lower than the experimental value ~\cite{Volkov:2020dvz}. The discrepancy of the theoretical and experimental results indicates the need of taking into account additional effects such as the interactions of mesons in the final state. This interaction is beyond the NJL model because it leads to higher degrees of $1/N_{c}$ than the NJL model allows.
        	
            In this Section, we present the results of studying on the possibility of considering such contributions in addition to the results obtained in the standard NJL model for a number of processes.
            
\begin{figure}[h]
        		\center{\includegraphics[scale = 0.4]{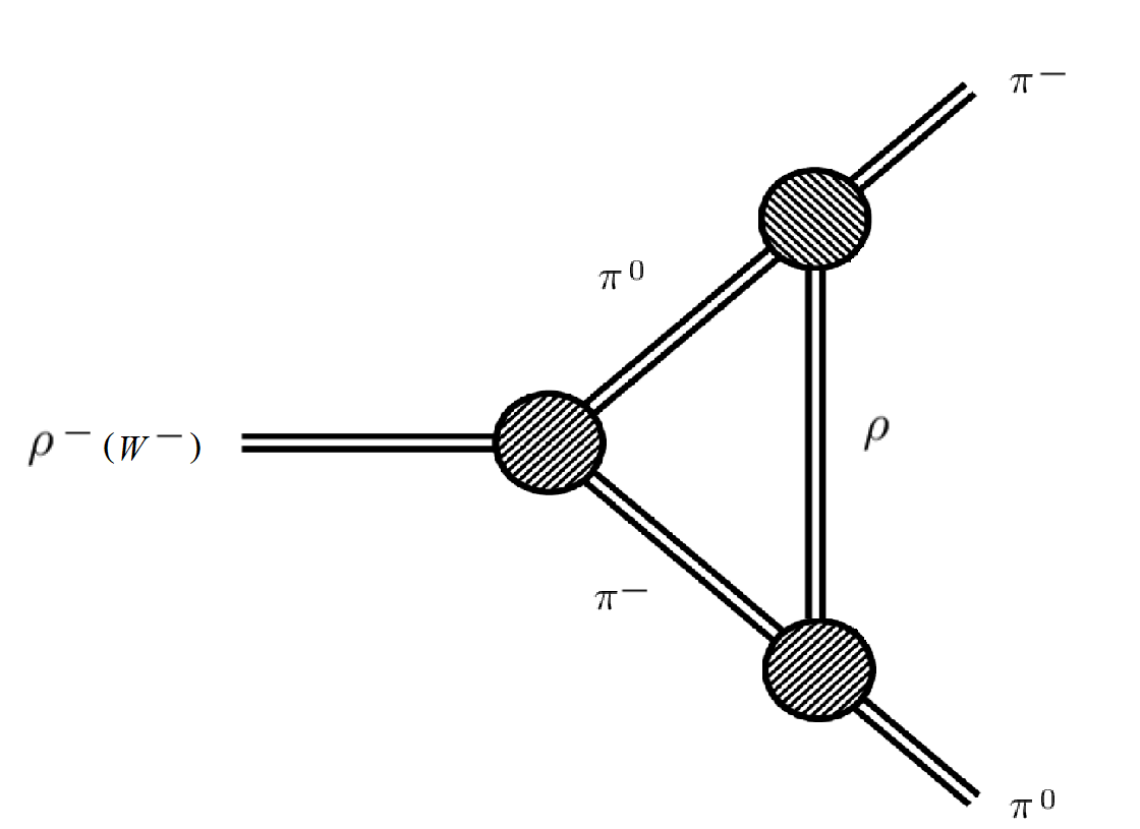}}
        		\caption{Diagram describing mesons interactions in the final state.}
        		\label{rho(W)pi0pi}
        	\end{figure}

            Those interactions in the final state for the process $\tau \to \pi \pi \nu_\tau$ can be represented as a triangle shown in Figure ~\ref{rho(W)pi0pi}.
        	
        	This triangle can be described with the integral of the following form \cite{Volkov:2020dvz}:
        	\begin{eqnarray}
        		g_{\rho}^{3} \int \frac{\left(k - 2p_{0}\right)^{\lambda}\left(k + 2p_{-}\right)^{\nu}
        		\left(2k + p_{-} - p_{0}\right)^{\mu}}{\left[k^{2} - M_{\rho}^{2}\right]
        		\left[(k - p_{0})^{2} - M_{\pi}^{2}\right]\left[(k + p_{-})^{2} - M_{\pi}^{2}\right]} 
		\left(g_{\nu\lambda} 
        		- \frac{k_{\nu}k_{\lambda}}{M_{\rho}^{2}}\right)
        	    \frac{d^{4}k}{(2\pi)^{4}}.
        	\end{eqnarray}
	
        	Expanding this integral in external momenta and leaving only divergent terms similarly to the method applied to the quark loops in the NJL model, one can obtain the following results:
        	\begin{eqnarray}
        		i g_{\rho}^{3} \left[\frac{I_{\rho}}{M_{\rho}^{2}} + I_{\rho\pi}\right] \left(p_{-} - p_{0}\right)^{\mu},
        	\end{eqnarray}
        	where $I_{\rho}$ and $I_{\rho\pi}$ are the quadratic and logarithmic divergent integrals, respectively:
        	\begin{eqnarray}
        	\label{meson_integ}
        	    I_{\rho} & = & \frac{-i}{(2\pi)^{4}}\int\frac{\Theta(\Lambda_{\pi\pi}^{2} + k^2)}{(M_{\rho}^{2} - k^2)} \mathrm{d}^{4}k 
        		= \frac{1}{(4\pi)^{2}} \left[\Lambda_{\pi\pi}^{2} - M_{\rho}^{2}\ln\left(\frac{\Lambda_{\pi\pi}^{2}}{M_{\rho}^{2}} + 1\right)\right], \nonumber\\
        		I_{\rho\pi} & = &
        		\frac{-i}{(2\pi)^{4}}\int\frac{\Theta(\Lambda_{\pi\pi}^{2} + k^2)}{(M_{\rho}^{2} - k^2)(M_{\pi}^{2} - k^2)} \mathrm{d}^{4}k \nonumber\\
        		& = & \frac{1}{(4\pi)^{2}}\frac{1}{M_{\rho}^{2} - M_{\pi}^{2}}\left[M_{\rho}^{2}
        		\ln\left(\frac{\Lambda_{\pi\pi}^{2}}{M_{\rho}^{2}} + 1\right) 
        		- M_{\pi}^{2}\ln\left(\frac{\Lambda_{\pi\pi}^{2}}{M_{\pi}^{2}} + 1\right)\right], 
        	\end{eqnarray}
        	were $\Lambda_{\pi\pi}$ is the cut-off parameter of the meson loop.
        	
        	Then the amplitude of the considered decay takes the form
        	\begin{eqnarray}
        	\mathcal{M}(\tau \to \pi \pi \nu) = -G_{F} V_{ud} \left[1 + \frac{s}{M_{\rho}^{2} - s - i \sqrt{s}\Gamma_{\rho}}\right]
        		  \\ \qquad  \nonumber
		\times \left\{1 + g_{\rho}^{2}\left[\frac{I_{\rho}}{M_{\rho}^{2}} + 
        		I_{\rho\pi}\right]\right\} L_{\mu} \left(p_{-} - p_{0}\right)^{\mu}.
        	\end{eqnarray}
	
        	The first term in the squared brackets describes the contact diagram. The second term describes the diagram with the intermediate $\rho$ meson. The first term in the curly brackets corresponds to the amplitude obtained in the standard NJL model. The second term is a correction taking into account the interactions in the final state. Without this correction the result for the considered process is
        	\begin{eqnarray}
        		\mathrm{Br}(\tau \to \pi \pi \nu) = (17 \pm 0.85)\%.
        	\end{eqnarray}
        	It is 30\% lower than the experimental value ~\cite{ParticleDataGroup:2020ssz}
        	\begin{eqnarray}
        	\label{tau_pipi_exp}
        		\mathrm{Br}(\tau \to \pi \pi \nu)_{exp} = (25.49\pm 0.09)\%.
        	\end{eqnarray}
        	
        	For calculation of the correction from the interactions in the final state in this process, it is necessary to know the value of the new cut-off parameter appearing in the meson loop $\Lambda_{\pi\pi}$. For its fixation, one can consider the process $e^+ e^- \to \pi^+ \pi^-$ whose structure is close to the structure of the process $\tau \to \pi \pi \nu_\tau$. The hadron currents in these processes are related to each other by the rotation in the isotopic space. This gives grounds to use the same set of parameters for both processes including the parameter $\Lambda_{\pi\pi}$. 
        	
        	The amplitude of the process $e^+ e^- \to \pi^+ \pi^-$ takes the form~\cite{Volkov:2020dvz}:
        	\begin{eqnarray}
        		M(e^{+}e^{-} \to \pi \pi) = -\frac{4 \pi \alpha_{em}}{s} 
        		\biggl[1 + \frac{s}{M_{\rho}^{2} - s - i \sqrt{s}\Gamma_{\rho}} 
		\nonumber\\
        		+ \frac{s^{2}}{9} \frac{g_{\rho}^{2}\left[I_{2}(u) - I_{2}(d)\right]}{\left[M_{\rho}^{2} 
        			- s - i \sqrt{s}\Gamma_{\rho}\right] \left[M_{\omega}^{2} - s - i \sqrt{s}\Gamma_{\omega}\right]} \biggl]
        		 \nonumber\\
        	         \times \left\{1 + g_{\rho}^{2}\left[\frac{I_{\rho}}{M_{\rho}^{2}} 
        		+ I_{\rho\pi}\right]\right\} 
        		l_{\mu} (p_{+} - p_{-})^{\mu}.
        	\end{eqnarray}
	
\begin{figure}[h]
        		\center{\includegraphics[scale = 0.6]{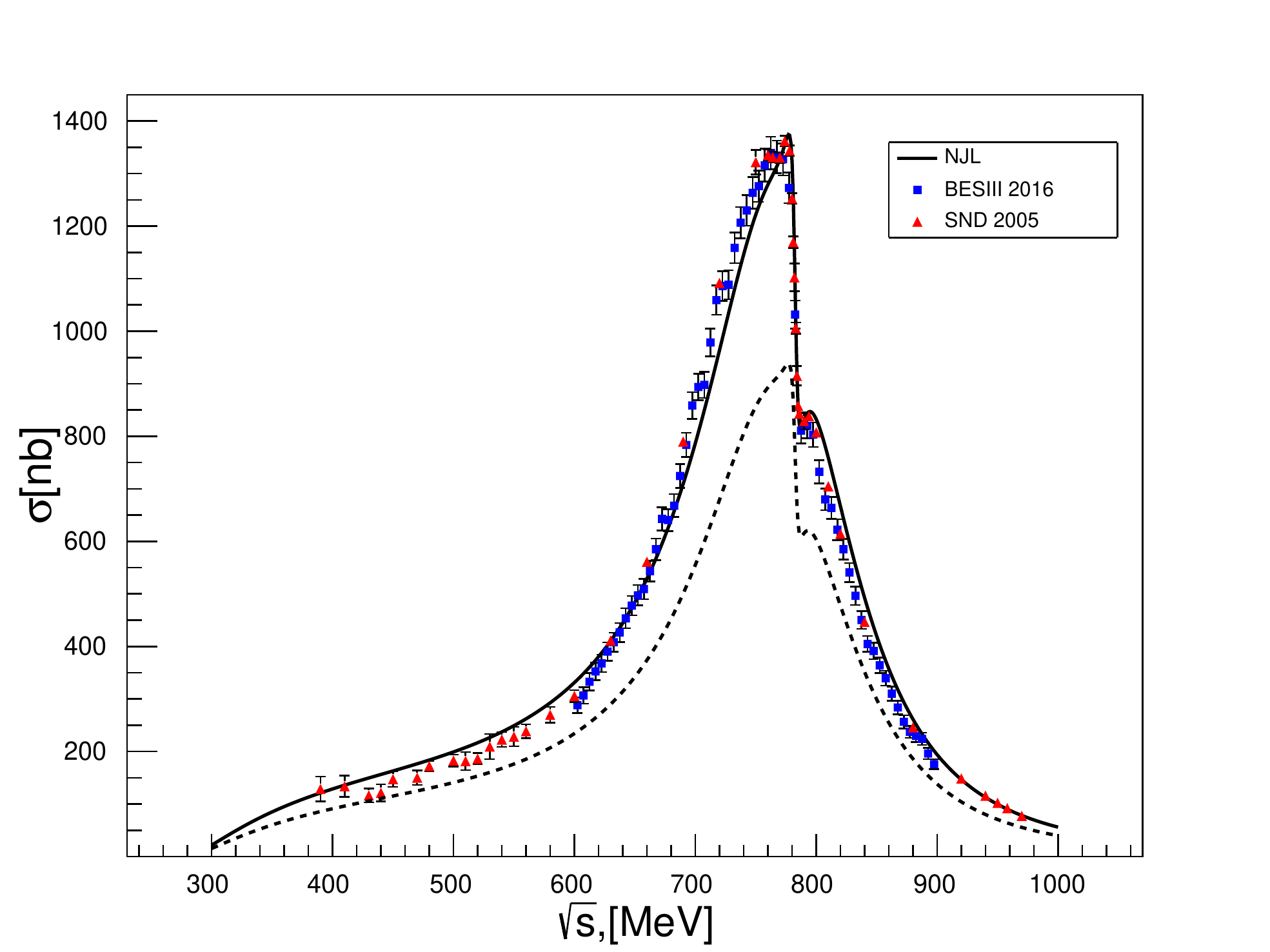}}
        		\caption{Cross section for the process $e^{+}e^{-} \to \pi^+ \pi^-$ versus the c.m. energy. The experimental points are taken from~\cite{Achasov:2005rg, BESIII:2015equ}. The solid and dashed lines are obtained with and without the contribution of interactions in the final state, respectively.}
        		\label{CrossSection}
        	\end{figure}

        	The third term in the squared brackets describes the diagram with the intermediate $\omega$ meson. By the known dependence of the cross-section of this process on the energy of the colliding leptons, one can fix the cut-off parameter of the meson loop $\Lambda_{\pi\pi} = 740$ MeV \cite{Volkov:2020dvz}. The appropriate diagram is shown in Fig.~\ref{CrossSection}. One can see that taking into account the interaction in the final state is especially important near the resonance.
        	
        	By using the obtained value of $\Lambda_{\pi\pi}$, the branching fraction of the decay $\tau \to \pi \pi \nu_\tau$ has been calculated~\cite{Volkov:2020dvz}:
        	\begin{eqnarray}
        		\mathrm{Br}(\tau \to \pi \pi \nu) = (25.1\pm 1.2)\%.
        	\end{eqnarray}
        	 
        	This result is in satisfactory agreement with the experimental data presented in (\ref{tau_pipi_exp}).
	
        \subsubsection{The processes $\tau \to [\eta, \eta'] \pi \nu_\tau$}
            The processes $\tau \to [\eta, \eta'] \pi \nu_\tau$ refer to the decays that include the second-class currents. These decays are suppressed by G-parity violation and can occur only due to the mass difference between light $u$ and $d$ quarks. These processes were researched within different phenomenological models \cite{Bramon:1987zb, Neufeld:1994eg, Nussinov:2008gx, Nussinov:2009sn, Paver:2010mz, Paver:2011md, Descotes-Genon:2014tla, Escribano:2016ntp, Hernandez-Tome:2017pdc, Garces:2017jpz}.
            
            For the description of the process $\tau \to \eta \pi \nu_\tau$ one can use the results obtained for the process $\tau \to \pi \pi \nu_\tau$ described above by adding the transition $\pi^{0}-\eta$ to it. Then the amplitude takes the form
            \begin{eqnarray}
                \mathcal{M}(\tau \to \pi \eta \nu_{\tau}) = G_F V_{ud} T_{\pi \eta} L_{\mu} M^2_{\rho} \left(1-\frac{i\sqrt{s}\Gamma_{\rho}}{M^2_{\rho}} \right) BW_{\rho} 
                \nonumber \\
               \times \left[ (p_{\eta}-p_{\pi})^{\mu}+ g^2_{\rho} \left( a(s) p^{\mu}_{\eta} -b(s) p^{\mu}_{\pi} \right) \right],
            \end{eqnarray}
            where $a(s)$ and $b(s)$ are the functions appearing as a result of taking into account the interactions in the final state:
            \begin{eqnarray}
            \label{afunc}
                a(s) = \frac{I_{\rho}}{M^2_{\rho}} + I_{\rho\pi}+I_{\rho 2\pi}\frac{M^2_{\pi}(M^2_{\eta} - M^2_{\pi})}{M^2_{\rho}} 
                - I_{\rho3\pi}\frac{M^4_{\pi}( -M^2_{\eta}+7M^2_{\pi}+6M^2_{\rho}+s)}{6M^2_{\rho}} \\ \nonumber
                - I_{\rho4\pi}\frac{M^6_{\pi}( 23M^2_{\eta}+M^2_{\pi}-5s)}{6M^2_{\rho}} 
                + 4 I_{\rho5\pi}\frac{M^8_{\pi}( 4M^2_{\eta}+2M^2_{\pi}-s)}{6M^2_{\rho}},
            \end{eqnarray} 
            \begin{eqnarray}
            \label{bfunc}
                b(s) = \frac{I_{\rho}}{M^2_{\rho}} + I_{\rho\pi}\frac{M^2_{\eta}-M^2_{\pi}+M^2_{\rho}}{M^2_{\rho}}
                - I_{\rho3\pi}\frac{M^4_{\pi}(13M^2_{\eta}-7M^2_{\pi}+6M^2_{\rho}+s)}{6M^2_{\rho}} \\ \nonumber
                - I_{\rho4\pi}\frac{M^6_{\pi}(M^2_{\eta}+23 M^2_{\pi}-5s)}{6M^2_{\rho}} 
                + 4 I_{\rho5\pi}\frac{M^8_{\pi}( 2M^2_{\eta}+4M^2_{\pi}-s)}{6M^2_{\rho}}.
            \end{eqnarray} 
            
            Here the integrals over the meson loops are
            \begin{eqnarray}
                \label{meson_integral_eta}
                && I_{\rho n \pi} = \frac{-i}{(2\pi)^{4}}\int\frac{\Theta(\Lambda_{\pi\eta}^{2} + k^2)}
                {(M^2_{\rho} - k^2)({M^2_{\pi}} - k^2)^n}\mathrm{d}^{4}k.
            \end{eqnarray}
            
            The Breit-Wigner propagator takes the standard form
            \begin{eqnarray}
            \label{BreitWigner}
                BW_{\rho} = \frac{1}{M_{\rho}^{2} - s - i\sqrt{s}\Gamma_{\rho}}.
            \end{eqnarray}
            
            The process $\tau \to \pi \eta \nu_{\tau}$ differs from the process $\tau \to \pi \pi \nu_{\tau}$ by the fact that it has mesons with different masses in the final state. This leads to the necessity of taking into account additional terms with the convergent integrals for the elimination of uncertainties.
            
            The factor $T_{\pi \eta}$ describes the transition $\pi^{0}-\eta$
            \begin{eqnarray}
                T_{\pi\eta(\eta')} = 2g^2_{\pi} \left[ \left(2I_1(m_d) + M^2_{\eta(\eta')}I_2(m_d) \right) - \left( 2I_1({m_u}) + M^2_{\eta(\eta')}I_2({m_u}) \right) \right] \frac{\sin(\bar{\alpha})(\cos(\bar{\alpha})) }{M^2_{\pi}-M^2_{\eta(\eta')}}, 
            \end{eqnarray}
            
            By using the same cut-off parameter as in the case of the process $\tau \to \pi \pi \nu_{\tau}$, one can obtain the following result:
            \begin{eqnarray}
                Br(\tau \to \pi \eta \nu_{\tau}) = 1.87 \times 10^{-5}.
            \end{eqnarray}
            It is not beyond the experimental restrictions:
            \begin{eqnarray}
                Br(\tau \to \pi \eta \nu_{\tau})_{exp} & < & 9.9 \times 10^{-5} \textrm{ \cite{BaBar:2010bul, BaBar:2012zfq}}, \nonumber\\
                Br(\tau \to \pi \eta \nu_{\tau})_{exp} & < & 7.3 \times 10^{-5} \textrm{ \cite{Hayasaka:2009zz}}
            \end{eqnarray}
            
            While describing the process $\tau \to \eta' \pi \nu_\tau$, it is necessary to apply the extended NJL model for taking into account the excited mesons in the intermediate state because of the higher value of the threshold of the final meson production.
            
            Then after taking into account the interactions in the final state the amplitude takes the form
            \begin{eqnarray}
            \label{amplitude1}
                \mathcal{M}(\tau \to \pi \eta' \nu_{\tau}) = G_{F} V_{ud} Z_{\pi} T_{\pi\eta'} L_{\mu} \left\{ \left[\mathcal{M}_{c} + \mathcal{M}_{\rho} + \mathcal{M}_{\hat{\rho}}\right]^{\mu\nu} \left(p_{\eta} - p_{\pi}\right)_{\nu} \right. \nonumber\\ 
                 \left. + \left[\mathcal{M}_{c(loop)} + \mathcal{M}_{\rho(loop)} + \mathcal{M}_{\hat{\rho}(loop)}\right]^{\mu\nu}\left(a(s) p_{\eta} - b(s) p_{\pi}\right)_{\nu} \right\}.
            \end{eqnarray}
            
            Here the functions $a(s)$ and $b(s)$ have been constructed by the exchange $M^2_{\eta} \to M^2_{\eta'}$ in the definitions (\ref{afunc}) and (\ref{bfunc}). The terms in the squared brackets in the amplitude (\ref{amplitude1}) describe the contact diagram and the diagram with the intermediate $\rho$ mesons in the ground and first radially excited states:
            \begin{eqnarray}
              &&  \mathcal{M}_{c}^{\mu\nu}= \left[1 - C^2_{\rho} \frac{6 m_{u}^{2}}{M_{a_{1}}^{2}}\right] g^{\mu\nu}, \nonumber \\
              &&  \mathcal{M}_{\rho}^{\mu\nu} = C_{\rho}^{2} \left[1 - 4 I_{20}^{\rho a_{1}} \frac{m_{u}^{2}}{M_{a_{1}}^{2}}\right] \frac{g^{\mu\nu}q^{2} - q^{\mu}q^{\nu}}{M_{\rho}^{2} - q^{2} - i \sqrt{q^{2}}\Gamma_{\rho}}, \nonumber
            \end{eqnarray}  
            \begin{eqnarray}
                \mathcal{M}_{\hat{\rho}}^{\mu\nu} = e^{i \pi} C_{\hat{\rho}}^{2} \left[1 - 4 I_{20}^{\hat{\rho} a_{1}} \frac{C_{\rho}}{C_{\hat{\rho}}} \frac{m_{u}^{2}}{M_{a_{1}}^{2}}\right] \frac{g^{\mu\nu}q^{2} - q^{\mu}q^{\nu}}{M_{\hat{\rho}}^{2} - q^{2} - i \sqrt{q^{2}}\Gamma_{\hat{\rho}}}, \nonumber
            \end{eqnarray}  
            \begin{eqnarray}    
                \mathcal{M}_{c(loop)}^{\mu\nu} = g_{\rho}^{2} Z_{\pi}^{2} C_{\rho}^{2} \left[1 - C_{\rho}^{2} \frac{6 m_{u}^{2}}{M_{a_{1}}^{2}}\right]\left[1 - 4 I_{20}^{\rho a_{1}} \frac{m_{u}^{2}}{M_{a_{1}}^{2}}\right]^{2} g^{\mu\nu}, \nonumber
            \end{eqnarray}    
            \begin{eqnarray}     
                \mathcal{M}_{\rho(loop)}^{\mu\nu} = g_{\rho}^{2} Z_{\pi}^{2} C_{\rho}^{4} \left[1 - 4 I_{20}^{\rho a_{1}} \frac{m_{u}^{2}}{M_{a_{1}}^{2}}\right]^{3} \frac{g^{\mu\nu}q^{2} - q^{\mu}q^{\nu}}{M_{\rho}^{2} - q^{2} - i \sqrt{q^{2}}\Gamma_{\rho}}, \nonumber
            \end{eqnarray} 
            \begin{eqnarray}      
                \mathcal{M}_{\hat{\rho}(loop)}^{\mu\nu} = e^{i \pi} g_{\rho}^{2} Z_{\pi}^{2} C_{\hat{\rho}}^{2} C_{\rho}^{2} \left[1 - 4 I_{20}^{\hat{\rho} a_{1}} \frac{C_{\rho}}{C_{\hat{\rho}}} \frac{m_{u}^{2}}{M_{a_{1}}^{2}}\right]\left[1 - 4 I_{20}^{\rho a_{1}} \frac{m_{u}^{2}}{M_{a_{1}}^{2}}\right]^{2} \frac{g^{\mu\nu}q^{2} - q^{\mu}q^{\nu}}{M_{\hat{\rho}}^{2} - q^{2} - i \sqrt{q^{2}}\Gamma_{\hat{\rho}}}, \nonumber
            \end{eqnarray}
        	where $q$ is the momentum of the intermediate meson; the integrals over quark loops $I_{20}^{\rho a_{1}}$ and $I_{20}^{\hat{\rho} a_{1}}$ are defined in (\ref{integral_ext}).
	
            The branching fraction of this process takes the value
            \begin{eqnarray}
                Br(\tau \to \pi \eta' \nu_{\tau}) = 1.25 \times 10^{-7}.
            \end{eqnarray}
            It is also within the experimental restrictions~\cite{BaBar:2010bul, BaBar:2012zfq}:
            \begin{eqnarray}
                Br(\tau \to \pi \eta' \nu_{\tau})_{exp} < 40 \times 10^{-7}.
            \end{eqnarray}
            
            The Belle II experimental collaboration presented the research program in recent works 
\cite{Belle-II:2010dht, Belle-II:2018jsg}. The upcoming experiment will allow to study the 
second class current decays more accurately. We hope that our results will receive 
experimental confirmation. 

       \subsubsection{The process $\tau \to K \pi \nu_\tau$}
            Interest in the process $\tau \to K^{-} \pi \nu_{\tau}$ is due to the fact that it is applied in studying the vacuum polarization and that it includes strange and non-strange particles simultaneously. This process was investigated in different theoretical works \cite{Finkemeier:1996dh,Jamin:2006tk,Boito:2008fq}.
            
\begin{figure}[h]
        		\center{\includegraphics[scale = 0.2]{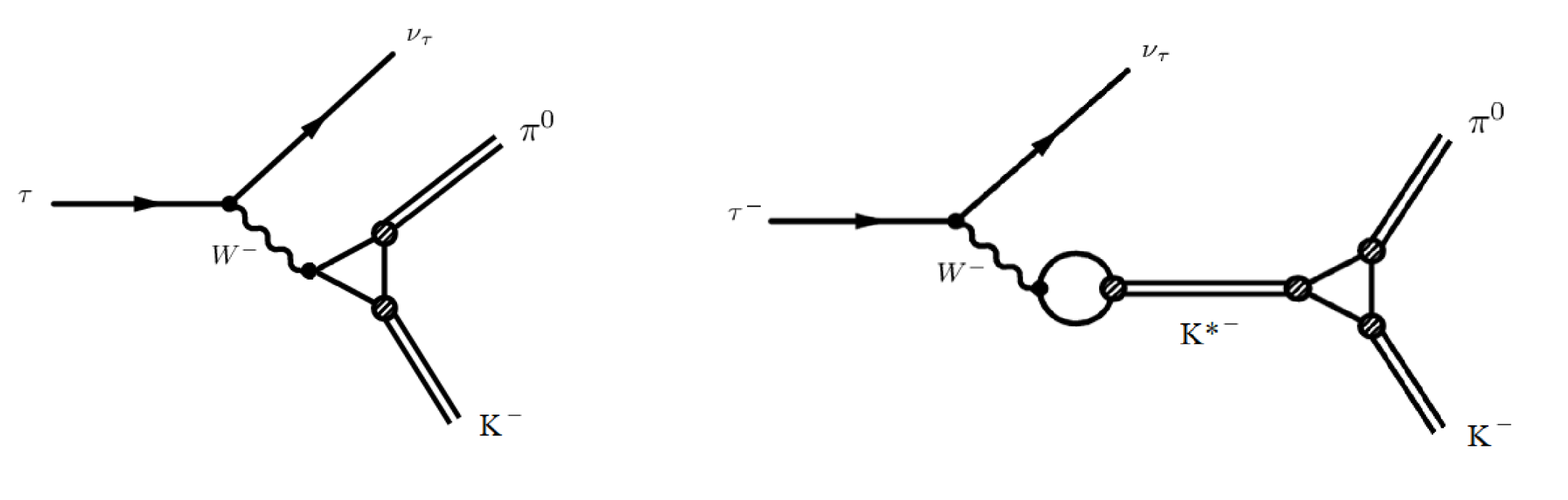}}
        		\caption{Contact diagram and diagram with the intermediate meson of the process $\tau \to K^{-} \pi^{0} \nu_{\tau}$.}
        		\label{ContactKPI}
        	\end{figure}

            The diagrams of this process are shown in Figure ~\ref{ContactKPI}.
        	
        	Due to the low energy threshold of the meson production in the process $\tau \to K^{-} \pi \nu_{\tau}$ the contribution of the excited mesons state is negligible and one can use the standard NJL model.
        	
        	The amplitude of this process in the standard NJL model takes the form
        	\begin{eqnarray}
        		\mathcal{M}(\tau \to K^{-} \pi^{0} \nu_{\tau})_{tree} = -3 G_{f} V_{us} \frac{g_{K}g_{\pi}}{g_{K^{*}}^{2}} L_{\mu} \biggl[ g^{\mu\nu} 
		\nonumber \\
		+ \frac{g^{\mu\nu}q^{2}-q^{\mu}q^{\nu}}{M_{K^{*}}^{2} - q^{2} - i\sqrt{q^{2}}\Gamma_{K^{*}}} \biggl] \cdot \left(T_{K} p_{K\nu} - T_{\pi} p_{\pi\nu}\right),
        	\end{eqnarray}
        	where the factors $T_{K}$ and $T_{\pi}$ describe the  $a_{1}-\pi$ and $K_{1}-K$ transitions:
        	\begin{eqnarray}
        		&& T_{\pi} = 1 - 3\frac{m_{u}\left(3m_{u} - m_{s}\right)}{M_{a_{1}}^{2}}, \\
        		&& T_{K} = 1 - 3\frac{m_{s}\left(m_{u} + m_{s}\right)}{M_{K_{1A}}^{2}}.
        	\end{eqnarray}
	
	Here the value $M_{K_{1A}}$ is determined in (\ref{Mk1A}). 
        	
        	This amplitude leads to the following value of the branching fraction:
        	\begin{eqnarray}
        		Br(\tau \to K^{-} \pi^{0} \nu_{\tau})_{tree} = 2.92 \times 10^{-3}.
        	\end{eqnarray}
        	This result is significantly lower than the experimental data~\cite{ParticleDataGroup:2020ssz}:
        	\begin{eqnarray}
        		Br(\tau \to K^{-} \pi^{0} \nu_{\tau})_{exp} = (4.33 \pm 0.15) \times 10^{-3}.
        	\end{eqnarray}
	
        	This may be caused by the necessity of taking into account the interactions in the final state.
        	
\begin{figure}[h]
        		\center{\includegraphics[scale = 1.0]{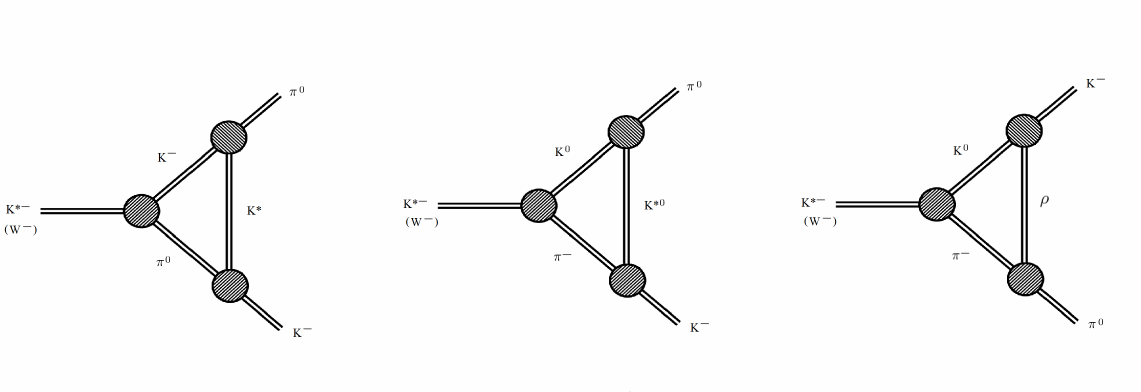}}
        		\caption{The interactions of the kaon and pion in the final state.}
        		\label{Kv_in_Kpi}
        	\end{figure}

        	To take into account the interactions in the final state one can consider three possible diagrams of the meson exchange given in Figure ~\ref{Kv_in_Kpi}.
        	
        	These meson triangles lead to the following integrals~\cite{Volkov:2021bfx}:
        	\begin{eqnarray}
        	\label{meson_loops}
        		F^{K^{*\pm}}_{\mu} = \int \frac{\left(T_{K}k - \left(T_{K} + T_{\pi}\right)p_{\pi}\right)_{\lambda}\left(T_{\pi}k + \left(T_{K} + T_{\pi}\right)p_{K}\right)_{\nu}
        		}{\left[k^{2} - M_{K^{*}}^{2}\right]
        		\left[(k + p_{K})^{2} - M_{\pi}^{2}\right]\left[(k - p_{\pi})^{2} - M_{K}^{2}\right]} 
		\nonumber \\
        		\times \left(\left(T_{K} + T_{\pi}\right)k + T_{\pi}p_{K} - T_{K}p_{\pi}\right)_{\mu} \left(g^{\nu\lambda} 
        		- \frac{k^{\nu}k^{\lambda}}{M_{K^{*}}^{2}}\right) \frac{d^{4}k}{(2\pi)^{4}},
		\end{eqnarray} 
		\begin{eqnarray}
		\label{meson_loops_2}
        		F^{\rho}_{\mu} = \int \frac{\left(k - 2p_{\pi}\right)_{\lambda}\left(k + 2p_{K}\right)_{\nu}
        		\left(\left(T_{K} + T_{\pi}\right)k + T_{K}p_{K} - T_{\pi}p_{\pi}\right)_{\mu}}{\left[k^{2} - M_{\rho}^{2}\right]
        		\left[(k + p_{K})^{2} - M_{K}^{2}\right]\left[(k - p_{\pi})^{2} - M_{\pi}^{2}\right]}
		\nonumber \\
		\times \left(g^{\nu\lambda} - \frac{k^{\nu}k^{\lambda}}{M_{\rho}^{2}}\right)
        		\frac{d^{4}k}{(2\pi)^{4}},
		\end{eqnarray} 
		\begin{eqnarray}
        		F^{K^{*0}}_{\mu} = F^{K^{*\pm}}_{\mu}.
         	\end{eqnarray} 
        	
        	These integrals are divergent and can be regularized with the cut-off parameter $\Lambda_{K\pi}$.
        	
        	The expressions for the integrals (\ref{meson_loops}) and (\ref{meson_loops_2}) are presented in the work~\cite{Volkov:2021bfx}.
        	
        	As a result, the correction to the amplitude from these triangles takes the form
        	\begin{eqnarray}
        		\mathcal{M}(\tau \to K^{-} \pi^{0} \nu_{\tau})_{loop} =  -3iG_{f} V_{us} \frac{g_{K}g_{\pi}}{g_{K^{*}}^{2}} L_{\mu} \left[g^{\mu\nu} + \frac{g^{\mu\nu}q^{2}-q^{\mu}q^{\nu}}{M_{K^{*}}^{2} - q^{2} - i\sqrt{q^{2}}\Gamma_{K^{*}}}\right] \nonumber\\
        		 \times \left\{-\left(3\frac{g_{K}g_{\pi}}{g_{K^{*}}}\right)^{2}F^{K^{*\pm}}_{\nu} + g_{\rho}^{2}F^{\rho}_{\nu} + 2\left(3\frac{g_{K}g_{\pi}}{g_{K^{*}}}\right)^{2}F^{K^{*0}}_{\nu}\right\}
        	\end{eqnarray}
        	
        	The comparison of the branching fraction calculated by using this amplitude with the experimental value \cite{ParticleDataGroup:2020ssz} leads to the cut-off parameter $\Lambda_{K\pi} = 950$ MeV. The obtained result is higher than the result obtained for the process $\tau \to \pi^{-} \pi^{0} \nu_{\tau}$ ($\Lambda_{\pi\pi} = 740$ MeV). This may be caused by the replacement of the pion with a more massive kaon.
        	
        	The meson triangles with the exchange of the scalar state give the result lower by orders of magnitude and it can be neglected.
        	
        	As one can see, in this process, in the absence of radially excited mesons in the intermediate state, the interaction of mesons in the final state plays an important role.
 
     \subsubsection{The process $\tau \to K^{-} \eta \nu_{\tau}$}
            In this process, the energy threshold of meson production is higher than the mass of the ground state $K^{*}(892)$. That is why the first radially excited mesons in the intermediate state give a significant contribution and they may not be neglected. This leads to the necessity of applying the extended NJL model.
            
            In the extended NJL model the amplitude of the process $\tau \to K^{-} \eta \nu_{\tau}$ takes the form
            
            \begin{eqnarray}
                \mathcal{M}(\tau \to K^{-} \eta \nu_{\tau})_{tree} = -2 G_{f} V_{us} \left(I_{11}^{K \eta^{u}} + \sqrt{2} I_{11}^{K \eta^{s}}\right) L_{\mu} \left[\left(T_{K}^{(c)}p_{K} - T_{\eta}^{(c)}p_{\eta}\right)^{\mu} \right. \nonumber\\
                 + \frac{C_{K^{*}}}{g_{K^{*}}} \frac{I_{11}^{K K^{*} \eta^{u}} + \sqrt{2} I_{11}^{K K^{*} \eta^{s}}}{I_{11}^{K \eta^{u}} + \sqrt{2} I_{11}^{K \eta^{s}}} \frac{g^{\mu\nu} q^{2} - q^{\mu}q^{\nu}}{M_{K^{*}}^{2} - q^{2} - i \sqrt{q^{2}}\Gamma_{K^{*}}} \left(T_{K}^{(K^{*})}p_{K} - T_{\eta}^{(K^{*})}p_{\eta}\right)_{\nu} \nonumber\\
                 \left. + \frac{C_{\hat{K}^{*}}}{g_{K^{*}}} \frac{I_{11}^{K \hat{K}^{*} \eta^{u}} + \sqrt{2} I_{11}^{K \hat{K}^{*} \eta^{s}}}{I_{11}^{K \eta^{u}} + \sqrt{2} I_{11}^{K \eta^{s}}} \frac{g^{\mu\nu} q^{2} - q^{\mu}q^{\nu}}{M_{\hat{K}^{*}}^{2} - q^{2} - i \sqrt{q^{2}}\Gamma_{\hat{K}^{*}}} \left(T_{K}^{(\hat{K}^{*})}p_{K} - T_{\eta}^{(\hat{K}^{*})}p_{\eta}\right)_{\nu} \right],
            \end{eqnarray}
            where the integrals over quark loops $I_{11}$ are defined in (\ref{integral_ext}); the constants $C$ are determined in (\ref{C_const}); the factors $T$ describe the transitions between axial vector and pseudoscalar mesons:      
            \begin{eqnarray}
              &&  T_{K}^{(c)} = 1 - 2 \frac{m_{s} I_{11}^{K_{1} \eta^{u}} + \sqrt{2} m_{u} I_{11}^{K_{1} \eta^{s}}}{I_{11}^{K \eta^{u}} + \sqrt{2} I_{11}^{K \eta^{s}}} I_{11}^{K_{1} K} \frac{m_{s} + m_{u}}{M_{K_{1A}}^{2}},  \\
              &&  T_{\eta}^{(c)} = 1 - 2 \frac{I_{11}^{K f^{u}} I_{20}^{f^{u} \eta^{u}}}{I_{11}^{K \eta^{u}} + \sqrt{2} I_{11}^{K \eta^{s}}} \frac{m_{u}\left(3m_{u} - m_{s}\right)}{M_{f_{1}^{u}}^{2}} - 2 \sqrt{2} \frac{I_{11}^{K f^{s}} I_{02}^{f^{s} \eta^{s}}}{I_{11}^{K \eta^{u}} + \sqrt{2} I_{11}^{K \eta^{s}}} \frac{m_{s}\left(3m_{s} - m_{u}\right)}{M_{f_{1}^{s}}^{2}}, \\
              &&  T_{K}^{(K^{*})} = 1 - 2 \frac{m_{s} I_{11}^{K^{*} K_{1} \eta^{u}} + \sqrt{2} m_{u} I_{11}^{K^{*} K_{1} \eta^{s}}}{I_{11}^{K K^{*} \eta^{u}} + \sqrt{2} I_{11}^{K K^{*} \eta^{s}}} I_{11}^{K_{1} K} \frac{m_{s} + m_{u}}{M_{K_{1A}}^{2}}, \\
              &&  T_{\eta}^{(K^{*})} = 1 - 2 \frac{I_{11}^{K^{*} K f^{u}} I_{20}^{f^{u} \eta^{u}}}{I_{11}^{K^{*} K \eta^{u}} + \sqrt{2} I_{11}^{K^{*} K \eta^{s}}} \frac{m_{u}\left(3m_{u} - m_{s}\right)}{M_{f_{1}^{u}}^{2}} \\
              && \qquad \qquad \qquad- 2 \sqrt{2} \frac{I_{11}^{K^{*} K f^{s}} I_{02}^{f^{s} \eta^{s}}}{I_{11}^{K^{*} K \eta^{u}} + \sqrt{2} I_{11}^{K^{*} K \eta^{s}}} \frac{m_{s}\left(3m_{s} - m_{u}\right)}{M_{f_{1}^{s}}^{2}}, \nonumber
\\
              &&  T_{K}^{(\hat{K}^{*})} = 1 - 2 \frac{m_{s} I_{11}^{\hat{K}^{*} K_{1} \eta^{u}} + \sqrt{2} m_{u} I_{11}^{\hat{K}^{*} K_{1} \eta^{s}}}{I_{11}^{K \hat{K}^{*} \eta^{u}} + \sqrt{2} I_{11}^{K \hat{K}^{*} \eta^{s}}} I_{11}^{K_{1} K} \frac{m_{s} + m_{u}}{M_{K_{1A}}^{2}},  \\
              &&  T_{\eta}^{(\hat{K}^{*})} = 1 - 2 \frac{I_{11}^{\hat{K}^{*} K f^{u}} I_{20}^{f^{u} \eta^{u}}}{I_{11}^{\hat{K}^{*} K \eta^{u}} + \sqrt{2} I_{11}^{\hat{K}^{*} K \eta^{s}}} \frac{m_{u}\left(3m_{u} - m_{s}\right)}{M_{f_{1}^{u}}^{2}} \\ \nonumber  
          && \qquad \qquad \qquad- 2 \sqrt{2} \frac{I_{11}^{\hat{K}^{*} K f^{s}} I_{02}^{f^{s} \eta^{s}}}{I_{11}^{\hat{K}^{*} K \eta^{u}} + \sqrt{2} I_{11}^{\hat{K}^{*} K \eta^{s}}} \frac{m_{s}\left(3m_{s} - m_{u}\right)}{M_{f_{1}^{s}}^{2}}.
            \end{eqnarray}
            where $M_{K_{1A}}$ is defined in (\ref{Mk1A}); $M_{f_{1}^{u}}$ and $M_{f_{1}^{s}}$ are the masses of the mesons $f_{1}(1285)$ and $f_{1}(1420)$.
            
            As a result, for the branching fraction of this process one can obtain the value
            \begin{eqnarray}
            	Br(\tau^{-} \to K^{-} \eta \nu_{\tau})_{tree} = 1.35  \times 10^{-4}.
            \end{eqnarray}
            It is 13\% lower than the experimental result~\cite{ParticleDataGroup:2020ssz}:
            \begin{eqnarray}
            	Br(\tau^{-} \to K^{-} \eta \nu_{\tau})_{exp} = (1.55 \pm 0.08) \times 10^{-4}.
            \end{eqnarray}
            
\begin{figure}[h]
            	\center{\includegraphics[scale = 0.35]{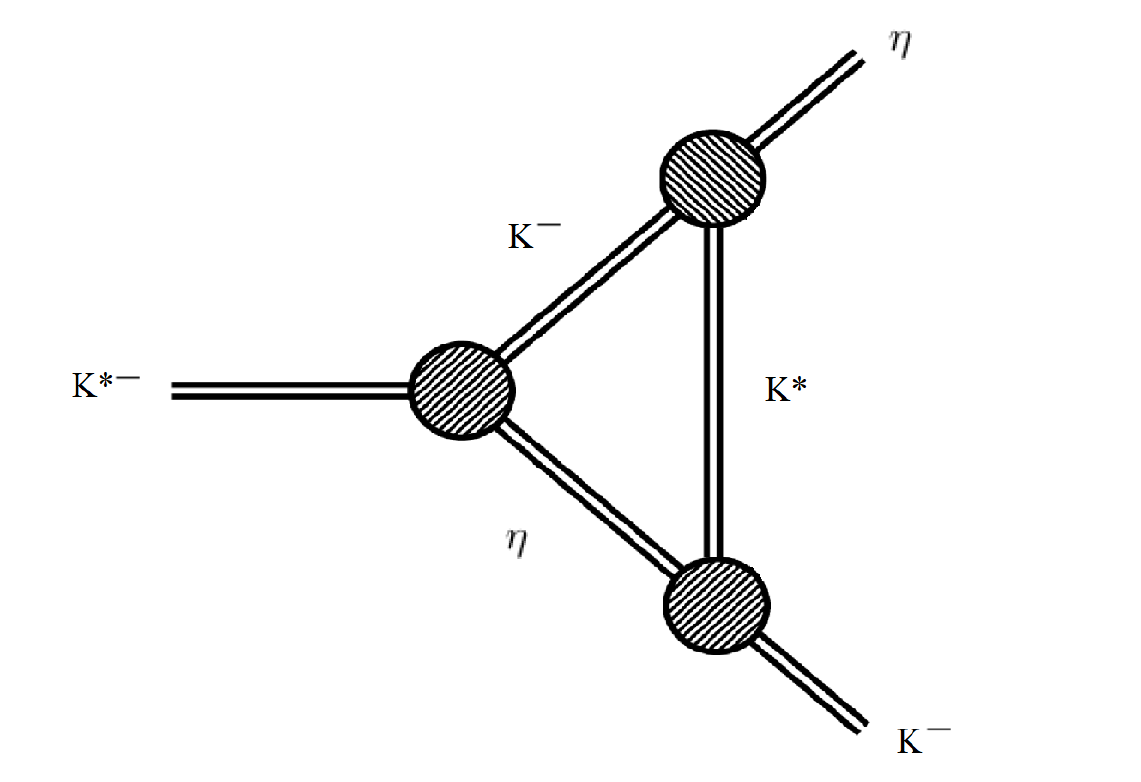}}
            	\caption{Interactions of the kaon and $\eta$ meson in the final state}
            	\label{KvKeta}
            \end{figure}

            Therefore, probably, taking into account the interactions in the final state matters in this process as well. This interaction can be considered through the exchange of the meson $K^{*}$ between the final particles. This leads to the meson triangle presented in Figure ~\ref{KvKeta}.
            
            This meson triangle can be described by the integral~\cite{Volkov:2021eny}
            \begin{eqnarray}
               F_{\mu} =  \int \frac{\left(T_{K}^{(K^{*})}k - \left(T_{K}^{(K^{*})} + T_{\eta}^{(K^{*})}\right)p_{\eta}\right)_{\lambda}\left(T_{\eta}^{(K^{*})}k + \left(T_{K}^{(K^{*})} + T_{\eta}^{(K^{*})}\right)p_{K}\right)_{\nu}
            	}{\left[k^{2} - M_{K^{*}}^{2}\right]
            	\left[(k + p_{K})^{2} - M_{\eta}^{2}\right]\left[(k - p_{\eta})^{2} - M_{K}^{2}\right]} 
	\nonumber\\
            	\times \left(\left(T_{K}^{(K^{*})} + T_{\eta}^{(K^{*})}\right)k + T_{\eta}^{(K^{*})}p_{K} - T_{K}^{(K^{*})}p_{\eta}\right)_{\mu} \left(g^{\nu\lambda} 
            	- \frac{k^{\nu}k^{\lambda}}{M_{K^{*}}^{2}}\right) \frac{d^{4}k}{(2\pi)^{4}}.
            \end{eqnarray}
            
            This integral is of a similar structure as the respective integral for the process $\tau \to K \pi \nu_{\tau}$ described above.
            
            The correction to the amplitude describing the interaction in the final state takes the form
            \begin{eqnarray}
              &&  \mathcal{M}_{loop}(\tau \to K \eta \nu_{\tau}) = 8i G_{f} V_{us} \left(I_{11}^{K \eta^{u}} + \sqrt{2} I_{11}^{K \eta^{s}}\right)^{3} L_{\mu}  \nonumber\\
              && \times \biggl[ g^{\mu\nu} 
                + \frac{C_{K^{*}}}{g_{K^{*}}} \left(\frac{I_{11}^{K K^{*} \eta^{u}} + \sqrt{2} I_{11}^{K K^{*} \eta^{s}}}{I_{11}^{K \eta^{u}} + \sqrt{2} I_{11}^{K \eta^{s}}}\right)^{3} \frac{g^{\mu\nu} q^{2} - q^{\mu}q^{\nu}}{M_{K^{*}}^{2} - q^{2} - i \sqrt{q^{2}}\Gamma_{K^{*}}} \nonumber\\
              &&  + \frac{C_{\hat{K}^{*}}}{g_{K^{*}}} \left(\frac{I_{11}^{K \hat{K}^{*} \eta^{u}} + \sqrt{2} I_{11}^{K \hat{K}^{*} \eta^{s}}}{I_{11}^{K \eta^{u}} + \sqrt{2} I_{11}^{K \eta^{s}}}\right)^{3} \frac{g^{\mu\nu} q^{2} - q^{\mu}q^{\nu}}{M_{\hat{K}^{*}}^{2} - q^{2} - i \sqrt{q^{2}}\Gamma_{\hat{K}^{*}}} \biggl] F_{\nu}.
            \end{eqnarray}
            
            While using the cut-off parameter obtained in the process $\tau \to K \pi \nu_{\tau}$ ($\Lambda_{K\pi} = 950$ MeV), the result for the process $\tau^{-} \to K^{-} \eta \nu_{\tau}$ is consistent with the experimental data:
            \begin{eqnarray}
            	Br(\tau^{-} \to K^{-} \eta \nu_{\tau}) & = & 1.56  \times 10^{-4}.
            \end{eqnarray}
            
            As one can see, the excited meson in the considered process plays a significant role, but the contribution of the interactions in the final state has decreased noticeably compared to the previous cases.
            
            This process was studied in other theoretical works by using the Vector Dominance Model, Chiral Perturbation Theory with Resonances, and others~\cite{Li:1996md,Escribano:2013bca}.
   
          \subsubsection{The process $\tau \to K^{-} K^{0} \nu_{\tau}$}
            Similarly to the previous case, in the process $\tau \to K^{-} K^{0} \nu_{\tau}$, it is necessary to take into account the first radially excited mesons in the intermediate states and, hence, to apply the extended NJL model.
            
\begin{figure}[h]
            	\center{\includegraphics[scale = 0.25]{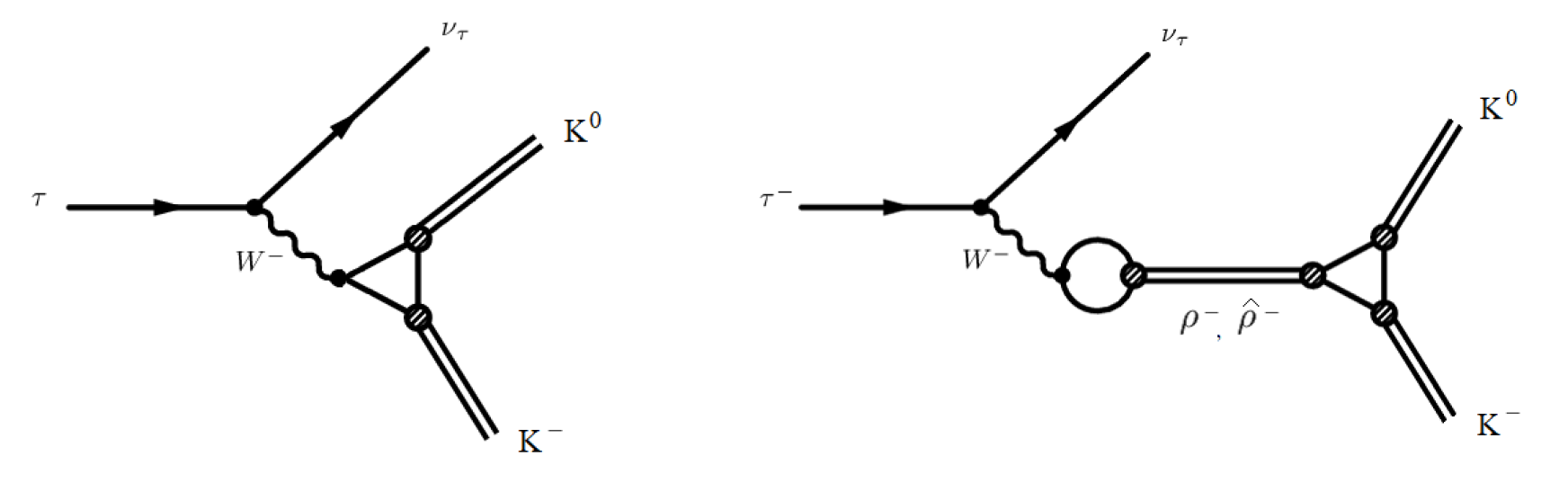}}
            	\caption{Contact diagram and diagram with the intermediate mesons of the process $\tau \to K^{-} K^{0} \nu_{\tau}$}
            	\label{tauKKContact}
            \end{figure}

            The diagrams of this process are presented in Figure ~\ref{tauKKContact}.
            
\begin{figure}[h]
            	\center{\includegraphics[scale = 0.4]{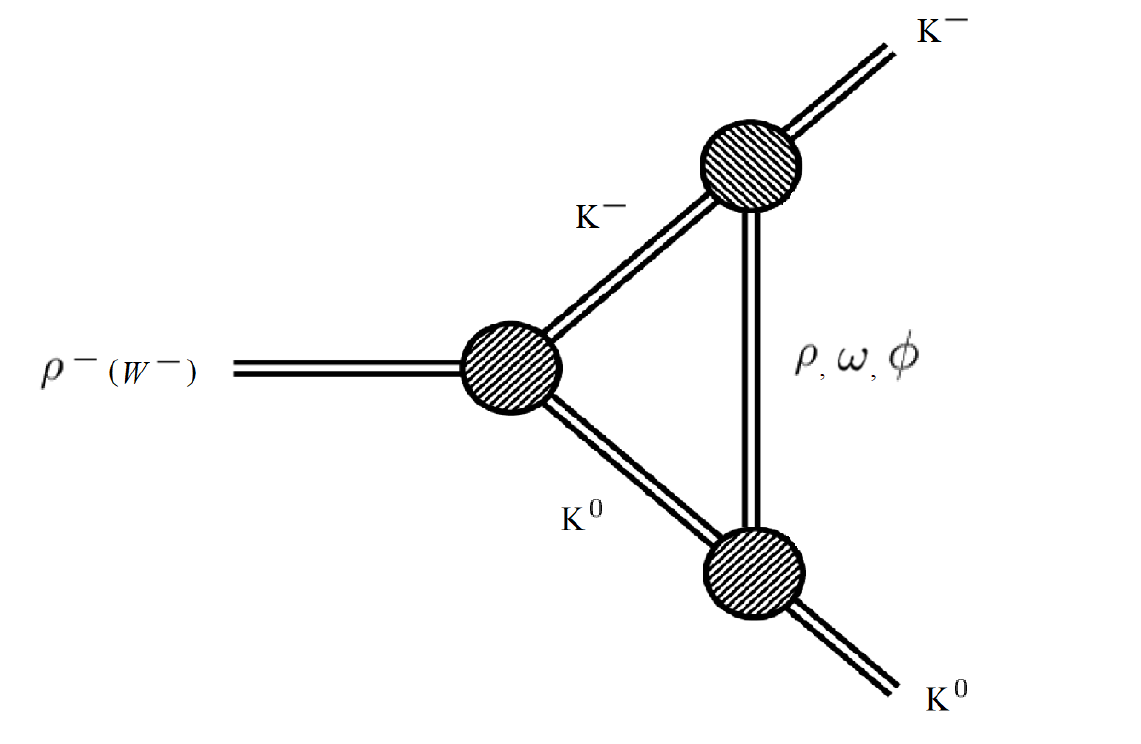}}
            	\caption{Interactions of the kaons in the final state.}
            	\label{rhoKK}
            \end{figure}

            To take into account the interaction in the final state in this process, one can consider the exchange of the neutral vector mesons between the final kaons. The appropriate diagrams are given in Figure ~\ref{rhoKK}.
            
            These diagrams can be described with the integrals \cite{Volkov:2021miu}
            \begin{eqnarray}
            	F_{\mu}^{(\rho)} = \int \frac{\left(k - 2p_{K^{-}}\right)_{\lambda}\left(k + 2p_{K^{0}}\right)_{\nu}
            	\left(2k + p_{K^{0}} - p_{K^{-}}\right)_{\mu} \left(g^{\nu\lambda} 
            	- \frac{k^{\nu}k^{\lambda}}{M_{\rho}^{2}}\right)}{\left[k^{2} - M_{\rho}^{2}\right]
            	\left[(k + p_{K^{0}})^{2} - M_{K}^{2}\right]\left[(k - p_{K^{-}})^{2} - M_{K}^{2}\right]} \frac{d^{4}k}{(2\pi)^{4}},\\
            	F_{\mu}^{(\omega)} = \int \frac{\left(k - 2p_{K^{-}}\right)_{\lambda}\left(k + 2p_{K^{0}}\right)_{\nu}
            	\left(2k + p_{K^{0}} - p_{K^{-}}\right)_{\mu} \left(g^{\nu\lambda} 
            	- \frac{k^{\nu}k^{\lambda}}{M_{\omega}^{2}}\right)}{\left[k^{2} - M_{\omega}^{2}\right]
            	\left[(k + p_{K^{0}})^{2} - M_{K}^{2}\right]\left[(k - p_{K^{-}})^{2} - M_{K}^{2}\right]} \frac{d^{4}k}{(2\pi)^{4}}, \\
            	F_{\mu}^{(\phi)} = \int \frac{\left(k - 2p_{K^{-}}\right)_{\lambda}\left(k + 2p_{K^{0}}\right)_{\nu}
            	\left(2k + p_{K^{0}} - p_{K^{-}}\right)_{\mu} \left(g^{\nu\lambda} 
            	- \frac{k^{\nu}k^{\lambda}}{M_{\phi}^{2}}\right)}{\left[k^{2} - M_{\phi}^{2}\right]
            	\left[(k + p_{K^{0}})^{2} - M_{K}^{2}\right]\left[(k - p_{K^{-}})^{2} - M_{K}^{2}\right]} \frac{d^{4}k}{(2\pi)^{4}}.
            \end{eqnarray}
            
            These integrals are similar to the respective integral obtained for the process $\tau \to \pi \pi \nu_{\tau}$:
            \begin{eqnarray}
            	F_{\mu}^{(\rho)} = i \left[\frac{I_{\rho}}{M_{\rho}^{2}} + I_{\rho K}\right] \left(p_{K^{0}} - p_{K^{-}}\right)_{\mu}, \\
            	F_{\mu}^{(\omega)} = i \left[\frac{I_{\omega}}{M_{\omega}^{2}} + I_{\omega K}\right] \left(p_{K^{0}} - p_{K^{-}}\right)_{\mu}, \\
            	F_{\mu}^{(\phi)} = i \left[\frac{I_{\phi}}{M_{\phi}^{2}} + I_{\phi K}\right] \left(p_{K^{0}} - p_{K^{-}}\right)_{\mu}.
            \end{eqnarray}
            where the integrals $I_{V}$ and $I_{VK}$ ($V = \rho, \omega, \phi$) are similar to the integrals defined in (\ref{meson_integ}).
            
            The full amplitude taking into account the interaction in the final state takes the form \cite{Volkov:2021miu}
            
            \begin{eqnarray}
                \mathcal{M}(\tau \to K^{-} K^{0} \nu_{\tau})_{tot} = -2 \sqrt{2} G_{f} V_{ud} I_{11}^{KK} \left[T_{K}^{(c)} + \frac{C_{\rho}}{g_{\rho}} \frac{I_{11}^{KK\rho}}{I_{11}^{KK}} T_{K}^{(\rho)} \frac{q^2}{M_{\rho}^{2} - q^{2} - i\sqrt{q^{2}}\Gamma_{\rho}} \right. \nonumber\\
                 \left.+\frac{C_{\hat{\rho}}}{g_{\rho}} \frac{I_{11}^{KK\hat{\rho}}}{I_{11}^{KK}} T_{K}^{(\hat{\rho})} \frac{q^2}{M_{\hat{\rho}}^{2} - q^{2} - i\sqrt{q^{2}}\Gamma_{\hat{\rho}}}\right] \left\{1 - 4 \left(I_{11}^{KK\rho}\right)^{2} \left(T_{K}^{(\rho)}\right)^{2} \left[\frac{I_{\rho}}{M_{\rho}^{2}} + I_{\rho K}\right] +4 \left(I_{11}^{KK\omega}\right)^{2} \right. \nonumber\\
                 \left. \times \left(T_{K}^{(\omega)}\right)^{2} \left[\frac{I_{\omega}}{M_{\omega}^{2}} + I_{\omega K}\right] + 4 \left(I_{11}^{KK\phi}\right)^{2} \left(T_{K}^{(\phi)}\right)^{2} \left[\frac{I_{\phi}}{M_{\phi}^{2}} + I_{\phi K}\right]\right\} L_{\mu} \left(p_{K^{0}} -p_{K^{-}}\right)^{\mu},
            \end{eqnarray}
            where
            \begin{eqnarray}
              &&  T_{K}^{(\rho)} = 1 - \frac{I_{11}^{K_{1}K\rho}I_{11}^{K_{1}K}}{I_{11}^{KK\rho}} \frac{\left(m_{s} + m_{u}\right)^{2}}{M_{K_{1A}}^{2}}, \\
              &&  T_{K}^{(\omega)} = 1 - \frac{I_{11}^{K_{1}K\omega}I_{11}^{K_{1}K}}{I_{11}^{KK\omega}} \frac{\left(m_{s} + m_{u}\right)^{2}}{M_{K_{1A}}^{2}}, \\
              &&  T_{K}^{(\phi)} = 1 - \frac{I_{11}^{K_{1}K\phi}I_{11}^{K_{1}K}}{I_{11}^{KK\phi}} \frac{\left(m_{s} + m_{u}\right)^{2}}{M_{K_{1A}}^{2}}.
        	\end{eqnarray}
            
            The first term in the curly brackets describes the diagrams in the tree approximation of the meson fields interaction. The branching fraction of this process in this approximation takes the value
            \begin{eqnarray}
            	Br(\tau \to K^{-} K^{0} \nu_{\tau}) & = & 13.95  \times 10^{-4}.
            \end{eqnarray}
            It is in satisfactory agreement with the experimental value~\cite{ParticleDataGroup:2020ssz}:
            \begin{eqnarray}
            	Br(\tau \to K^{-} K^{0} \nu_{\tau})_{exp} & = & (14.86 \pm 0.34) \times 10^{-4}.
            \end{eqnarray}
            
            Taking into account the interactions in the final state leads to the appearance of the cut-off parameter in the meson loop. Full agreement with experimental data can be achieved with the value of this parameter $\Lambda_{KK} = 610$ MeV.
            
            As one can see, taking into account the interactions in the final state does not play an important role in this process and gives the correction within the model uncertainties. The correction of the same level can be achieved by variation of the width of the intermediate radially excited meson within the experimental errors. When describing a similar process $e^{+}e^{-} \to K^{+}K^{-}$ in the extended NJL model \cite{Volkov:2018cqp}, a satisfactory result was obtained without taking into account the interactions in the final state. All these facts indicate the decreasing role of the interactions in the final state, while the role of excited mesons in the intermediate state increases. They also indicate that there is no need to take into account the interactions in the final state in the case when the energy threshold of the final meson production is higher than 1 GeV. For this reason, in the following Section \ref{NJL_pv}, where the $\tau$ lepton decays into vector and pseudoscalar particles, the interactions in the final state will not be taken into account.
            
            The decay $\tau \to K^{-} K^{0} \nu_{\tau}$ was considered in many theoretical works~\cite{Li:1996md,Dubnicka:2010grh,Dai:2018thd,Gonzalez-Solis:2019lze}.
 
     \subsection{The decays $\tau \to P V \nu_{\tau}$}
     \label{NJL_pv}
         \subsubsection{The decay $\tau \to \pi \omega \nu_\tau$}
            The decay of $\tau \to \pi \omega \nu_\tau$ has been repeatedly investigated in various theoretical \cite{LopezCastro:1996xh, Flores-Tlalpa:2007hbv, Guo:2008sh} and experimental works \cite{ALEPH:1996kok, CLEO:1999heg}. In particular, in the paper \cite{LopezCastro:1996xh}, a phenomenological model of the vector dominance type was used, where the channels with intermediate mesons $\rho(770)$, $\rho(1450)$ and $\rho(1700)$ were considered. Wherein, for good agreement with experimental data, additional arbitrary parameters were used. 
            \begin{figure}[h]
            	\center{\includegraphics[scale = 0.08]{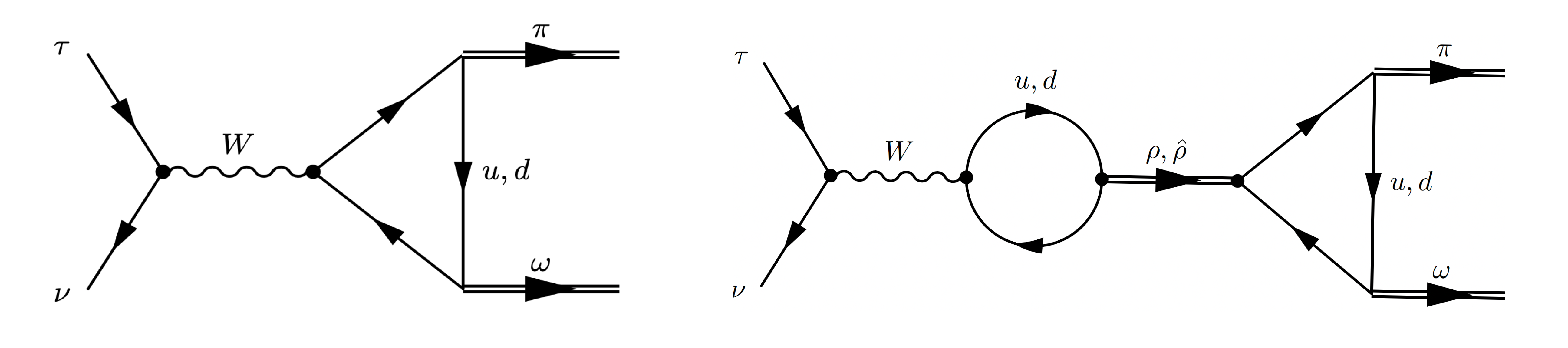}}
            	\caption{Contact diagram and diagram with the intermediate mesons describing the decay $\tau \to \pi \omega \nu_{\tau}$}
            	\label{contact_pi_om}
            \end{figure}
            
            In the NJL model, this decay is described by the diagrams of two types. In the first diagram, the intermediate $W$ boson directly generates $\pi \omega$ meson pairs through the quark triangle. The second diagram is related to the $W$ boson which transits into vector mesons and also generates $\pi \omega$ meson pair. These diagrams are shown in Figure \ref{contact_pi_om}. The amplitude of the process under consideration takes the form \cite{Volkov:2012gv}
            \begin{eqnarray}
            &&\mathcal{M}(\tau \to \pi \omega \nu_\tau) = G_F L^\mu V_{ud} 4 m_u g_\pi \left( B_c + B_{(\rho +\hat{\rho})} \right) \varepsilon_{\mu\nu \lambda \delta} e^{*\nu}(p_{\omega}) p^\lambda_\omega p^\delta_\pi.
            \end{eqnarray}
            
            The corresponding contribution from the contact diagram to the amplitude is $B_c = I^\rho_{30}(m_u)$. The contributions from intermediate mesons are defined as 
            \begin{eqnarray}
            &&B_{(\rho +\hat{\rho})} = \frac{C_\rho}{g_\rho}I^{\rho \omega}_{30} \frac{s}{M^2_\rho - s- i M_\rho \Gamma_\rho} +
            e^{i\pi}\frac{C_{\hat{\rho}}}{g_\rho}I^{\hat{\rho} \omega}_{30} \frac{s}{M^2_{\hat{\rho}} - s- i M_{\hat{\rho}} \Gamma_{\hat{\rho}}(q)},
            \end{eqnarray}
            where the integrals $I_{30}$ are defined in (\ref{integral_ext}); the width of the radially excited meson $\Gamma_{\hat{\rho}}(q)$ is taken from \cite{Arbuzov:2010xi}. The constants $C_\rho$ and $C_{\hat{\rho}}$ are defined in (\ref{C_const}).
            
            As a result, the branching fraction of the decay $\tau \to \pi \omega \nu_\tau$ turns out to be equal to 
            \begin{eqnarray}
                Br(\tau \to \omega \pi \nu_{\tau}) = 1.83 \%.
            \end{eqnarray}
                        
           This is in satisfactory agreement with the experimental data \cite{ParticleDataGroup:2020ssz}:
            \begin{eqnarray}
                Br(\tau \to \omega \pi \nu_{\tau})_{exp} = (1.95 \pm 0.06) \%.
            \end{eqnarray}
                        
            Note that we do not take into account the contribution to the amplitude from the heavier $\rho(1700)$ meson since its contribution is strongly suppressed due to the phase volume factor. 

         \subsubsection{The decay $\tau \to \rho \pi \nu_{\tau}$}
            The process $\tau \rightarrow \rho^{0}(770) \pi^{-} \nu_{\tau}$ is actively investigated theoretically in various phenomenological models \cite{Guo:2008sh,Dai:2018thd}. In the framework of the NJL model, this process was considered in the work \cite{Volkov:2019udu}. Its amplitude in the extended NJL model takes the following form: 
            \begin{eqnarray}
                \mathcal{M}(\tau \to \rho \pi \nu_{\tau}) = -i F_{\pi} G_{F} V_{ud} g_{\rho} Z_{\pi} L_{\mu} \left\{ \mathcal{M}_{c} + \mathcal{M}_{A} + \mathcal{M}_{\hat{A}} + \mathcal{M}_{P} + \mathcal{M}_{\hat{P}} \right\}^{\mu\nu} e_{\nu}(p_{\rho}),
            \end{eqnarray}
            where the contributions from the contact diagram and the contributions from diagrams with intermediate mesons are in the curly brackets
            \begin{eqnarray}
              \mathcal{M}_{c}^{\mu\nu} = C_{\rho} g^{\mu\nu} - \frac{2}{3} \frac{C_{a_{1}} I_{20}^{\rho a_{1}}}{M_{a_{1}}^{2}}\left\{(q^2 - M_{\rho}^2) g^{\mu\nu} - q^{\mu}q^{\nu}\right\}, 
             \end{eqnarray}
             \begin{eqnarray}  
              &&  \mathcal{M}_{A}^{\mu\nu} = \frac{2}{3} C_{a_{1}} BW_{a_{1}} \left[(q^{2} - 6m_u^{2})g^{\mu\lambda} - \frac{q^{\mu}q^{\lambda}}{Z_{\pi}}\right] 
              \nonumber \\
              && \times \left\{I_{20}^{a_{1}\rho}g_{\lambda\delta} - \frac{C_{a_{1}} I_{20}^{a_{1}\rho a_{1}}}{g_{\rho} M_{a_{1}}^2}\left[(q^2 - M_{\rho}^2) g_{\lambda\delta} - q_{\lambda}q_{\delta}\right]\right\} g^{\delta\nu}
                \nonumber\\
                && -4m_u^{2}Z_{\pi}C_{a_{1}}\frac{1}{M_{a_{1}}^{2}}\frac{M_{a_{1}}^{2} - q^{2}}{M_{\pi}^{2} - q^{2}} BW_{a_{1}}\left[I_{20}^{a_{1}\rho} + \frac{C_{a_{1}}I_{20}^{a_{1}\rho a_{1}}M_{\rho}^{2}}{g_{\rho}M_{a_{1}}^{2}}\right] q^{\mu} q^{\nu},  
                \end{eqnarray}
                \begin{eqnarray} 
              &&  \mathcal{M}_{\hat{A}}^{\mu\nu} = \frac{2}{3} C_{\hat{a}_{1}} BW_{\hat{a}_{1}} \left[(q^{2} - 6m_u^{2})g^{\mu\lambda} - \frac{q^{\mu}q^{\lambda}}{Z_{\pi}}\right] 
              \nonumber \\
              && \times \left\{I_{20}^{\hat{a}_{1}\rho}g_{\lambda\delta} - \frac{C_{a_{1}} I_{20}^{\hat{a}_{1}\rho a_{1}}}{g_{\rho} M_{a_{1}}^2}\left[(q^2 - M_{\rho}^2) g_{\lambda\delta} - q_{\lambda}q_{\delta}\right]\right\} g^{\delta\nu},
                \end{eqnarray}
                \begin{eqnarray} 
               && \mathcal{M}_{P}^{\mu\nu} = 2 Z_{\pi} C_{\rho}\left[1 - 4 I_{20}^{\rho a_{1}}  \frac{m_u^{2}}{M_{a_{1}}^{2}}\right] \left[1 - 6\frac{m_u^{2}C_{a_{1}}^{2}}{M_{a_{1}}^{2}}\right] BW_{\pi} q^{\mu} p_{\pi}^{\nu},
                \\
               && \mathcal{M}_{\hat{P}}^{\mu\nu} = 8 \frac{g_{\pi}}{g_{\rho}} C_{\hat{\pi}} I_{20}^{\rho\hat{\pi}}\left[1 - 6 \frac{I_{20}^{\rho a_{1}\hat{\pi}}C_{a_{1}}}{I_{20}^{\rho\hat{\pi}}g_{\rho}} \frac{m_u^{2}}{M_{a_{1}}^{2}}\right] BW_{\hat{\pi}} q^{\mu} p_{\pi}^{\nu},
            \end{eqnarray}
            where $M_{a_{1}} = 1230 \pm 40$ MeV, $\Gamma_{a_{1}} = 425$ MeV are the mass and total width of the $a_{1}(1260)$ meson \cite{ParticleDataGroup:2020ssz}; $BW_{M}$ are the Breit-Wigner propagators defined in (\ref{BreitWigner}); $q=p_{\rho}+p_{\pi}$. The amplitude is given with allowance for the $a_{1}-\pi$ transitions in all channels. 
            
            Taking into account the $a_{1}-\pi$ transitions the branching fraction of this process is 
            \begin{eqnarray}
                Br(\tau \rightarrow \rho^{0}(770) \pi^{-} \nu_{\tau}) = 1.1 \%.
            \end{eqnarray}
            
            If we exclude the $a_{1}-\pi$ transitions in the axial-vector channel, we get the following result: 
            \begin{eqnarray}
                Br(\tau \rightarrow \rho^{0}(770) \pi^{-} \nu_{\tau}) = 4.96 \%.
            \end{eqnarray}

            In the case of the decay into an excited state, the branching fraction, taking into account the $a_{1}-\pi$ transitions, takes the following value: 
            \begin{eqnarray}
                Br(\tau \rightarrow \rho^{0}(1450) \pi^{-} \nu_{\tau}) = 1.2 \times 10^{-4}.
            \end{eqnarray}
            
            Without taking into account the $a_{1}-\pi$ transitions, the decay into an excited state gives 
            \begin{eqnarray}
                Br(\tau \rightarrow \rho^{0}(1450) \pi^{-} \nu_{\tau}) = 1.8 \times 10^{-4}.
            \end{eqnarray}
            
            At present there are no satisfactory numerical estimates for the decay width $\tau \rightarrow \rho^{0}(770) \pi^{-} \nu_{\tau}$, therefore we can compare our results with the process $\tau \rightarrow \pi^{-} \pi^{-} \pi^{+} \nu_{\tau}$.         
            The branching fraction of this process is $Br(\tau \rightarrow \pi^{-} \pi^{+} \pi^{-} \nu_{\tau}) = 9.31 \pm 0.05 \%$ \cite{ParticleDataGroup:2020ssz}. The process $\tau \rightarrow \pi^{-} \pi^{-} \pi^{+} \nu_{\tau}$ was also considered theoretically in the work \cite{Ivanov:1989qw} with the participation of one of the authors. Due to the fact that this process can also contain channels with a scalar meson, as well as channels with a box diagram, there is reason to believe that the process under consideration $\tau \rightarrow \rho^{0}(770) \pi^{-} \nu_{\tau}$ should have a width smaller than $\tau \rightarrow \pi^{-} \pi^{+} \pi^{-} \nu_{\tau}$.
            
            This process was also considered in the NJL model in the paper \cite{Osipov:2018lnl}. There, the $a_{1}-\pi$ transitions in the axial-vector channel were taken into account, but a different approach was used to take into account the excited meson states than the one used in the extended NJL model. As a result, the branching fraction $Br(\tau \rightarrow \rho^{0}(770) \pi^{-} \nu_{\tau}) = 5.65 \%$ was obtained. 

 \subsubsection{The decays $\tau \to [\omega, \phi] K \nu_\tau$} 
\begin{figure}[h]
\center{\includegraphics[scale = 0.23]{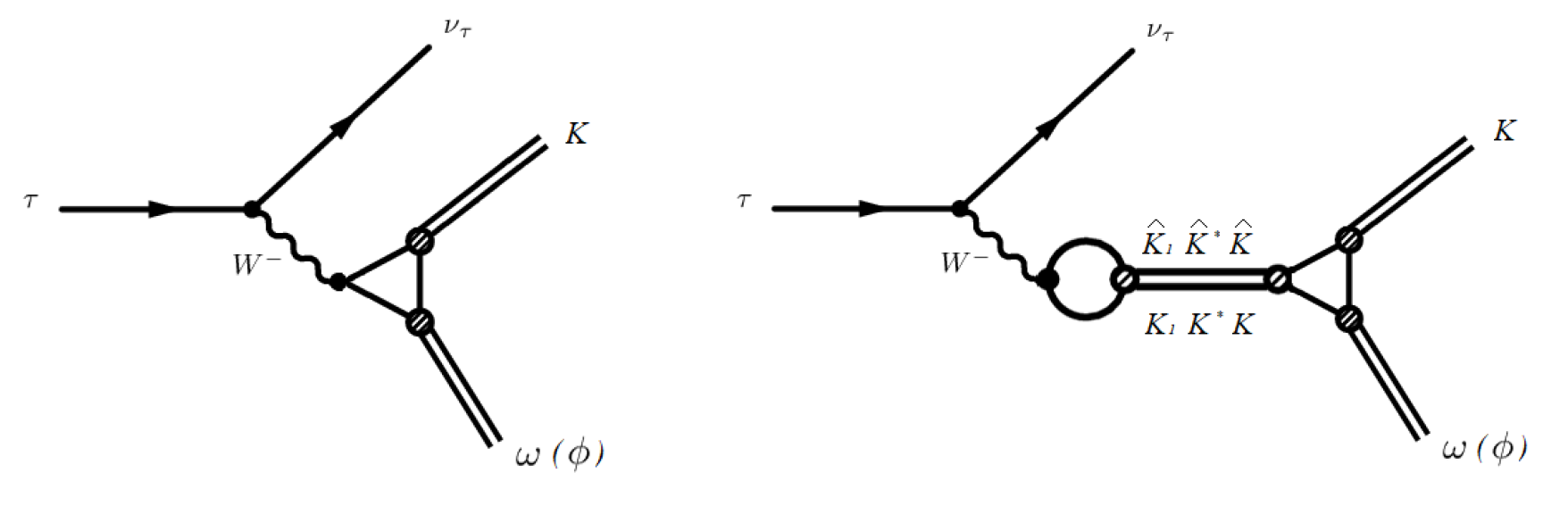}}
\caption{Contact diagram and diagram with intermediate mesons of the decays $\tau \to [\omega, \phi] K \nu_\tau$}
\label{k_o1}
\end{figure}

The $\tau \to [\omega, \phi] K \nu_\tau$ decays will be considered in the extended NJL model, following the paper \cite{Volkov:2019cja}. Unlike the previous $\tau \to \rho [\pi, \hat{\pi}] \nu_\tau$ decay, this decay is interesting because all four channels work in it: contact, axial-vector, vector and pseudoscalar channels. Note that in the axial vector channel we take into account the mixing of the mesons $K_{1A}$ and $K_{1B}$. The diagrams describing the considered decay are given in Figure \ref{k_o1}. 

The corresponding amplitude in the extended NJL model takes the form
            	\begin{eqnarray}
	         \label{amplitude_o_k}
            	\mathcal{M}(\tau \to K \omega \nu_{\tau}) = -i G_{F} V_{us} L_{\mu} \left\{ \mathcal{M}_{c} + \mathcal{M}_{A} + \mathcal{M}_{V} + \mathcal{M}_{P} \right. \nonumber\\
                 \left. + \mathcal{M}_{\hat{A}} + \mathcal{M}_{\hat{V}} + \mathcal{M}_{\hat{P}} \right\}^{\mu\nu} e_{\nu}^{*}(p_{\omega}),
            	\end{eqnarray}
where $e_{\nu}^{*}(p_{\omega})$ is the polarization vector of the $\omega$ meson. The contributions from the diagrams with intermediate mesons $K_{1A}$, $K^*$, $K$, $\hat{K}_{1}$, $\hat{K}^{*}$, and $\hat{K}$ have the form 
            	\begin{eqnarray}
            	&& \mathcal{M}_{c}^{\mu\nu} = (m_{s} + m_{u}) I_{11}^{K\omega} g^{\mu\nu} + i 2 m_{u} \left[I_{21}^{K\omega} + (m_{s} - m_{u}) m_{u} I_{31}^{K\omega}\right] \varepsilon^{\mu\nu\lambda\delta} p_{K\lambda} p_{\omega\delta}, \nonumber
            	\end{eqnarray}
            	\begin{eqnarray}	
            	&& \mathcal{M}_{A}^{\mu\nu} = \frac{C_{K_{1}}}{g_{K_{1}}} (m_{s} + m_{u}) I_{11}^{K\omega K_{1}} \nonumber\\
            	&& \times \left\{\left[g^{\mu\nu}\left[q^{2} - \frac{3}{2}(m_{s} + m_{u})^{2}\right] - q^{\mu}q^{\nu}\left[1 - \frac{3}{2}\frac{(m_{s} + m_{u})^{2}}{M_{K_{1(1270)}}^{2}}\right]\right]BW_{K_{1(1270)}}\sin^{2}{\beta} \right. \nonumber\\
            	&& \left. + \left[g^{\mu\nu}\left[q^{2} - \frac{3}{2}(m_{s} + m_{u})^{2}\right] - q^{\mu}q^{\nu}\left[1 - \frac{3}{2}\frac{(m_{s} + m_{u})^{2}}{M_{K_{1(1400)}}^{2}}\right]\right]BW_{K_{1(1400)}}\cos^{2}{\beta}\right\}, \nonumber
            	\end{eqnarray}
            	\begin{eqnarray}
            	&& \mathcal{M}_{V}^{\mu\nu} = i 2 m_{u} \frac{C_{K^{*}}}{g_{K^{*}}} \left[I_{21}^{K\omega K^{*}} + (m_{s} - m_{u}) m_{u} I_{31}^{K\omega K^{*}}\right] \left[q^{2} - \frac{3}{2}(m_{s} - m_{u})^{2}\right] BW_{K^{*}} \varepsilon^{\mu\nu\lambda\delta} p_{K\lambda} p_{\omega\delta}, \nonumber
            	\end{eqnarray}
            	\begin{eqnarray}
            	&& \mathcal{M}_{P}^{\mu\nu} = 2 (m_{s} + m_{u}) \frac{Z_{K}}{g_{K}} C_{K} I_{11}^{\omega K K} q^{\mu}q^{\nu} BW_{K}, \nonumber\\
            	&& \mathcal{M}_{\hat{A}}^{\mu\nu} = \frac{C_{\hat{K}_{1}}}{g_{K_{1}}} (m_{s} + m_{u}) I_{11}^{K\omega \hat{K}_{1}} \left\{g^{\mu\nu}\left[q^{2} - \frac{3}{2}(m_{s} + m_{u})^{2}\right] - q^{\mu}q^{\nu}\left[1 - \frac{3}{2}\frac{(m_{s} + m_{u})^{2}}{M_{K_{1}(1650)}^{2}}\right]\right\} BW_{\hat{K}_{1}}, \nonumber
            	\end{eqnarray}
            	\begin{eqnarray}
            	&& \mathcal{M}_{\hat{V}}^{\mu\nu} = i 2 m_{u} \frac{C_{\hat{K}^{*}}}{g_{K^{*}}} \left[I_{21}^{K\omega \hat{K}^{*}} + (m_{s} - m_{u}) m_{u} I_{31}^{K\omega \hat{K}^{*}}\right] \left[q^{2} - \frac{3}{2}(m_{s} - m_{u})^{2}\right] BW_{\hat{K}^{*}}  \varepsilon^{\mu\nu\lambda\delta} p_{K\lambda}p_{\omega\delta}, \nonumber
            	\end{eqnarray}
            	\begin{eqnarray}
            	&& \mathcal{M}_{\hat{P}}^{\mu\nu} = 2 (m_{s} + m_{u}) \frac{Z_{K}}{g_{K}} C_{\hat{K}} I_{11}^{K\omega \hat{K}} q^{\mu}q^{\nu} BW_{K}. 
            	\end{eqnarray}
	
Here the contribution from the contact diagram contains the axial-vector and vector parts. The transition constants of the $W$ boson to the intermediate mesons $C_M$ and $C_{\hat{M}} $ are defined above. The integrals containing vertices from the quark-meson interaction Lagrangian of the extended NJL model are defined in \ref{integral_ext}. Intermediate mesons are described by the Breit-Wigner propagators. The mixing angle ($\beta$) of the mesons $K_{1}(1270)$ and $K_{1}(1400)$ is defined in Section \ref{NJL_U3U3}.

The decay amplitude $\tau \to \phi K \nu_\tau$ is obtained from (\ref{amplitude_o_k}) by replacing the corresponding vertices $\omega \to \phi$ in loop integrals, the mass of the light quark $m_u$ is replaced by the mass $m_s$ in the vector channel and vector part of the contact diagram; an additional factor of 2 also appears. 

\begin{figure}[h]
\center{\includegraphics[scale = 1.0]{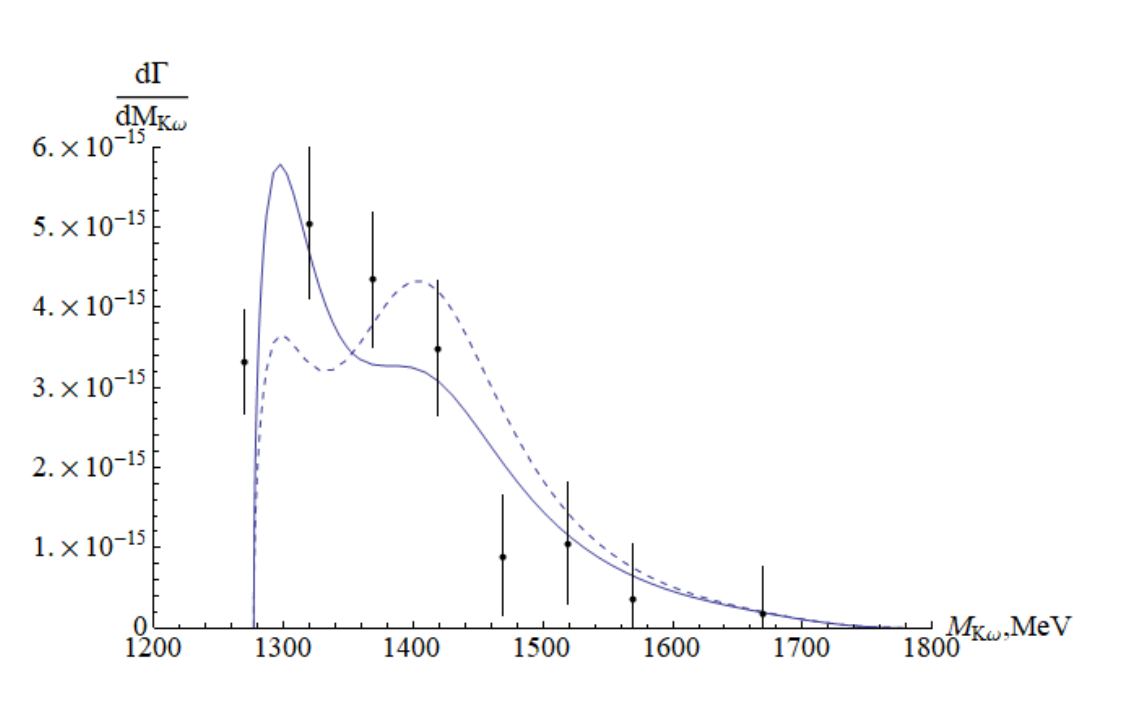}}
\caption{ The differential decay width for the process $\tau \rightarrow \omega K \nu_\tau$. The solid line corresponds to the case $\beta=57^\circ$, the dashed line corresponds to the case $\beta=45^\circ$, the experimental points are taken from \cite{CLEO:2005qyl}.}
\label{width_omega}
\end{figure}

\begin{figure}[h]
\center{\includegraphics[scale = 1.0]{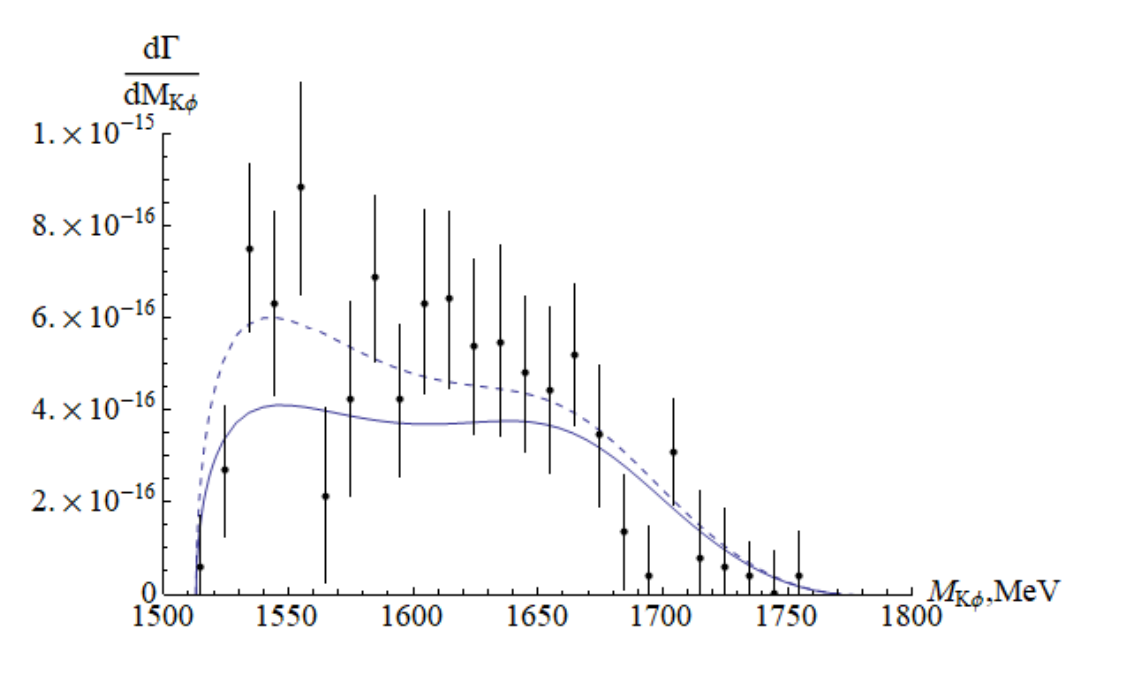}}
\caption{The differential decay width for the process $\tau \rightarrow \omega K \nu_\tau$. The solid line corresponds to the case $\beta=57^\circ$, the dashed line corresponds to the case $\beta=45^\circ$, the experimental points are taken from \cite{Belle:2006enh} }
\label{width_phi}
\end{figure}

There are different ways to choose the mixing angle $\beta$. The PDG gives the value $45^\circ$ \cite{ParticleDataGroup:2020ssz}. At the same time, in the work \cite{Volkov:1984gqw}, the value of $57^\circ$ was obtained in the NJL model. Therefore, here we present the results for the partial and differential decay widths depending on two values of the mixing angle $45^\circ$ and $57^\circ$. The results obtained for the branching fractions and differential decay widths are given in the Table \ref{tab_4} and Figures \ref{width_omega} and \ref{width_phi}.   	

\begin{table}[h!]
\begin{center}
\begin{tabular}{ccccc}
\hline
  & \multicolumn{4}{c}{\textbf{Br($\times 10^{-4}$)}} \\
 \hline
    & \multicolumn{2}{c}{\textbf{$\tau \to \omega(782) K \nu_{\tau}$}}  & \multicolumn{2}{c}{\textbf{$\tau \to \phi(1020) K \nu_{\tau}$}} \\
 \hline
    & $\beta = 57^{\circ}$ & $\beta = 45^{\circ}$ & $\beta = 57^{\circ}$ & $\beta = 45^{\circ}$ \\
\hline
  W$_{A}$	 	 & 0.54 				& 0.52				   & 2.02				  & 1.97  \\
   A        	 & 4.73 				& 4.47 				   & 7.49				  & 8.59  \\
  W$_{A}$ + A	 & 3.08 				& 3.27				   & 2.12				  & 2.94  \\
  W$_{V}$	 	 & 0.66 				& 0.64				   & 0.84				  & 0.82  \\
     V        	 & 1.99 				& 1.94         		   & 0.89				  & 0.86	\\
    W$_{V}$ + V	 & 0.37 				& 0.36         		   & 2.6 $\times 10^{-3}$ & 2.5 $\times 10^{-3}$	\\
      P        	 & 0.57 				& 0.53 				   & 0.66				  & 0.61    \\
    Ground		 & 3.83 				& 3.96 				   & 2.57				  & 3.34    \\
 $\hat{\textrm{A}}$   	 & 5.4 $\times 10^{-3}$ & 5.2 $\times 10^{-3}$ & 0.74				  & 0.72  \\
  $\hat{\textrm{V}}$   	 & 0.31 				& 0.3 				   & 18.1 $\times 10^{-3}$& 17.7 $\times 10^{-3}$  \\
  $\hat{\textrm{P}}$   	 & 8.6 $\times 10^{-4}$ & 6.1 $\times 10^{-4}$ & 1.6 $\times 10^{-4}$ & 1.2 $\times 10^{-4}$ \\
     Excited 	 & 0.32 				& 0.31 				   & 0.76				  & 0.74  \\
\hline
   Total      	 & 3.79 				& 3.95 				   & 3.15				  & 4.04    \\
\hline 
   Experiment & \multicolumn{2}{c}{\textbf{$4.1 \pm 0.9$ \cite{ParticleDataGroup:2020ssz}}}  & \multicolumn{2}{c}{\textbf{$4.4 \pm 1.6$ \cite{ParticleDataGroup:2020ssz}}} \\
                      &  \multicolumn{2}{c}{}    & \multicolumn{2}{c}{\textbf{$4.05 \pm 0.51$ \cite{Belle:2006enh}}} \\
                      &   \multicolumn{2}{c}{}   & \multicolumn{2}{c}{\textbf{$3.39 \pm 0.48$ \cite{BaBar:2007chl}}} \\
\hline 
\end{tabular}
\end{center}
\caption{Predictions of the extended NJL model for the branching fractions of $\tau \to [\omega, \phi] K \nu_\tau$. The contributions of different channels are given with different lines. The lines $W_{A}$ and $W_{V}$ correspond to the axial-vector and vector parts of the contact channel. The Ground line contains the summed results of all channels with intermediate mesons in the ground state and contact channel. The Excited line contains the results for the contributions of all excited intermediate mesons.}
\label{tab_4}
\end{table}

It is interesting to compare the results obtained at different values of the mixing angle $\beta$. The obtained branching fractions of the process $\tau \to \omega K \nu_\tau$ for $\beta = 57^\circ$ and $\beta = 45^\circ$ are within the errors of experimental values. The obtained branching fractions of the process $\tau \to \phi K \nu_\tau$ with different mixing angles agree with different experimental results. Note that the value $\beta = 57^\circ$ leads to better agreement of the form of the invariant mass distribution of the process $\tau \to \omega K \nu_\tau$ with the experimental points. However, the invariant mass distribution of the decay $\tau \to \phi K \nu_\tau$ with $\beta=45^\circ$ is in better agreement with the experimental data. In any case, our results are consistent with the experimental data with allowance for the precision of the model which is expected to be near $17\%$.

Note that from a theoretical point of view the $\tau \to [\omega, \phi] K \nu_\tau$ decays were described using Resonant Chiral Theories and angular momentum algebra in \cite{Guo:2008sh, Dai:2018thd}. 
	
 \subsubsection{The decay $\tau \to K^* \eta \nu_\tau$} 
In the NJL model, the process $\tau \to K^* \eta \nu_\tau$ was described in the work \cite{Volkov:2019izp}. The decisive role in this process is played by the axial-vector channel with the intermediate $K_1(1270)$ and $K_1(1400)$ mesons. To describe the considered decay, we will use the quark-meson Lagrangians of the standard NJL model. Note that due to the participation of the $\eta$ meson in this decay, it is necessary to take into account the influence of gluon anomalies in order to correctly describe the $\eta$ and $\eta'$ meson masses. This problem was described in the Section \ref{tHooft} of this review.

\begin{figure}[h]
\center{\includegraphics[scale = 0.3]{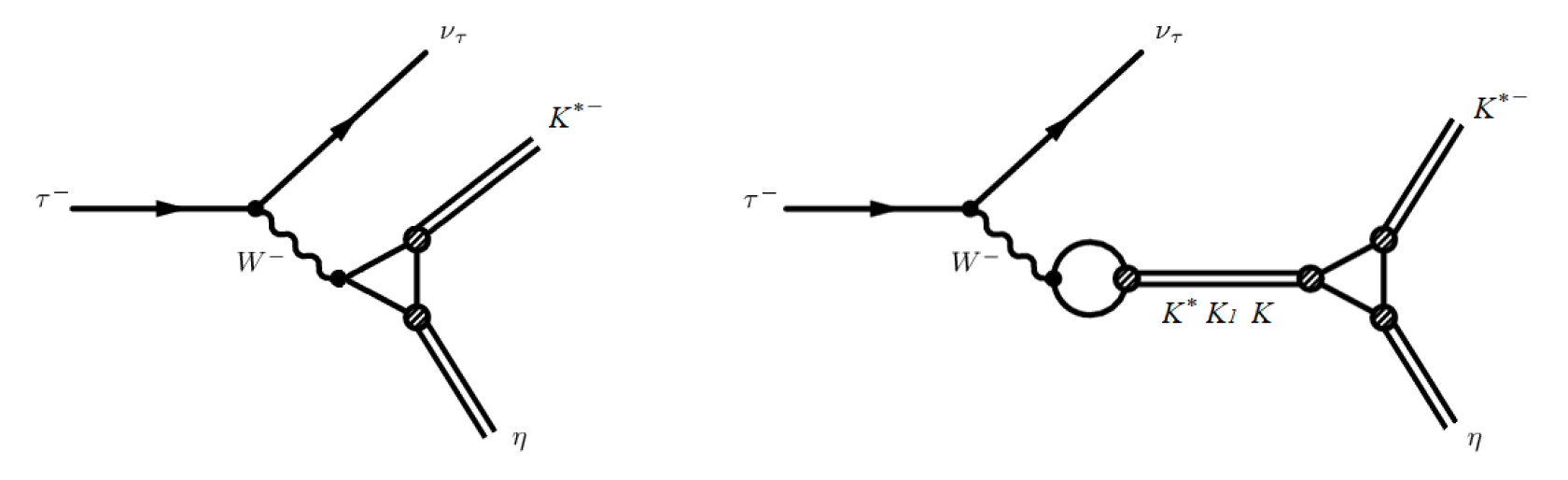}}
\caption{Contact diagram and diagram with intermediate mesons of the decay $\tau \to K^* \eta \nu_\tau$}
\label{kveta_1}
\end{figure}

Diagrams, describing the decay of $\tau \to K^* \eta \nu_\tau$ are shown in Figure \ref{kveta_1}.

As a result, for the total decay amplitude in the NJL model, we obtain 
            	\begin{eqnarray}
            	&& \mathcal{M}(\tau \to K^* \eta \nu_\tau)  =  2i G_{F} V_{us} L_{\mu} \biggl[ \mathcal{M}_{c} + \mathcal{M}_{A(1270)} \nonumber \\ && \qquad 
            	+ \mathcal{M}_{A(1400)} + \mathcal{M}_{V} + \mathcal{M}_{P} \biggl]^{\mu\nu} e_{\nu}^{*}(p_{K^{*}}) ,
            	\end{eqnarray}
where $e_{\nu}^{*}(p_{K^{*}})$ is the polarizations vector of the vector meson $K^{*}(892)$. In the square brackets, the contributions of separate channels are given, which have the form
            \begin{eqnarray}
            	&& \mathcal{M}_{c}^{\mu\nu}  =  \frac{3}{2 g_{K^{*}}}\left( m_{s} g_{\eta_{u}}\sin(\bar{\alpha}) + \sqrt{2} m_{u}g_{\eta_{s}}\cos(\bar{\alpha}) \right) g^{\mu\nu} \nonumber \\ && \qquad
            	- i \biggl[ m_{u} g_{K^{*}} g_{\eta_{u}}\sin(\bar{\alpha}) \left[I_{21} + m_{u} (m_{s} - m_{u}) I_{31}\right] \\ && \qquad \nonumber
                     - \sqrt{2} m_{s} g_{K^{*}} g_{\eta_{s}}\cos(\bar{\alpha}) \left[I_{12} - m_{s} (m_{s} - m_{u}) I_{13}\right] \biggl] \times \varepsilon^{\mu\nu\lambda\delta} p_{\eta\lambda} p_{K^{*}\delta},
            	\end{eqnarray}	
            	\begin{eqnarray}
            	&& \mathcal{M}_{A(1270)}^{\mu\nu}  =  \frac{3}{2 g_{K^{*}}}  \left(m_{s}g_{\eta_{u}}\sin(\bar{\alpha}) + \sqrt{2} m_{u}g_{\eta_{s}}\cos(\bar{\alpha}) \right) \nonumber \\ && \qquad 
            	\times \biggl[ g^{\mu\nu}\left[q^{2} - \frac{3}{2}(m_{s} + m_{u})^{2}\right] - q^{\mu}q^{\nu}\biggl] 
            	\times BW_{K_{1}(1270)} \sin^{2}(\beta),	
            	\end{eqnarray}
            	\begin{eqnarray}
            	&& \mathcal{M}_{A(1400)}^{\mu\nu}  =  \frac{3}{2 g_{K^{*}}}  \left(m_{s}g_{\eta_{u}}\sin(\bar{\alpha}) + \sqrt{2} m_{u}g_{\eta_{s}}\cos(\bar{\alpha})\right) \nonumber \\ && \qquad 
            	\times \biggl[ g^{\mu\nu}\left[q^{2} - \frac{3}{2}(m_{s} + m_{u})^{2}\right] - q^{\mu}q^{\nu}\biggl] 
            	\times BW_{K_{1}(1400)} \cos^{2}(\beta), 
            	\end{eqnarray}
            	\begin{eqnarray}
            	&& \mathcal{M}_{V}^{\mu\nu}  =  -i g_{K^{*}} \biggl[ m_{u} g_{\eta_{u}}\sin(\bar{\alpha}) \left[I_{21} + m_{u} (m_{s} - m_{u}) I_{31}\right] \nonumber  \\ && \qquad 
            	- \sqrt{2} m_{s} g_{\eta_{s}}\cos(\bar{\alpha}) \left[I_{12} - m_{s} (m_{s} - m_{u}) I_{13}\right]\biggl] \nonumber \\ && \qquad 
            	\times \biggl[ g^{\mu\xi}\left[ q^{2} - \frac{3}{2}(m_{s} - m_{u})^{2}\right] - q^{\mu}q^{\xi} \biggl]
            	\times BW_{K^{*}} \varepsilon_{\xi\zeta\lambda\delta} p_{\eta}^{\lambda} p_{K^{*}}^{\delta}g^{\zeta\nu},
            	\end{eqnarray}
            	\begin{eqnarray}
            	&& \mathcal{M}_{P}^{\mu\nu}  =  -\frac{3}{2 g_{K^{*}}}(m_{s} + m_{u}) Z_{K} \left(g_{\eta_{u}}\sin(\bar{\alpha}) + \sqrt{2}g_{\eta_{s}}\cos(\bar{\alpha}) \right) \nonumber \\ && \qquad 
                           \times \biggl[ 1 - \frac{3}{2} (m_{s} + m_{u})^{2} \left(\frac{\sin^{2}(\beta)}{M_{K_{1}(1270)}^{2}} + \frac{\cos^{2}(\beta)}{M_{K_{1}(1400)}^{2}}\right) - \frac{3}{2}(m_{s} + m_{u}) \nonumber \\  && 
                           \times \frac{m_{s} g_{\eta_{u}}\sin(\bar{\alpha}) + \sqrt{2} m_{u} g_{\eta_{s}}\cos(\bar{\alpha})}{g_{\eta_{u}}\sin(\bar{\alpha}) + \sqrt{2} g_{\eta_{s}}\cos(\bar{\alpha})} 
                            \times \left(\frac{\sin^{2}(\beta)}{M_{K_{1}(1270)}^{2}} + \frac{\cos^{2}(\beta)}{M_{K_{1}(1400)}^{2}}\right) \biggl] q^{\mu}q^{\nu}BW_{K},
            \end{eqnarray} 
where $q$ is the momentum of intermediate mesons; $\beta$ is the mixing angle of the mesons $K_{1}(1270)$ and $K_{1}(1400)$; $p_{K^{*}}$, $p_{\eta}$ are the momenta of $K^{*}$ and $\eta$ mesons, respectively; $\bar{\alpha}$ is the mixing angle of the $\eta$ and $\eta'$ mesons. The values of the loop integrals are taken from \cite{Volkov:2019izp}.  

\begin{table}[h!]
\begin{center}
\begin{tabular}{cc}
\hline
\textbf{Channels}	& \textbf{Br($\tau \to K^* \eta \nu_\tau$) $\times 10^{-4}$}  \\
\hline
$A$  & 1.21 \\
$V$    & $2 \times 10^{-3}$ \\
$P$  & 0.04 \\
\hline
Total & 1.23 \\
Experiment & $1.38 \pm 0.15$ \cite{ParticleDataGroup:2020ssz} \\
\hline
\end{tabular}
\end{center}
\caption{Branching fractions of the decay $\tau \to K^* \eta \nu_\tau$}
\label{tab_kveta} 
\end{table} 

The results of numerical calculations of the branching fractions of the considered decay using the obtained amplitude are given in the Table \ref{tab_kveta}. As we can see, in the definitions of the branching fractions, the dominant contribution is given by the axial-vector channel. The contribution from the axial-vector channel is noticeably enhanced by taking into account the two axial-vector poles $K_{1}(1270)$ and $K_{1}(1400)$. The vector channel gives a small contribution $10^{-3}$ compared to the axial vector channel. The amplitude of the vector channel is orthogonal and does not interfere with other channels. The contributions of the pseudoscalar channel are small and interfere only with the axial vector channel. 


In the paper \cite{Li:1996md}, the $U(3) \times U(3)$ chiral symmetric model and Vector Dominance Model (VDM) were used to describe the decay $\tau \to K^* \eta \nu_\tau$. As a result, using the mixing angle $\alpha = -20^{\circ}$ of the $\eta$ and $\eta'$ mesons, the branching fractions $Br (\tau \to K^* \eta \nu_\tau) = 1.01 \times 10^{-4}$ were obtained.

This process was also studied in the work \cite{Dai:2018thd}. However, it was used there to fix the parameters of the model based on experimental data and calculate other decay modes. 

\subsubsection{The decay $\tau \to K^* K \nu_\tau$} 
\begin{figure}[h]
\center{\includegraphics[scale = 0.3]{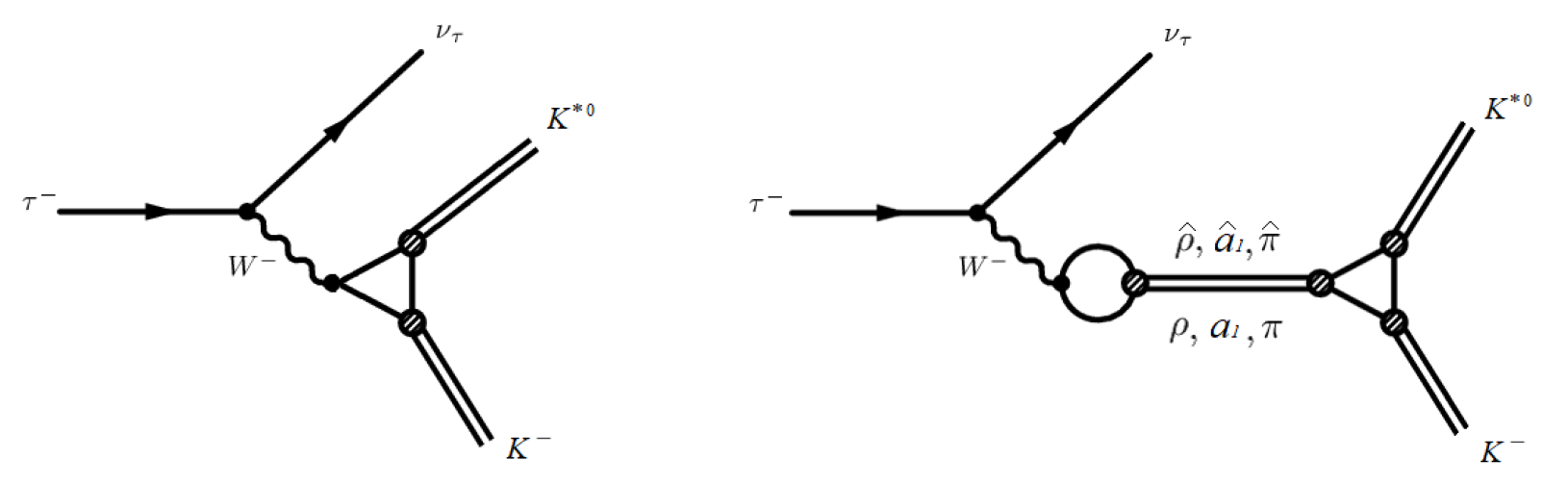}}
\caption{Contact diagram and diagram with intermediate mesons describing the decay $\tau \to K^* K \nu_\tau$}
\label{kvk_1}
\end{figure}

The process $\tau \to K^{*0}(892) K^- \nu_{\tau}$ will be described within the extended NJL model in accordance with the recent paper \cite{Volkov:2019yli}. When describing this process, it is also necessary to consider all four channels. The axial-vector and vector channels play the main role. It is interesting to note that the existing contribution in the vector channel comes from the intermediate radially excited meson $\rho(1450)$. The contact diagram and diagram with intermediate mesons are presented in Figure \ref{kvk_1}. 

The corresponding decay amplitude takes the form 
            	\begin{eqnarray}
            	\label{amplitude_kvk}
            	&& \mathcal{M}  =  i\sqrt{2} G_{F} V_{ud} L_{\mu} \biggl[ \mathcal{M}_{c} + \mathcal{M}_{a_1}  \nonumber 
            	+\mathcal{M}_{\hat{a}_1}  \nonumber \\ && \qquad
            	+ \mathcal{M}_{\rho} +e^{i\pi} \mathcal{M}_{\hat{\rho}} 
            	+ \mathcal{M}_{\pi} + \mathcal{M}_{\hat{\pi}} \biggl]^{\mu\nu} e_{\nu}^{*}(p_{K^{*}}).
            	\end{eqnarray}

The terms in the brackets in the amplitude (\ref{amplitude_kvk}) describe the contributions from the contact diagram and from diagrams with different intermediate mesons in the ground and first radially excited states: 
            	\begin{eqnarray}
                   && \mathcal{M}_{c}^{\mu\nu} = \left(3m_{u} - m_{s} \right)I^{K^{*}K}_{11} g^{\mu\nu} \nonumber\\
            		&& - 2i\biggl[ m_{s}I^{K^{*}K}_{21} -(m_{s} - m_{u})[I^{K^{*}K}_{21} + {m^2_{u}}I^{K^{*}K}_{31}] \biggl] \varepsilon^{\mu\nu\alpha\delta} p_{K\alpha} p_{K^{*}\delta}, 
            		\end{eqnarray}	
            		\begin{eqnarray}
            	  &&	\mathcal{M}_{a_1}^{\mu\nu} = \left(3m_{u} - m_{s}\right) \frac{C_{\rho}}{g_{\rho}}I^{a_{1}K^{*}K}_{11} \biggl[ g^{\mu\nu} \left(q^{2} - 6 m^2_{u}\right) - q^{\mu}q^{\nu}\biggl] BW_{a_{1}}, 
            		\end{eqnarray}	
            		\begin{eqnarray}
            	  &&	\mathcal{M}_{\hat{a}_1}^{\mu\nu} = \left(3m_{u} - m_{s}\right) \frac{C_{\hat{\rho}}}{g_{\rho}}I^{\hat{a}_{1}K^{*}K}_{11} \biggl[ g^{\mu\nu} \left(q^{2} - 6 m^2_{u}\right) - q^{\mu}q^{\nu}\biggl] BW_{\hat{a}_{1}}, 
            		\end{eqnarray}	
            		\begin{eqnarray}
            	  &&	\mathcal{M}_{\rho}^{\mu\nu} = -i2\frac{C_{\rho}}{g_{\rho}} \biggl[ m_{s}I^{\rho K^{*}K}_{12} -(m_{s} - m_{u}) [I^{\rho K^{*}K}_{12} + {m^2_{u}}I^{\rho K^{*}K}_{13}] \biggl] \nonumber\\
            		&& \times \left( g^{\mu\lambda}q^{2} - q^{\mu}q^{\lambda} \right) BW_{\rho} \varepsilon_{\lambda\sigma\alpha\delta} p_{K}^{\alpha} p_{K^{*}}^{\delta} g^{\sigma\nu}, 
            		\end{eqnarray}	
            		\begin{eqnarray}
            	  &&	\mathcal{M}_{\hat{\rho}}^{\mu\nu} = -i2\frac{C_{\hat{\rho}}}{g_{\rho}} \biggl[ m_{s}I^{\hat{\rho} K^{*}K}_{21} -(m_{s} - m_{u}) [I^{\hat{\rho} K^{*}K}_{21} + {m^2_{u}}I^{\hat{\rho} K^{*}K}_{31}] \biggl] \nonumber\\
            		&& \times \left( g^{\mu\lambda}q^{2} - q^{\mu}q^{\lambda} \right) BW_{\hat{\rho}} \varepsilon_{\lambda\sigma\alpha\delta} p_{K}^{\alpha} p_{K^{*}}^{\delta} g^{\sigma\nu}, 
            		\end{eqnarray}	
            		\begin{eqnarray}
            	  && \mathcal{M}_{\pi}^{\mu\nu} = -4 \biggl[ \left( \frac{m_{u} Z_{\pi}}{g_{\pi}} - \frac{6m^2_{u}}{M^2_{a_1}} \frac{C_{\rho}}{g_{\rho}}4m_{u}I^{a_1 \pi}_{20} \right)I^{\pi K^{*}K}_{11} \nonumber\\
            		&& - \frac{m^2_{u} (3m_{u}-m_{s})Z_{\pi}}{M^2_{a_1} g_{\pi}} I^{a_1 \pi}_{20} I^{a_1K^{*}K}_{11} \biggl] q^{\mu}q^{\nu}BW_{\pi}, \\
            	 && \mathcal{M}_{\hat{\pi}}^{\mu\nu} = -4 \frac{m_{u}}{g_{\pi}} Z_{\pi} C_{\hat{\pi}} I^{\hat{\pi} K^{*}K}_{11} q^{\mu}q^{\nu}BW_{\hat{\pi}}.
            	\end{eqnarray}
	
Here, the transition constants of the $W $ boson to the intermediate mesons $C_M$ and $C_{\hat{M}}$ are defined in (\ref{C_const}). Intermediate mesons are described by the Breit-Wigner propagators defined in (\ref{BreitWigner}). 	
	
Unfortunately, the extended NJL model cannot describe the relative phase between the ground and excited states. Therefore, here we will consider two versions of the phase for the $\rho$ and $\hat{\rho}$ mesons: the firstly, version is the $e^{i0}$ phase and the second version is $e^{i\pi}$. The second version can be justified by the results of the experimental work \cite{Achasov:2000wy}.

The obtained results for the branching fractions of $\tau \to K^* K \nu_\tau$ are given in Table \ref{tab_kvk}. 
	
\begin{table}[h!]
\begin{center}
\begin{tabular}{ccc}
\hline
\textbf{Channels}	& \multicolumn{2}{c}{Br($\tau \to K^* K \nu_\tau$) $\times 10^{-3}$} \\
\hline
$A$  & 1.01 & 1.01 \\
$\hat{A}$  & $1.18 \times 10^{-5}$ & $1.18 \times 10^{-5}$ \\
$V$    & $0.32$ & $0.32$  \\
$\hat{V}$    & $0.24$ & $0.24$  \\
$P$  & 0.09 & 0.09 \\
$\hat{P}$  & $3.50 \times 10^{-5}$ & $3.50 \times 10^{-5}$ \\
\hline
The phase & $\phi = 180^\circ$ & $\phi = 0^\circ$\\
\hline
Experiment   & \multicolumn{2}{c}{$2.1 \pm 0.4$ \cite{ParticleDataGroup:2020ssz}}   \\
\hline
\end{tabular}
\end{center}
\caption{Predictions of the NJL model for the branching fractions of $\tau \to K^* K \nu_\tau$}
\label{tab_kvk} 
\end{table} 

It is interesting to note that this decay differs from other $\tau$ lepton decay modes by the dominant contribution of not only axial-vector but also vector channels. 

Note that here it is possible to obtain satisfactory agreement with the experimental data using the phase factor in the $\hat{\rho}$ meson, similarly to how it was done earlier in the paper \cite{Arbuzov:2010xi}. 


Similar calculations of the $\tau \to K^* K \nu_\tau$ decay have been carried out in a number of papers by other authors. In Li's paper \cite{Li:1996md} the calculations were carried out within the $U(3) \times U(3)$ chiral-symmetric model where intermediate mesons were considered only in the ground state. As a result, for the branching fraction of the decay, the value $Br(\tau \to K^{*0}(892) K^- \nu_{\tau}) = 3.92 \times 10^{-3}$ was obtained. The main contribution came from the one vector channel with the intermediate $\rho$ meson.

In the work \cite{Guo:2008sh} theoretical value $Br(\tau \to K^{*0}(892) K^- \nu_{\tau}) = 1.5 \times 10^{-3}$ was obtained using the Chiral Theory with Resonances. Also, the theoretical result for the branching fraction $Br(\tau \to K^{*0}(892) K^- \nu_{\tau}) = 4.93 \times 10^{-3}$, exceeding the experimental value, was obtained in the paper \cite{Dai:2018thd}. 

\section{Conclusions}
\label{NJL_dis}
    There are a large number of different phenomenological models for studying  interactions of mesons at low energies. Chiral Perturbation Theory is especially popular \cite{Gasser:1983yg, Gasser:1984gg}. However, it allows one to describe meson states only in the energy range up to 1 GeV. Chiral Perturbation Theory with Resonances \cite{Ecker:1988te} made it possible to expand the energy range to 2 GeV and take into account the excited meson states. However, the introduction of new states each time leads to the appearance of additional parameters, which reduces the predictive power of the model. 

    The extended NJL model, used in this review, makes it possible to unambiguously describe the coupling constants of radially excited mesons with quarks. In the form factor introduced to describe radially excited mesons, the parameter $c$ affects only the values of the meson masses. The slope parameter is uniquely determined from the condition that the introduction of excited mesons does not affect the quark condensate. After constructing a free Lagrangian containing both ground and first radially excited states, the mixing angles occurring after the diagonalization of this Lagrangian are completely determined. As a result, we obtain the interaction Lagrangian of physical mesons with quarks where coupling constants are uniquely determined. This allows using quark loops to describe the interactions of mesons with each other without introducing additional parameters in the tree level approximation in meson fields corresponding to the lowest order of the $1/N_{c}$ expansion. Namely in this approximation the NJL model is formulated. 
     
 A more complicated situation takes place when describing processes such as the production of mesons on colliding $e^{+}e^{-}$ beams and $\tau$ lepton decays. Here when describing intermediate radially excited mesons, as the experiment shows (see Section \ref{NJL_ee}), it becomes necessary in a number of cases to introduce phase factors. Unfortunately, until now, the NJL model can not describe such factors. Therefore, those relative phase factors between the ground and radially excited intermediate meson states should be considered as additional parameters. 
    
The next problem is related to taking into account the interactions in the final state. In Section \ref{tau_PP}, this was made by considering meson triangle diagrams, leading to the appearance of an additional parameters. Those parameters are not universal and take different values for different processes. However, for structurally similar processes, such parameters turned out to be approximately equal to each other. The main problem with this approach is that these meson triangles are of a higher order in $1/N_{c}$ as compared to the approximation in which the NJL model is formulated. Therefore, such loops can only be considered as additional corrections to the results obtained in the framework of the NJL model. 

It is interesting to note that when considering these effects, a certain correlation appeared between the role of the final state interactions of mesons and excited mesons in the intermediate state. Namely, with an increase in the contribution from the diagrams with the first radial excitations in the intermediate state, the contributions from the corrections taking into account the interactions in the final state decreased. The existence of such a behaviors, as well as a deeper theoretical substantiation of the possibility of taking into account the interactions of mesons in the final state by means of meson loops using the NJL model, can be the subject of further research. 

\subsection*{Acknowledgments}
We are grateful to prof. A.B. Arbuzov for his interest in our work and important remarks which improved the paper. This research has been funded by the Science Committee of the Ministry of Education and Science of the Republic of Kazakhstan (Grant No. AP09057862).

\end{document}